\documentclass[mnsc,nonblindrev]{informs3_nojournal}

\usepackage{endnotes}
\let\footnote=\endnote

\usepackage{natbib}
 \bibpunct[, ]{(}{)}{,}{a}{}{,}%
 \def\bibfont{\small}%
 \def\bibsep{\smallskipamount}%
 \def\bibhang{24pt}%

\usepackage{bbm}

\usepackage[scaled=0.95]{helvet}
\DeclareMathAlphabet{\mathsfit}{T1}{\sfdefault}{\mddefault}{\sldefault}
\SetMathAlphabet{\mathsfit}{bold}{T1}{\sfdefault}{\bfdefault}{\sldefault}
\DeclareMathAlphabet{\mathcal}{OMS}{cmsy}{m}{n}
\usepackage{mathtools,accents}
\usepackage{xifthen,enumitem,comment}
\setlist[enumerate,1]{label=\normalfont{(\Roman*)},leftmargin=2em}
\usepackage{xcolor}
\usepackage{etoolbox}
\makeatletter
\patchcmd{\env@cases}{1.2}{0.96}{}{}
\makeatother
\usepackage{algorithm,algpseudocode}
\usepackage{caption}
\usepackage{subcaption}
\usepackage[hypertexnames=false]{hyperref}
\hypersetup{%
  colorlinks=true,
  linkcolor={blue!66!black},
  linkbordercolor=red,
  citecolor={blue!66!black},
}
 \usepackage{cleveref}
 \hypersetup{
 	colorlinks = true,
 	citecolor={blue!66!black},
 }
\definecolor{myblue}{RGB}{0, 0, 200}
\usepackage{booktabs,tikz}
\usetikzlibrary{matrix,decorations.pathreplacing}

\definecolor{longhorn}{rgb}{0.8, 0.33, 0.0}

\ifx\argmax\undefined\DeclareMathOperator*{\argmax}{arg\,max}\fi
\ifx\argmin\undefined\DeclareMathOperator*{\argmin}{arg\,min}\fi

\newcommand*{\QED}{%
\leavevmode\unskip\penalty9999 \hbox{}\nobreak\hfill
    \quad\hbox{$\square$}%
}

\providecommand{\allzero}{\boldsymbol{0}}

\renewcommand{\P}{\mathbbm{P}}

\newcommand{\hp}{\widehat{p}}

\providecommand{\R}{\mathbbm{R}}
\providecommand{\F}{\mathbbm{F}}

\providecommand{\cB}{\mathcal{B}}

\providecommand{\cF}{\mathcal{F}}

\providecommand{\cL}{\mathcal{L}}

\providecommand{\cP}{\mathcal{P}}

\providecommand{\cR}{\mathcal{R}}

\providecommand{\cV}{\mathcal{V}}

\providecommand{\cPi}{\mathcal{\boldsymbol{\Pi}}}

\providecommand{\Rev}{\mathrm{Rev}}
\providecommand{\CR}{\mathrm{CR}}
\providecommand{\OPT}{\mathsf{OPT}}

\newcommand{\bv}{\boldsymbol{v}}

\newcommand{\bx}{\boldsymbol{x}}

\newcommand{\bpi}{ \boldsymbol{\pi}}
\newcommand{\bPi}{ \boldsymbol{\Pi}}

\newcommand{\of}[1]{\left(#1\right)}

\newcommand{\bigof}[1]{\big(#1\big)}

\newcommand{\bigofff}[1]{\big\{#1\big\}}
\newcommand{\Bigof}[1]{\Big(#1\Big)}

\newcommand{\Bigoff}[1]{\Big[#1\Big]}
\newcommand{\Bigofff}[1]{\Big\{#1\Big\}}
\newcommand{\lv}{\underline{v}}
\newcommand{\uv}{\bar{v}}

\newcommand{\beqn}{\begin{eqnarray*}}
\newcommand{\eeqn}{\end{eqnarray*}}
\newcommand{\beq}{\begin{eqnarray}}
\newcommand{\eeq}{\end{eqnarray}}
\newcommand{\lm}{\lambda}

\newcommand{\lmb}{\boldsymbol{\lambda}}

\newcommand{\lphi}{ \underline{\phi}}
\newcommand{\ba}{\begin{aligned}}
\newcommand{\ea}{\end{aligned}}

\usepackage{cases}
\usepackage{multirow}

\OneAndAHalfSpacedXI

\TheoremsNumberedThrough  
\ECRepeatTheorems

\EquationsNumberedThrough    
\MANUSCRIPTNO{}

\begin{document}

\RUNAUTHOR{Wang, Shixin}
\RUNTITLE{The Power of Simple Menus in Robust Selling Mechanisms}

\TITLE{\Large The Power of Simple Menus in Robust Selling Mechanisms}
\ARTICLEAUTHORS{%
\AUTHOR{Shixin Wang}
\AFF{Department of Decisions, Operations and Technology, CUHK Business School, Chinese University of Hong Kong, \EMAIL{shixinwang@cuhk.edu.hk}} 
}

\ABSTRACT{
We study a robust selling problem where a seller attempts to sell one item to a buyer but is uncertain about the buyer's valuation distribution. Existing literature shows that robust screening provides a stronger theoretical guarantee than robust deterministic pricing, but at the expense of implementation complexity, as it requires a menu of \emph{infinite} options. 
Our research aims to find simple mechanisms to hedge against market ambiguity effectively. We develop a general framework for robust selling mechanisms with a finite menu (or randomization across finite prices). We propose a tractable reformulation that addresses various ambiguity sets of the buyer's valuation distribution, including support, mean, and quantile ambiguity sets. We derive optimal selling mechanisms and corresponding performance ratios for different menu sizes, showing that even a modest menu size can deliver benefits similar to those achieved by the optimal robust mechanism with infinite options, establishing a favorable trade-off between theoretical performance and implementation simplicity. Remarkably, a menu size of merely \emph{two} can significantly enhance the performance ratio compared to deterministic pricing.
}

\maketitle    
\section{Introduction}
Revenue maximization is a fundamental objective in mechanism design. A widely celebrated result in \cite{myerson1981optimal} indicates that when selling one product to one buyer, the optimal selling scheme is to post a take-it-or-leave-it price. This mechanism is elegant and easy to implement when the seller has full knowledge of the buyer's valuation distribution. Nevertheless, the real-world business environment is often more complex, with inherent uncertainties about buyer valuations. Acquiring precise information on a buyer's valuation distribution poses significant operational and financial challenges for a seller.
Sellers often only have limited statistical data on valuations, such as the range of possible valuations (support), average valuation (mean), or the conversion rate at some valuations (quantile). For instance, an e-commerce firm may estimate purchase probabilities at various prices by price experimentation, and then make price decisions armed with this sparse quantile information of the valuation distribution. Consider the following example. 
\begin{example}
   \textit{Suppose the seller knows the customers' valuation is between 0 and 100, and based on the historical price experimentation, they know that at least $60\%$ of the customers are willing to buy the product at a price of 40. Without other information on the valuation distribution, if the seller assumes a three-point valuation distribution with probabilities of $40\%,30\%,30\%$ at $0,40,100$, respectively, a posted price of 100 would be optimal, yielding an expected revenue of 30. However, if the true distribution is $30\%,60\%,10\%$ at these points respectively, the posted price of 100 based on the assumed distribution would only result in a revenue of $100\times 10\% = 10$ in the true valuation distribution, in contrast to the optimal revenue of 28 in the same distribution.}
    \label{example1}
\end{example}
\Cref{example1} illustrates that if the seller has only partial information about the buyer's valuation distribution, it is crucial to consider the robustness of pricing decisions across different valuation distributions. To address this issue, robust mechanism design has relaxed the strong distributional assumptions and proposed attractive robust mechanisms whose performance is less sensitive to the buyer's valuation distribution of the product. A robust mechanism can be implemented by offering a menu of lotteries to the buyer. Each lottery specifies an allocation probability of the product and requests a payment from the buyer. The buyer then selects the lottery that optimizes their utility based on their valuation of the product. Consider the mechanism in the following example.
\begin{example}
\textit{Consider a selling mechanism where the seller offers a menu of two options to the buyer: (a) pay 50 and get the product deterministically, or (b) pay $\frac{100}{3}$ and get the product with a probability of $\frac{5}{6}$. A buyer valuing the product at 100 would choose (a) because their utility for option (a) is $100 -50=50$ while their utility for option (b) is $\frac{5}{6}\times 100 - \frac{100}{3} = 50$. \footnotemark \footnotetext[1]{We assume that the buyer will purchase the product if indifferent between buying and not buying, and will choose the option with the higher probability of obtaining the product if two options are indifferent.} Conversely, a buyer with a valuation of 40 will select (b) since their utility for option (a) is $40-50=-10$ while for option (b), it is $\frac{5}{6}\times 40 - \frac{100}{3} = 0$. Therefore, the seller can earn a revenue of 50 (payment of option (a)) from a buyer with a valuation of 100 and  $\frac{100}{3}$ (payment of option (b)) from a buyer with a valuation of 40. Consider the distributions in \Cref{example1}. The seller achieves a revenue of $\frac{100}{3}\times 30\%+ 50\times 30\% = 25$ for the distribution with probabilities at 0, 40, and 100 being 40\%, 30\%, and 30\%, respectively, and a revenue of $\frac{100}{3}\times 60\%+ 50\times 10\% = 25$ if the distribution is $30\%,60\%,10\%$ at $0,40,100$ respectively.}
    \label{example2}
\end{example}

\Cref{example2} shows that the seller can obtain a more robust selling mechanism by offering a menu with different options to the buyer. 
This type of selling mechanism could be applied in blind box selling \citep{elmachtoub2015retailing,elmachtoub2021power} and loot box pricing \citep{chen2020loot}.
At the same time, this lottery selling mechanism can also be implemented as a randomized pricing mechanism over the range of the buyer's valuation. Consequently, the robust mechanism design problem can be interpreted as optimizing the price density function \citep{chen2023intertemporal}. 
Notice that in our \Cref{example2}, the possible valuation support only has \emph{three} different values: 0, 40, and 100, so the optimal robust selling mechanism only has two different options in the menu. In practice, the possible values of the buyer's valuation can be many or even \emph{infinite}. As the support of possible valuations expands to a continuous scale and as the number of possible distributions increases, the robust mechanism may similarly require infinite options in the menu.
As illustrated in previous literature on robust mechanism design \citep{bergemann2008pricing,carrasco2018optimal}, the optimal robust mechanism can be conceptualized as a \emph{continuous} menu of lottery options.

While the robust mechanism studied in the current literature has wide applications and appealing theoretical performances, its practical implementation poses three significant challenges. The first challenge stems from the complexity associated with managing infinite options in the menu or conducting randomization among an infinite set of prices. This intricate allocation requires sophisticated systems and processes that may be beyond the operational scope of many businesses. Furthermore, the continuum options in the menu imply a wide range of prices and corresponding allocations, which could complicate inventory management, demand forecasting, and other operational decisions. 
The second challenge arises from the potential consumer confusion engendered by the complexity of the menu. When presented with an overwhelming number of options, customers may experience decision paralysis. 
The third concern is that an extensive menu may lead to excessive discrimination among buyers, potentially raising fairness issues \citep{chen2023intertemporal,jin2020tight,cohen2022price}.
Consequently, these challenges have often impelled practitioners to endorse simpler mechanisms, even if they are suboptimal. 

Motivated by the trade-off between theoretical performance and practical implementation, in this paper, we aim to answer the following research questions: 
\textit{\begin{enumerate}
    \item Is it possible to develop simple selling mechanisms that offer a limited number of options yet maintain a strong performance guarantee? How effective are simple mechanisms compared to the optimal robust mechanism with a continuous menu?
    \item What is the optimal robust selling mechanism with a limited number of options in the menu? In other words, if the seller seeks to adopt a randomized pricing strategy over only a limited number of prices, what should be the optimal price levels and their corresponding probabilities?
\end{enumerate}}
To address these questions, we propose a general framework to investigate the effectiveness of simple mechanisms in robust mechanism design. We measure the complexity of a mechanism by its \emph{menu size}, defined as the number of possible outcomes of the mechanism, where an outcome consists of a probability of allocating the product and a payment from the buyer. Since a mechanism with a menu size $n$ can be implemented as a pricing mechanism that randomizes over only $n$ distinct prices, in the following, we refer to \emph{$n$-level pricing} as the selling mechanisms with menu size $n$, and $\infty$-level pricing as the optimal robust mechanism without menu-size constraints. 
In this \emph{$n$-level pricing} problem, the seller aims to find the optimal $n$ prices and the corresponding pricing probabilities, hedging against the most adversarial distribution that nature may choose from some given ambiguity set. We adopt the objective of competitive ratio, in which the performance of a mechanism is measured relative to the hindsight optimal expected revenue that a clairvoyant may obtain with full information on the buyer's valuation distribution. 

This $n$-level robust pricing problem is generally hard to solve due to the following reasons. First, in the $n$-level pricing problem where $n$ is finite, the seller's action set is nonconvex, which makes the problem more challenging compared to the robust mechanism design problem with continuous pricing, i.e., $n=\infty$. In particular, the $n$-level pricing problem involves two types of decision variables: the selection of possible prices and the probabilities allocated to each price level. 
For the robust mechanism design with continuous pricing, i.e., $n=\infty$, since the possible prices cover the entire feasible valuation set, the decision variable contains only the price density function, so the problem is convex. As for deterministic robust pricing, i.e., $n=1$, the price density function converges to a single point, and thus the decision variable reduces to a scalar variable representing the price. However, for the $n$-level pricing scheme, where $n\ge 2$, the decision variables involve both the price levels and the corresponding probabilities, interacting in a nonlinear manner in the objective, making the problem non-convex. Therefore, the solution approaches for $n=1$ and $n=\infty$ do not work for general finite $n$.
Second, even if the possible price levels are fixed, the maximin ratio objective is nonlinear. The denominator of the objective, representing the clairvoyant's revenue, encompasses a product of two decision variables: nature's selection of the worst-case distribution and the corresponding optimal posted price. Analyzing this nonlinear objective results in an optimization problem with an infinite number of variables and constraints. 

\subsection{Main Results}
Our main contribution is to provide a tractable formulation to the $n$-level pricing problem with fixed $n$-level prices. For given $n$ price levels, we first optimize the pricing probability corresponding to each price level. Then the robust maximin ratio problem for fixed price levels can be formulated into a linear programming problem with an infinite number of decision variables and constraints. However, by leveraging the geometric structure of the expected payment function corresponding to the $n$-level pricing mechanism, we are able to reduce this infinite linear programming to a finite linear programming problem where the number of variables and constraints are linear in the number of price levels. This transformation is based on a certain assumption on the ambiguity set of the valuation distributions, which is compatible with a variety of important and commonly applied ambiguity sets, including support ambiguity set, mean ambiguity set, and quantile ambiguity set.

Utilizing our finite LP formulation, we analytically characterize the optimal mechanism and the corresponding optimal competitive ratio for general $n$-level pricing under the support ambiguity set, for $n=1,2$ under the mean ambiguity set, and for $n=1,2,\infty$ under the quantile ambiguity set. For other ambiguity sets satisfying certain assumptions, one can approximate the optimal selling mechanism and the competitive ratio numerically by solving our finite linear programming problems for given price levels. In \Cref{tab:results}, we summarize our results and existing results on robust screening and pricing with the competitive ratio objective across different ambiguity sets. 
\begin{table}[htbp]
  \centering
  \footnotesize
  \caption{Summary of our and existing results on robust screening and pricing with the maximin ratio objective}
    \begin{tabular}{l|cccc}
    \toprule
    Ambiguity Set & $n=1$   & $n=2$   & General $n$ & $n=\infty$ \\
    \midrule
    \multirow{2}[1]{*}{Lower \&  Upper Bound} &   \multirow{2}[1]{*}{\textemdash }    &     \multirow{2}[1]{*}{\Cref{thm:support} }  &   \multirow{2}[1]{*}{\Cref{thm:support} }   & \cite{eren2010monopoly} \\
         &       &       &       & \cite{ball2009toward} \\
         \hline
    Mean \& Upper Bound &   \Cref{thm:meansupport-1}    &  \Cref{thm:meansupport-2}     &   \Cref{thm:finitelp0}$^\#$   & \cite{wang2024minimax} \\
    \hline
   \multirow{2}[1]{*}{Mean \&  Variance} &    \cite{chen2023distributionally}   & \multirow{2}[1]{*}{\Cref{thm:mean-var}$^*$ }     &       & \cite{wang2024minimax}$^*$ \\
    &   \cite{giannakopoulos2023robust}    &       &       &  \cite{giannakopoulos2023robust}$^*$\\
    \hline
    Quantile \& Regular or MHR &   \cite{allouah2023optimal}    &       &       &  \cite{allouah2023optimal}$^\#$ \\
    \hline
    Quantile \& Upper Bound &  \Cref{thm:quantile-1}     &   \Cref{thm:quantile-2}      &    \Cref{thm:finitelp0}$^\#$   &  \Cref{thm:quantile-inf}  \\
    \hline
    Others Satisfying \Cref{assume:phi} &  &  & \Cref{thm:finitelp0}$^\#$   &  \\
    \bottomrule
    \end{tabular}
  \label{tab:results}
  
     \raggedright{The sign $^*$ indicates a closed-form approximation rather than an optimal solution. The sign $^\#$ indicates tractable numerical approximation by finite linear programming. The sign $-$ suggests a straightforward problem. }
\end{table}

\begin{figure}[htbp]
\centering
\begin{minipage}[t]{0.49\textwidth} 
\caption{$\cR_n/\cR_{\infty}$ for $n=1,2,5$, with Different ${\lv}/{\uv}$ under Support Ambiguity Set}
\captionsetup{justification=centering}
\includegraphics[width=\textwidth,keepaspectratio]{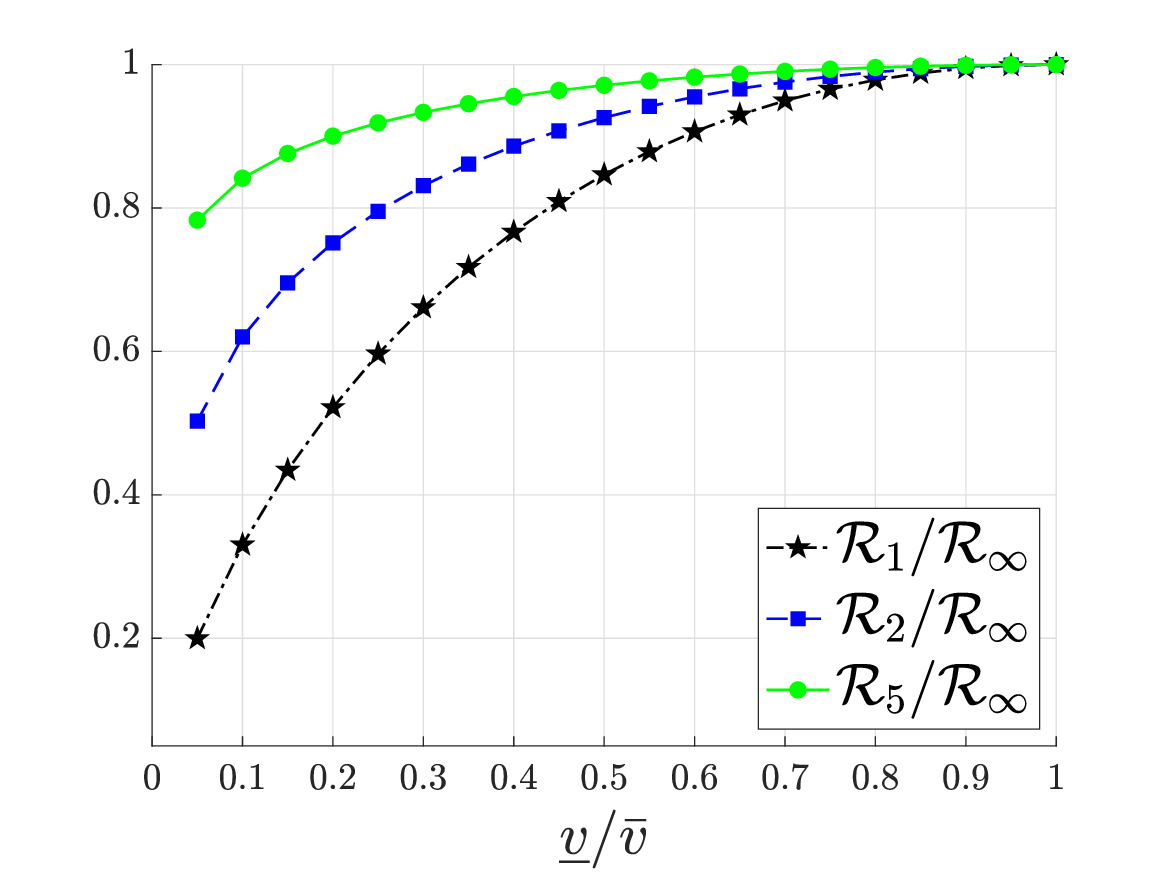}
\label{fig:intro_support}
\end{minipage}
\begin{minipage}[t]{0.49\textwidth}
\caption{$\cR_n/\cR_{\infty}$ for $n=1,2$ with Different $\mu/{\uv}$ under Mean Ambiguity Set}
\includegraphics[width=\textwidth,keepaspectratio]{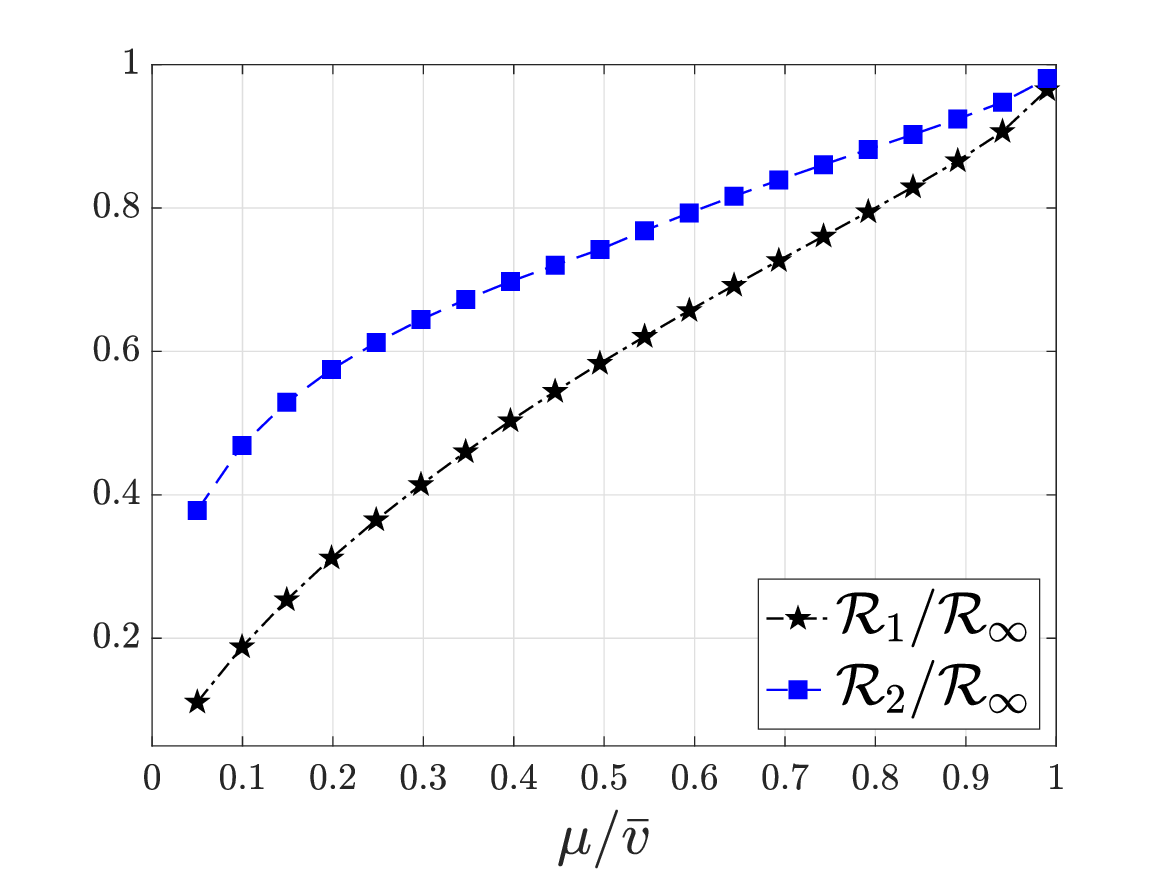}
\label{fig:intro_mean}
\end{minipage}
\end{figure}

\begin{figure}[htbp]
\centering
\captionsetup{justification=centering}
\caption{$\cR_1/\cR_{\infty}$ and $\cR_2/\cR_{\infty}$ Under the Quantile Ambiguity Set}
\label{fig:quantile_intro}
\begin{subfigure}[t]{0.328\textwidth}
\includegraphics[width=\textwidth,keepaspectratio]{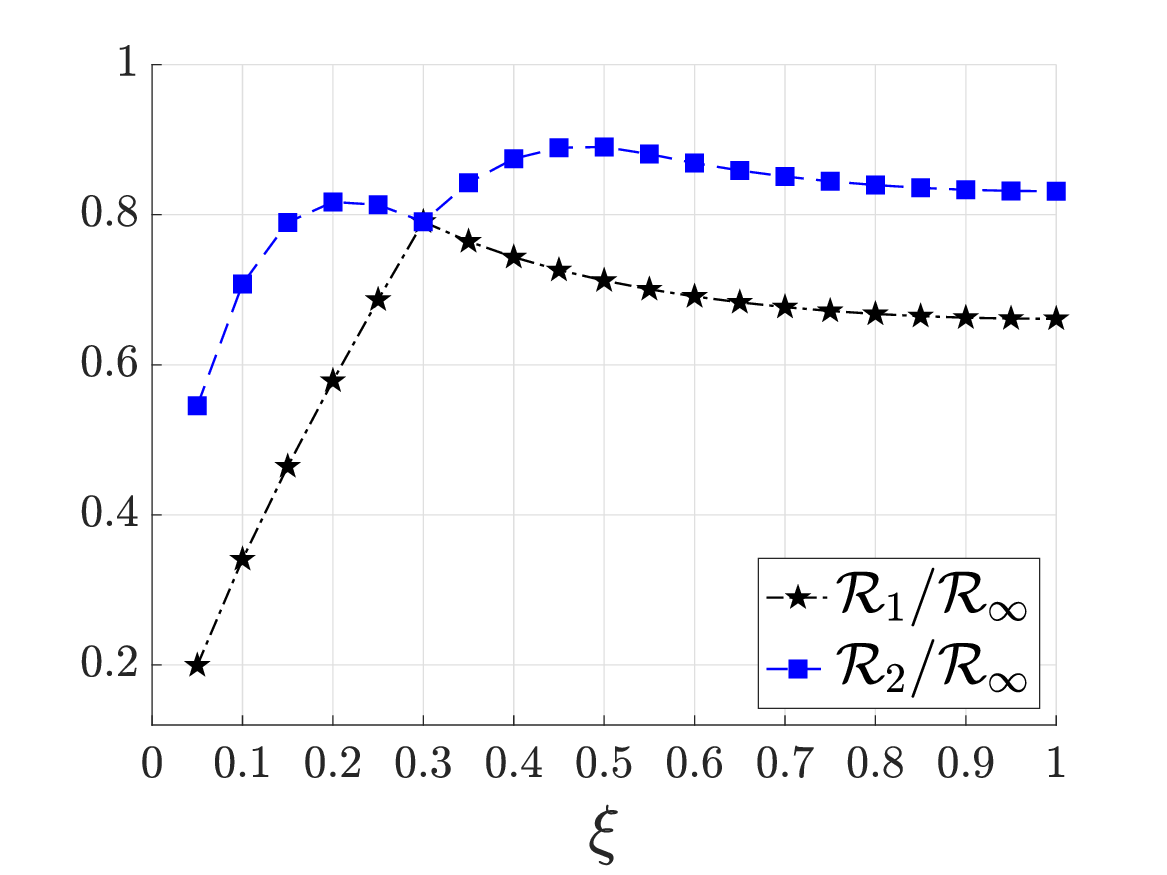}
\caption{$\omega/\uv=0.3$}
\label{fig:quantile-w03}
\end{subfigure}
\begin{subfigure}[t]{0.328\textwidth}
\includegraphics[width=\textwidth,keepaspectratio]{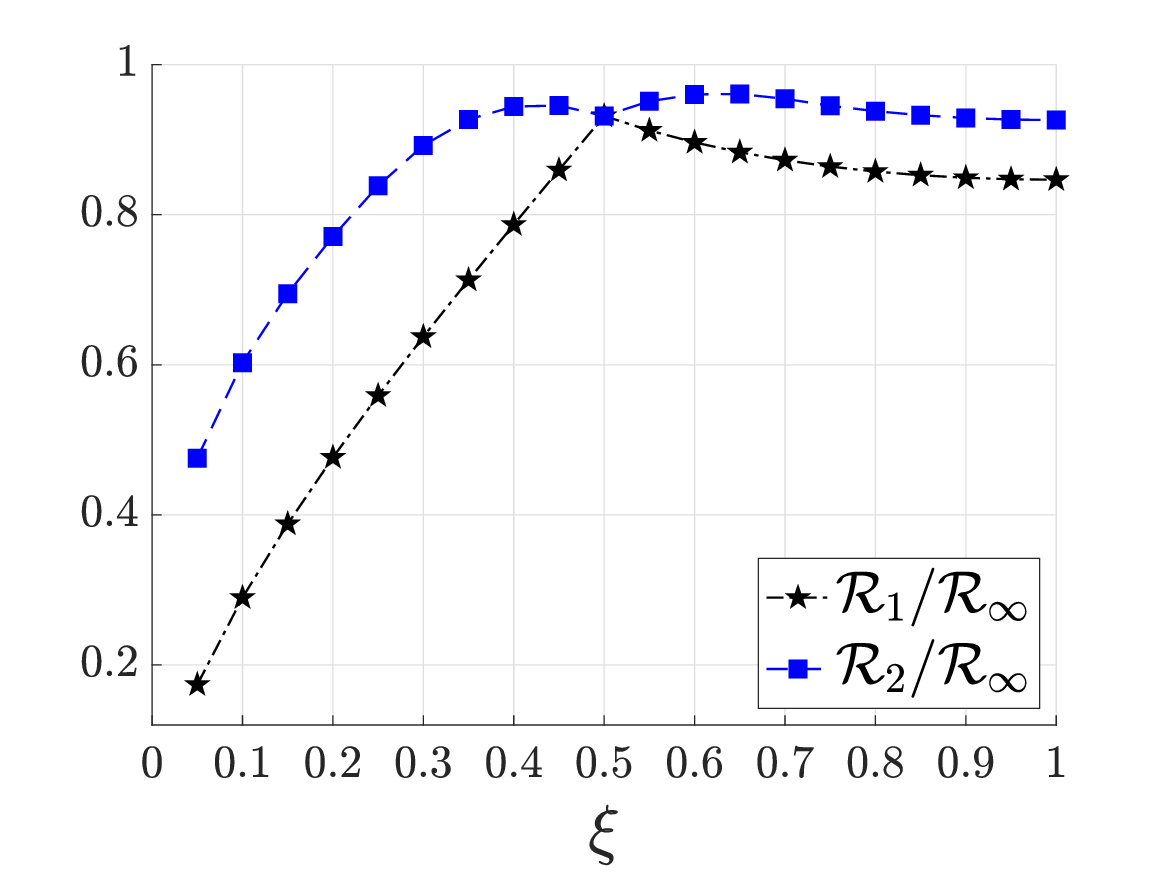}
\caption{$\omega/\uv=0.5$}
\label{fig:quantile-w05}
\end{subfigure}
\begin{subfigure}[t]{0.328\textwidth}
\includegraphics[width=\textwidth,keepaspectratio]{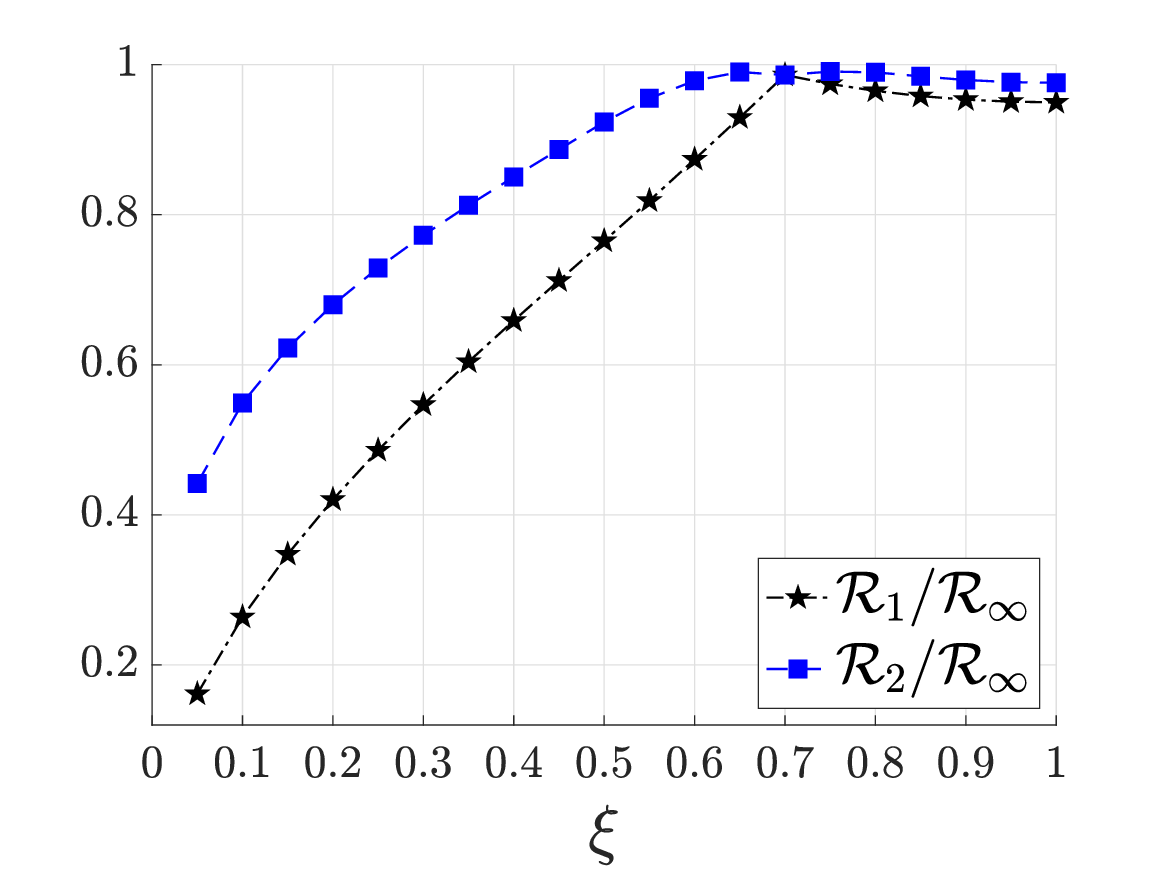}
\caption{$\omega/\uv=0.7$}
\label{fig:quantile-w07}
\end{subfigure}
\end{figure}

 In addition, we investigate the value of incorporating different price levels in various ambiguity sets. 
For the support ambiguity set where the buyer's valuation is assumed to be within $[\lv,\uv]$, 
\Cref{fig:intro_support} compares the competitive ratios for $n$-level pricing against that for $\infty$-level pricing, i.e. $\cR_n/\cR_{\infty}$, where $\cR_n$ stands for the competitive ratio achieved by $n$-level pricing.
For the mean ambiguity set, where the mean of the buyer's valuation is greater than or equal to $\mu$ and the upper bound of the valuation is $\uv$, 
\Cref{fig:intro_mean} depicts $\cR_1/\cR_{\infty}$ and  $\cR_2/\cR_{\infty}$ with $\mu/\uv$ ranging from 0.05 to 1.
For the quantile ambiguity set which suggests that, given the historical sales data or price experimentation, the seller estimates a lower bound of $\xi$ for the probability that the buyer's valuation equals or exceeds $\omega$, \Cref{fig:quantile_intro} illustrates the comparison of the competitive ratios obtained by 1-level and 2-level pricing to that of the $\infty$-level pricing under varying quantile information. 
All the figures show that the $2$-level pricing mechanism improves significantly from 1-level pricing, and reduces the gap between 1-level pricing and $\infty$-level remarkably. Furthermore, for the support ambiguity set, the competitive ratio of $5$-level pricing is already close to that of the optimal $\infty$-level pricing, as long as $\lv/\uv$ is not too small. 
Moreover, \Cref{fig:quantile_intro} shows the performance of 2-level pricing is not far from that of $\infty$-level pricing across various quantile information. Our closed-form characterization of the competitive ratios under different $n$ provides the insight that a menu with a limited number of options (or randomizing over a few prices) achieves much of the benefit from $\infty$-level pricing. To complement our findings, we also provide a closed-form approximation for 2-level pricing under the mean-variance ambiguity set. Even if this is only a feasible rather than optimal solution, the competitive ratio has already improved substantially from the 1-level pricing.

Beyond the maximin ratio objective, our framework extends to general robust metrics, such as maximin revenue and minimax absolute regret. Our findings emphasize the significant effectiveness of 2-level pricing mechanisms across various ambiguity sets and performance metrics. Notably, 2-level pricing strategies are practical and easily implementable. For example, e-commerce platforms can adopt a 2-level pricing strategy by setting a high regular price with intermittent promotions, where the duration and extent of the promotion can be based on historical price experimentation or market research. Our research could serve as a stepping stone for future research to explore the value of simple menus in robust mechanism design.

\subsection{Related Literature}
Our work belongs to the literature on robust mechanism design, focusing on maximizing worst-case revenue under different ambiguity sets. 
 \cite{bergemann2011robust} studies an ambiguity set under the Prohorov metric. 
 \cite{carrasco2018optimal} investigate the optimal robust selling mechanism under moments information, showing that it outperforms optimal deterministic pricing in terms of worst-case revenue. \cite{pinar2017robust} address moment information with discrete support and discuss the challenges of implementing the optimal mechanism compared to simpler deterministic pricing.
\cite{li2019revenue} study the Wasserstein ambiguity set for both single-buyer and multi-buyer problems. \cite{chen2024screening} introduce a geometric approach based on strong duality to solve the robust screening problem in various ambiguity sets. 
Beyond the maximin revenue criterion,
\cite{bergemann2008pricing} and \cite{ eren2010monopoly} study robust mechanism design with support information under the minimax absolute regret and maximin ratio criterion, respectively. \cite{caldentey2017intertemporal} consider dynamic pricing with support ambiguity under the minimax absolute regret. \cite{wang2024minimax} study robust screening under the moments and support information under the maximin ratio criterion. In multi-item setting, robust mechanisms are studied under support information \citep{bandi2014optimal}, marginal distributions \citep{carroll2017robustness,gravin2018separation}, partial information on marginal distributions \citep{koccyiugit2020distributionally,koccyiugit2022robust}, and moment conditions on the values of subsets of goods \citep{che2021robustly}. In auction settings, robust mechanisms are studied under partial information on joint distributions from the bidders \citep{azar2013parametric,azar2013optimal,fu2019vickrey,allouah2020prior}.

Our work is also closely related to the literature on robust deterministic pricing. 
\cite{chen2022distribution} study robust deterministic pricing under the mean-variance ambiguity set for single- and multi-product problems. \cite{chen2023distributionally} study robust pricing based on asymmetric distribution with semivariance information.
\cite{cohen2021simple} consider robust pricing without knowing the demand curve. 
\cite{elmachtoub2021value} investigate the advantage of personalized pricing over deterministic price. 
\cite{giannakopoulos2023robust} study the optimal deterministic pricing and approximation for screening under mean-variance ambiguity. \cite{allouah2023optimal} characterize the optimal pricing and screening with quantile information under regular and MHR distributions. Another important stream of this literature studies pricing with samples from the valuation distribution
\citep{cole2014sample,dhangwatnotai2015revenue,fu2015randomization,gonczarowski2017efficient,huang2018making,allouah2022pricing, guo2019settling,hu2021targeting}, which investigates the sample complexity to achieve a target competitive ratio.

 Since \cite{myerson1981optimal} established that \emph{simple auction is optimal} given buyers' valuation distribution, extensive research has explored optimal mechanisms for multi-item, multi-bidder problems. These problems often require infinite menu options, prompting a focus on the performance of simple mechanisms \citep{hartline2009simple,hart2013menu,wang2014optimal,babaioff2017menu,cai2017simple,hart2017approximate,hart2019selling,babaioff2020simple,eden2021simple,feng2023simple}. 
 However, the methodology in \emph{optimal} mechanism design with known valuation distribution differs significantly from our \emph{robust} mechanism design, without distribution information. 
In addition, sparse solutions and simple policy structures have been studied across various domains. For example,
\citet{Jordan1995,Bassamboo2010,chou2010,Chou2011,Simchi-Levi2012,Simchi-Levi2015,chou2014performance,WangandZhang2015,bidkhori2016analyzing, shi2019process, wang2022robust, wang2023impact} examine the effectiveness of long chain compared to full flexibility in supply chain. \cite{bassamboo2012little,tsitsiklis2013power, tsitsiklis2017flexible} study the efficiency of sparse structures for queueing systems, while \cite{chen2021scheduling,bergquist2023static} shows the effective of simple priority rules or pricing policies. \cite{besbes2019static, balseiro2023dynamic,elmachtoub2023power} study simple pricing policies, including static and two-price policies in reusable resource allocation. \cite{elmachtoub2023simple} study the effectiveness of simple policies in joint pricing and inventory management.
In our work, we shed light on the effectiveness of a menu with \emph{two} options in the robust selling mechanisms.

\subsection*{Notation} Throughout the paper, $v-$ represents the left-hand limit of $v$, i.e., for any function $f$, $f(v-)=\lim_{v'\uparrow v} f(v')$. This extends to any variable such as $p-,v_i-,\uv -$, representing the respective left-hand limits. For a given set $A$, $\cP(A)$ represents the collection of all probability measures on $A$.
\section{$n$-Level Robust Pricing Problem}
\label{sec:basic formulation}
We consider a monopolist seller selling a product to a risk-neutral, quasi-linear buyer, whose valuation of the product, $v$, falling within $[0, \uv]$, is unknown to the seller. 
By the revelation principle \citep{myerson1981optimal}, the seller can focus on a direct mechanism $(q,t)$ consisting of an allocation rule $q:[0, \uv]\rightarrow[0,1]$ and a payment rule: $t:[0, \uv]\rightarrow \R$, without loss of revenue. 
The feasible mechanisms must satisfy individual rationality (IR) and incentive compatibility (IC) constraints to ensure the buyer is motivated to participate truthfully.
Specifically, the admissible mechanisms are defined as:
$
\left\{ 
(q,t):[0,\uv]\rightarrow [0,1]\times \R \  \right \vert  vq(v)-t(v)\ge 0,\ \forall v\in [0,\uv]; \ 
vq(v)-t(v)\ge vq(v^{\prime})-t(v^{\prime}),\ \forall v,v^{\prime}\in [0, \uv]; \ \, 
t(0)=0 \mbox{ and } q(\cdot) \mbox{ is right-continuous.}\left.\right\}.
$
Accordingly, for a buyer with a valuation of $v$, the seller allocates the product with probability $q(v)$ and requests a payment of $t(v)$. It is shown in mechanism design literature that a direct mechanism is incentive compatible if and only if the allocation rule $q(v)$ is increasing in $v$ and the payment rule is specified by $t(v)=vq(v)-\int_{0}^v q(u)du$ \citep{myerson1981optimal}. This mechanism can be implemented by showing a menu of lotteries to the buyer. Each lottery, indexed by $v$, consists of an allocation probability of $q(v)$ and a requested payment $t(v)$. Moreover, this mechanism is also payoff-equivalent to a randomized price mechanism with a price density function $\bpi$, where the allocation rule $q(v)=\int_0^{v}\pi(u)du$ serves as the cumulative price distribution. The feasible mechanisms can be represented by the set of all price density functions over $[0,\uv]$, where the integral of $\pi(v)$ over $[0,\uv]$ is 1.

The above definition of feasible mechanisms is consistent with the previous literature \citep{myerson1981optimal, carrasco2018optimal, chen2024screening}. In this work, we investigate how the complexity, characterized by the menu size or the number of prices that the seller randomizes over, affects the effectiveness of the robust mechanism. 
In particular, we focus on \emph{$n$-level pricing mechanisms}, defined as the set of feasible mechanisms that offer a finite menu of lotteries with at most $n$ options, or equivalently, that randomize over at most $n$ prices, denoted by $\bPi_n$. 
Formally,
\small \beqn
\cPi_n := \Bigofff{ \bpi:[0,\uv]\rightarrow \R_+ \mid \exists\, v_1\le\dots\le v_n\in [0,\uv], \{x_1,\dots,x_n\}\subseteq [0,1], \mbox{s.t.}\,\sum_{i=1}^n x_i = 1, \mbox{and }\, \pi(v) = \sum_{i=1}^n x_i \delta (v-v_i), \forall v\in [0,\uv]
},
\eeqn \normalsize
where $\delta$ is the Dirac delta function.  Here $n$ can be any positive integer, and it follows that $\bPi_n\subseteq \bPi_{\hat{n}}$ for $\hat{n}\ge n$. Each $\bpi\in \bPi_n$ specifies a price density function that defines a randomized pricing strategy.

For any given buyer's valuation distribution $\F$, the expected revenue achieved by a mechanism $\bpi\in \bPi_n$ is characterized as $\Rev(\bpi,\F)=\int_0^{\uv}t(v)d\F(v)=  \int_{0}^{\uv}(\int_0^v \pi(u)u\, du)\, d\F(v)$, since a buyer with valuation $v$ makes an expected payment of $t(v)=\int_{0}^v \pi(u)u\, du$. 
While the seller may not know $\F$ precisely, it is assumed that $\F$ belongs to a known convex ambiguity set $\cF$. To assess the robustness of a mechanism, we compare its performance against the optimal revenue, achievable by a clairvoyant with complete knowledge of $\F$, denoted as $\Rev(\OPT, \F)=\max_{p\ge 0} \   p(1-\F(p-))$ \citep{riley1983optimal}.
The performance of a mechanism $\bpi\in\bPi_n$ is evaluated by the competitive ratio, defined as the minimum revenue ratio of $\bpi$ to the optimal mechanism across all distributions in $\cF$:
$\CR(\bpi)=\inf_{\F\in \cF}\frac{\Rev(\bpi,\F)}{\Rev(\OPT,\F)}.$
Our goal is to design an $n$-level pricing mechanism that maximizes the competitive ratio against all potential distributions within $\cF$. 
\small \begin{equation}
\sup_{\bpi\in \bPi_n} \CR(\bpi)=
\sup_{\bpi\in \bPi_n}\inf_{\F\in \cF}\frac{\int_{0}^{\uv}(\int_0^v \pi(u)u\, du)\, d\F(v)}{\max_{p\in [0,\uv]} \  p(1-\F(p-))}
\label{original}
\end{equation}\normalsize
Problem \eqref{original} can be formulated as a zero-sum game between the seller and nature. For any selling mechanism that the seller adopts, nature will adversarially choose the worst-case distribution $\F$ from ambiguity set $\cF$ and the corresponding optimal posted price~$p\in[0,\uv]$ so that the competitive ratio is minimized. The ambiguity set we consider takes the following form:
{\small
\beq
\cF=\left\{\F\in \cP([0,\uv]): \,\int_{v\in[0,\uv]}\phi_k(v)\,d\F(v) \ge 0, \ 
\forall k=1,2\ldots,K. \right\}.
\label{worstf}
\eeq}
where $K$ is a given number,  $\cP([0,\uv])$ represents the set of all probability measures on $[0,\uv]$, and $\phi_k$ is a predetermined function of $v$  satisfying the following assumption.
\begin{assumption}
\label{assume:phi}
In the ambiguity set definition given by \eqref{worstf}, each $\phi_k(v)$ is a bounded and piecewise continuous on $[0,\uv]$,  with finite segments and is nondecreasing in $v$ within each segment.
\end{assumption}
\Cref{assume:phi} encompasses several ambiguity sets commonly found in the literature or in practice, including the following examples.
\begin{enumerate}
\item \emph{Support ambiguity set:} If there is only one constraint in \eqref{worstf} that $\phi(v) = \mathbbm{1}[v\ge \lv]-1$, the resulting ambiguity set can be represented as  $\cF=\left\{\F\in \cP([0,\uv]): \, \int_{v=0}^{\uv}  \mathbbm{1}[v\ge \lv]\, d\F(v)\ge 1\right\}$. This corresponds to the support ambiguity set, focusing on distributions where the buyer's valuation falls between $\lv$ and $\uv$. Here $\phi(v) = \mathbbm{1}[v\ge \lv]-1$ is a step function, satisfying \Cref{assume:phi}.
 \item   \emph{Mean ambiguity set:} In this case, $\phi(v)=v-\mu$, which is a linear function of $v$, satisfying \Cref{assume:phi}. The ambiguity set becomes $\cF=\left\{\F\in \cP([0,\uv]): \, \int_{v\in[0,\uv]}v \,d\F(v) \ge \mu\right\}$, where $\mu$ is the estimated mean of the buyer's valuation.
\item \emph{Quantile ambiguity set:} If we define $\phi_k(v)= \mathbbm{1}[v\ge \omega_k]-\xi_k$ where $\omega_k \in [0,\uv]$, the ambiguity set \eqref{worstf} corresponds to the quantile ambiguity set, 
represented as $\cF=\left\{\F\in \cP([0,\uv]): \P[v\ge \omega_k] \ge \xi_k, \
\forall k=1,2, \dots,K\right\}$. This ambiguity set includes distributions where the purchase probability for a posted price $\omega_k$ is no less than $\xi_k$ for all $k=1,\dots, K$. In practice, the seller can estimate these purchase probabilities $\xi_k$ based on historical sales data. 
\item \emph{Multi-Segment ambiguity set:} If we define $\phi_k(v) = \mathbbm{1}[v \in I_k]-\xi_k$, for $k=1,\dots, K$, where $I_k=[a_1^k,b_1^k]\cup [a_2^k,b_2^k]\cup\dots \cup [a_{l_k}^k, b_{l_k}^k]$, with $l_k=1,2,\dots$, the ambiguity set in \eqref{worstf} corresponds to what we call a multi-segment ambiguity set. In this scenario, the seller assumes a lower bound $\xi_k$ on the probability that a buyer's valuation falls within the union of specific intervals, denoted by $I_k$.  
For instance, if $\phi_k(v) = \mathbbm{1}[v \in [a_k,b_k]]-\xi_k$ for all $k=1,\dots, K$, the seller posits that the probability of the buyer's valuation falling within the interval $[a_k,b_k]$ is at least $\xi_k$, for every $k=1,\dots,K$, where intervals $[a_k,b_k]$ can overlap with each other. Here $\phi_k(v) = \mathbbm{1}[v \in I_k]-\xi_k$ is a step function with value $1-\xi_k$ within interval $I_k$ and $-\xi_k$ outside $I_k$, satisfying \Cref{assume:phi}. 
\item \emph{Segmented-Mean ambiguity Set:} Let $\phi_k(v) =\mathbbm{1}[v \in I_k]\cdot (v-\mu_{k})$, where $I_k\subseteq [0,\uv]$ is defined as in the previous example. Here $\phi_k(v)$ is a piecewise linear function of $v$, satisfying \Cref{assume:phi}. 
This ambiguity set suggests the mean of the valuation in market segment $I_k$ is at least $\mu_k$. 
\end{enumerate}
The above illustrations highlight the broad applicability of \Cref{assume:phi}, which is introduced for the tractability of the $n$-level pricing problem. While \Cref{assume:phi} may not encompass certain ambiguity sets, such as the mean-variance set, we believe the main insights of the paper extend to more general ambiguity sets. 
To solve the maximin ratio problem in \eqref{original} under \Cref{assume:phi}, we first transform  \eqref{original} into a maximization problem as per \eqref{primal}. 
This transformation results in a problem formulation with only linear constraints, except for the constraints defined within $\bPi_n$. 
\begin{proposition}
\label{prop:lpformulation_inf}
The $n$-level robust pricing Problem \eqref{original} is equivalent to Problem \eqref{primal}, where the optimal solution $r$ is the optimal competitive ratio, and $\bpi$ is the corresponding optimal price density function.
\small \begin{equation}
\label{primal}
\ba
\displaystyle
\sup\limits_{\lmb \in \R^{K\times [0,\uv]}_+,\ \bpi\in\bPi_n, \ r\in \R} & r \\
 \mathrm{s.t.}\quad  &   \sum_{k=1}^{K}\phi_k(v)\lm_{k}(p)\le \int_0^v \pi(u)u\, du\quad \forall p\in[0,\uv], \ \forall v\in[0,p)  \\
 & r\cdot p+ \sum_{k=1}^{K}\phi_k(v)\lm_{k}(p)\le \int_0^v \pi(u)u\, du\quad \forall p\in[0,\uv], \ \forall v\in[p, \uv]
\ea
\end{equation}\normalsize
\end{proposition}
The proof of \Cref{prop:lpformulation_inf} primarily employs two key techniques: (1) the transformation of the linear fractional programming to handle the hindsight optimal revenue appearing in the denominator, and (2) the application of Sion's Minimax Theorem, which enables the conversion of the adversary's inner minimization problem into a maximization problem. 
\Cref{prop:lpformulation_inf} is an extension of Theorem 1 of \cite{wang2024minimax} from moment ambiguity set to general ambiguity sets satisfying \Cref{assume:phi}.
\cite{wang2024minimax} considers the robust screening problem without price level constraints, so Problem \eqref{primal} becomes an infinite linear programming problem. However, in our problem, the decision variables in $\bPi_n$ involve both the price levels and the corresponding probabilities, interacting in a nonlinear manner in the objective, making the problem non-convex, so Problem \eqref{primal} is generally more difficult to solve. Besides the nonconvexity due to $\bPi_n$, the complexity is compounded by the infinite-dimensional decision variables and the infinite number of constraints. 
In the following, we first fix the set of possible price levels $\cV=\{v_1,\dots,v_n\}$ and further simplify and analyze problem \eqref{primal} based on \Cref{assume:phi}.
Notice that according to \Cref{assume:phi}, each $\phi_k$ is piecewise continuous for $k=1,\dots,K$, so there are finite breakpoints for each $\phi_k(\cdot)$. Hence, the aggregate of all breakpoints in $\phi_k(\cdot)$ for all $k=1,\dots, K$ are also finite. We denote the set of all breakpoints for $\phi_k(\cdot)$'s as $\cB=\{b_1,\dots,b_m\}$, where $0<b_1<\dots<b_m<\uv$.
For any price level set $\cV=\{v_1,\dots,v_n\}$, we arrange all the valuations in $\cB$ and $\cV$ in ascending order, i.e., $u_1\le u_2\le \dots\le u_{m+n}$, where $u_j\in \cV\cup \cB$ for all $j=1,\dots, m+n$. For notational simplicity, we represent $u_{n+m+1}=\uv$.  \Cref{thm:finitelp0} establishes a finite linear programming formulation for the $n$-level pricing problem with fixed price levels. 
Hereafter, when the values of price levels: $v_1,\dots, v_n$ are specified, we denote $\bx=[x_1,x_2,\dots, x_n]$ as the probability associated with each price level.
For further notational clarity, we define $\lphi_{k,j} =\lim_{v\rightarrow u_j-} \phi_k(v)$ for $k=1,\dots,K,\,  j=1,\dots,n+m+1$, and denote $S_{j}=\{l =1,\dots,n\mid v_l<u_j\}$ for $j=1,\dots,n+m+1$. 
\begin{theorem}[Finite-LP]
\label{thm:finitelp0}
Under \Cref{assume:phi},  for any given price level set $\cV$ and breakpoint set $\cB$, the seller's $n$-level pricing problem is equivalent to the following finite linear programming problem, where $x_i$ is the probability that the seller sets price $v_i$ for $i=1,\dots, n$, and $r$ is the corresponding competitive ratio.
\small \begin{equation}
\label{primal-finite0}
\ba
\displaystyle
\sup\limits_{\substack{\lmb \in \R^{K\times (m+n+1)}_+\\ \bx\in\R_+^n, r\in \R^+}} & r \\
 \mathrm{s.t.}\quad  &   \sum_{k=1}^{K}\lphi_{k,j}\lm_{k,i}\le \sum_{l\in S_{j}}v_lx_l, \ \forall \  i\in [n+m+1], \, j\in[i-1]  \\
 & r\cdot u_i+ \sum_{k=1}^{K}\lphi_{k,j}\lm_{k,i}\le \sum_{l\in S_{j}}v_lx_l, \ \forall \  i\in [n+m+1], \, j=i,\dots,n+m+1 \\
 & \sum_{i=1}^n x_i = 1
\ea
\end{equation}\normalsize
\end{theorem}

\Cref{thm:finitelp0} reduces the number of variables and constraints in \eqref{primal} to finite, which simplifies the analysis significantly. In light of \Cref{thm:finitelp0}, we can characterize the performance ratio of robust mechanisms with different price levels (or menu sizes) under some important ambiguity sets that satisfy \Cref{assume:phi}.
Notice that this finite LP differs from previous literature, such as Proposition 1 in \cite{wang2024minimax}.
One critical distinction is that our model permits the seller to assign pricing probabilities within a limited set of valuations, $\cV=\{v_1, v_2, \dots, v_n\}$, which is a proper subset of the whole range $[0, \uv]$ that the adversary, or nature, can exploit. 
This contrasts with the models in the current literature, where both the seller and the adversary are constrained to the same finite set of valuations $\cV$.
For example, Proposition 1 in \cite{wang2024minimax} provides a finite LP to approximate the optimal competitive ratio for the robust screening problem ($\infty$-level pricing) under the moment ambiguity set. This formulation implies that there are in total $n$ different possible valuations $\cV=\{v_1,v_2,\dots,v_n\}$ and assumes that both the seller and adversary operate within the finite set $\cV$. It provides an upper bound of the robust screening problem ($\infty$-level pricing). As $n$ goes to infinity and $\cV$ becomes dense enough within $[0,\uv]$, it approximates well the $\infty$-level pricing, while when $n$ is small, it solves a different problem from ours. 
In our model, though the seller can only randomize among a small finite set $\cV$,  the adversary can assign positive probabilities to arbitrary valuations within $[0,\uv]$. This discrepancy allows the adversary to potentially leverage more adverse distributions against the seller's strategy. Denoting $\bpi$ as the seller's pricing policy and $\Rev(\bpi,\F)$ as the revenue achieved by pricing strategy $\bpi$ under valuation distribution $\F$, the difference between the objective of the finite LP in \Cref{thm:finitelp0} (denoted as $\cR^\dag$) and that of the finite LP in Proposition 1 of \cite{wang2024minimax} (represented as $\cR^\ddag$) is characterized as follows:
{\small
\begin{align*}
\cR^\ddag &= \max_{\bpi\in \cP(\cV)} \inf_{\F\in \cP(\cV)\cap \cF}\frac{\Rev(\bpi,\F)}{\Rev(\OPT,\F)} =   \max_{\bpi\in \cP([0,\uv])} \inf_{\F\in \cP(\cV)\cap \cF}\frac{\Rev(\bpi,\F)}{\Rev(\OPT,\F)}\\ & \ge \max_{\bpi\in \cP([0,\uv])} \inf_{\F\in \cP([0,\uv])\cap \cF}\frac{\Rev(\bpi,\F)}{\Rev(\OPT,\F)}\ge \max_{\bpi\in \cP(\cV)} \inf_{\F\in \cP([0,\uv])\cap \cF}\frac{\Rev(\bpi,\F)}{\Rev(\OPT,\F)}  = \cR^\dag
\end{align*}
}
where $\cF$ represents some ambiguity set of interest. The second equality is because the seller can not increase the performance by pricing outside $\cV$ if the adversary's decision on valuation's support is restricted to $\cV$. The first inequality is because expanding the adversary's feasible set of valuation distribution from $\F\in \cP(\cV)\cap \cF$ to $\F\in \cP([0,\uv])\cap \cF$ can decrease the objective, and the term after the first inequality is exactly the optimal competitive ratio for the $\infty$-level pricing. The second inequality is because the objective can decrease if the seller shrinks the decision space from $\cP([0,\uv])$ to $\cP(\cV)$. Therefore, the linear programming formulation in Proposition 1 in \cite{wang2024minimax} can serve as an upper bound of the $\infty$-level pricing as $n$ goes to infinity, while our problem in \Cref{thm:finitelp0} is a lower bound of $\infty$-level pricing as $n$ goes to infinity. 
In the remaining sections, we will develop tractable simplifications of \eqref{primal-finite0} for different specifications on the ambiguity sets.

\section{Support Ambiguity Set}
\label{sec:support}
\begin{figure}[htb]
\centering
\captionsetup{justification=centering}
\caption{Payment Function $\sum_{j=1}^{i}x_jv_j$ for $x\in[v_i,v_{i+1})$ together with $r\cdot v$}
\begin{subfigure}[t]{0.48\textwidth}
\includegraphics[width=\textwidth,keepaspectratio]{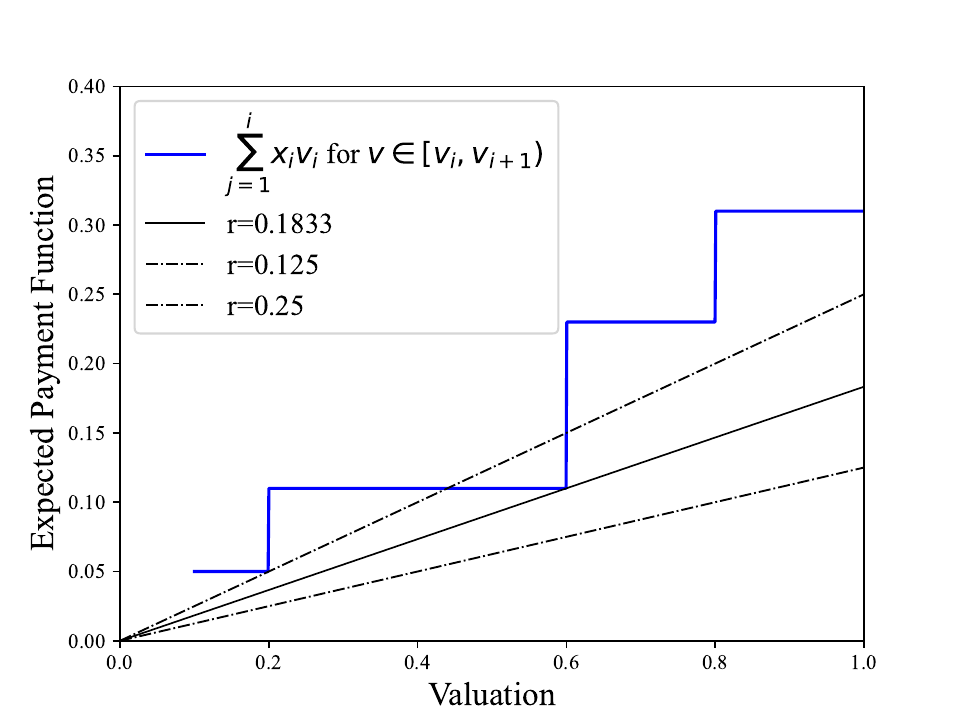}
\caption{Arbitrary $\bx$ for Fixed $\cV$}
\label{fig:stepm1}
\end{subfigure}
\quad
\begin{subfigure}[t]{0.48\textwidth}
\includegraphics[width=\textwidth,keepaspectratio]{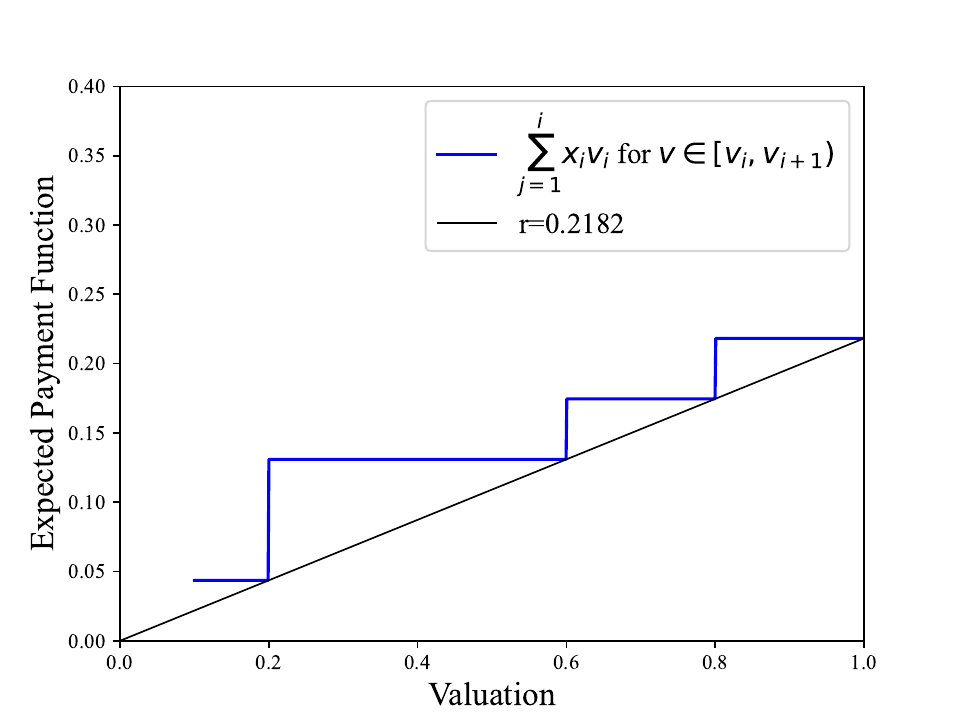}
\caption{Optimal $\bx$ for Fixed $\cV$}
\label{fig:stepm2}
\end{subfigure}
\end{figure}

In this section, we assume the seller knows there is a lower bound on the buyer's valuation, denoted as $\lv$. Hence, the ambiguity set is specified by
$
\cF=\left\{\F\in \cP([0,\uv]): \,\int_{\lv}^{\uv}\,d\F(v) \ge 1\right\}.
$
In this specification, there is only one constraint in $\cF$, which is $\phi(v)=\mathbbm{1}[v\ge \lv]-1$. For ease of reference, we omit the constraint subscript $k=1$. Since the buyer's valuation cannot fall below $\lv$, posting a price lower than $\lv$ is suboptimal, so we can focus on prices at or above $\lv$. Within the domain $v\ge \lv$, the term $\phi(v)$ consistently equals $0$. Thus, the coefficients for $\lmb$ are zero in Problem \eqref{primal-finite0}, allowing us to eliminate the decision variables $\lmb$.  Given price level set $\{v_1,\dots, v_n\}$, Problem \eqref{primal-finite0} is specified as 
\small \begin{subequations}
\begin{align}
\displaystyle
\addtocounter{equation}{-1}
\sup_{\bx\in \R_+^n, r\in \R^+} & \, r \label{eq:supportobj} \\
 \mathrm{s.t.}\quad
 & r\cdot v_i \le \sum_{j=1}^{i-1} x_j\cdot v_j \quad \forall i=2,\dots,n+1. \label{eq:supportcon1}\\
 & r\cdot v_1 \le x_1\cdot v_1\cdot \mathbbm{1}[v_1= \lv] \label{eq:supportcon3}\\
 & \sum_{i=1}^n x_i = 1 \label{eq:supportcon4}
\end{align}
\end{subequations} \normalsize
where $v_{n+1}=\uv$ for notational simplicity.
From constraint \eqref{eq:supportcon3}, we observe that if $v_1 > \lv$, it would result in a non-positive competitive ratio, i.e., $r\le 0$. Thus, we must have $v_1=\lv$. 
\Cref{fig:stepm1} offers a graphical representation of the constraints in Problem \eqref{eq:supportobj} to provide an intuitive understanding. 
As depicted in \Cref{fig:stepm1}, given any feasible $v_1,\dots,v_n$ and $x_1,\dots,x_n$, the optimal feasible $r^*$ represents the largest possible slope of the function $r\cdot v$ which remains below the step function $\sum_{j=1}^{i} x_j\cdot v_j$, where $i=\max\{i': v_{i'}\le v\}$. To maximize $r$, one can adjust $\bx$ so that the breakpoints of the step function align with the ray $r\cdot v$, as illustrated in \Cref{fig:stepm2}. Therefore, for any feasible price level set $\cV=\{v_1,\dots,v_n\}$, where $v_1=\lv$, it is possible to enhance $r$ by adjusting $\bx$ to satisfy constraints \eqref{eq:supportcon1} and \eqref{eq:supportcon3} as equalities. Hence, the optimal $x_1,\dots, x_n$ and $r$ can be solved by 
$r =  \frac{\sum_{j=1}^{i}x_j\cdot v_j}{v_{i+1}},\,\forall \ i=1,\dots,n.
$ \Cref{lemma:fixv-support} formalizes the optimal competitive ratio $r$ and the corresponding pricing probability $\bx$ for a given price level set $\cV$.
\begin{lemma}
\label{lemma:fixv-support}
For any given price level set $\cV=\{v_1,v_2 \ \dots,\ v_n\}$, where $v_1=\lv$, the optimal competitive ratio is $r= \big( \frac{v_2}{v_1} +\sum_{i=2}^n \frac{v_{i+1}-v_{i}}{v_{i}} \big)^{-1}$, and the
optimal pricing probability is determined by $x_1=r \cdot \frac{v_2}{\lv} ,\,  x_i = r\cdot\frac{v_{i+1}-v_{i}}{v_i},\, \forall \ i=2,\dots,n$. 
\end{lemma}

\Cref{lemma:fixv-support} can be intuitively explained as follows. For any given price levels $\cV=\{v_1,\dots,v_n\}$ and seller's pricing policy $x_1,\dots,x_n$, nature's strategy is to focus on valuations just below each $v_i$ (for $i=2,\dots,n$), and identify the $i$ that minimizes the ratio between the seller's revenue and its true valuation, specifically $\frac{\sum_{j=1}^{i-1}v_jx_j}{v_i-}$. To hedge against this adversary, the seller adopts pricing probabilities $x_1,\dots, x_n$ such that the ratio of the expected payment at each $v_i-$ relative to its valuation $v_i-$, remains equal across all $v_i-$, for $i=2,\dots,n$.

Next, based on the optimal competitive ratio and the corresponding pricing probability $\bx$ assigned to each price level defined in \Cref{lemma:fixv-support}, we proceed to optimize the selection of pricing levels $\cV$. In \Cref{thm:support}, we characterize the optimal $n$-level pricing mechanism and the competitive ratio for each $n$. As illustrated in \Cref{thm:support}, the optimal price levels form a geometric sequence, and the optimal pricing probability is equal for each price level except the first one at $\lv$.

\begin{proposition}
\label{thm:support}
 Suppose the buyer's valuation is within $[\lv,\uv]$. Then the optimal competitive ratio the seller achieves by adopting an $n$-level randomized pricing mechanism is $\cR_n = \big(n(\uv/\lv)^{1/n}-(n-1)\big)^{-1}$. The optimal price levels form a geometric sequence defined as:
$
v_i=\lv^{\frac{n+1-i}{n}}\cdot \uv^{\frac{i-1}{n}}, \quad i=1,\dots,n.
$
The optimal pricing probabilities assigned to all price levels are equal except for the first price level at $\lv$, i.e.,
$
x_1 = (\uv/\lv)^{1/n}\cdot \cR_n, \quad x_i = ((\uv/\lv)^{1/n}-1)\cdot \cR_n, \quad i=2,\dots,n.
$
\end{proposition}

\begin{figure}[htbp]
    \centering
 \caption{Competitive Ratio of $n$-Level Pricing for n=1, 2, 5, and infinity}
\includegraphics[width=0.5\textwidth]{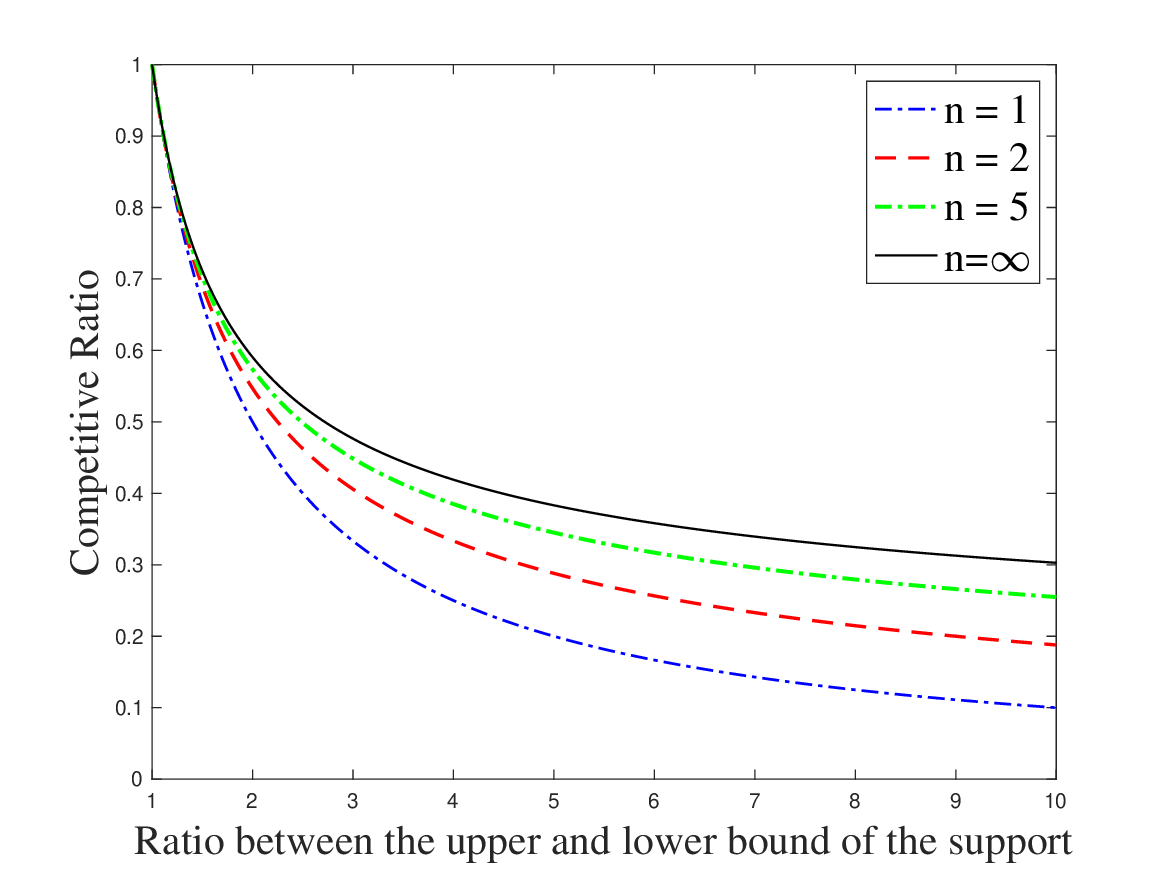}
        \label{fig:leveln}
\end{figure}

As the number of price levels increases, the competitive ratio approaches that of the optimal robust mechanism facilitated by a continuous randomized pricing scheme: 
\small \beqn 
\lim_{n\rightarrow\infty} \cR_n = \lim_{n\rightarrow\infty}\big(n(\uv/\lv)^{1/n}-(n-1)\big)^{-1} = \lim_{n\rightarrow\infty}\big(n((\uv/\lv)^{1/n}-1)+1\big)^{-1} =\big(\ln(\uv/\lv)+1\big)^{-1}
\eeqn \normalsize 
In \Cref{fig:leveln}, we compare the competitive ratio for different price levels ranging from 1, 2, 5, 100, and infinity under different ratios between the upper and lower bounds of the buyer's valuation support. \Cref{fig:leveln} shows that the competitive ratio decreases with the ratio between the upper and lower bounds, and the decreasing rate of the 1-level pricing scheme is steeper compared to other mechanisms with higher price levels.
\Cref{fig:leveln} emphasizes the substantial improvement in the competitive ratio when evolving from a 1-level pricing scheme to a 2-level pricing scheme. Specifically, by merely introducing one additional price level, the discrepancy between the competitive ratio of the 1-level pricing scheme and that of the optimal robust mechanism is cut approximately in half.
Furthermore, a 5-level pricing scheme yields a competitive ratio close to that achievable with optimal robust screening. This suggests that a finite, and notably modest, set of price levels can capture a substantial proportion of the benefit associated with continuous randomized pricing. This observation provides practical implications for sellers who are concerned about the complexity of menu sizes, and it offers a clear path for sellers to balance profitability and operational practicality.
\begin{remark}
    Notice that the robust screening problem can also be implemented as a price-skimming policy, where $x_i$ represents the duration for which the firm posts price $v_i$ \citep{eren2010monopoly}. \cite{eren2010monopoly} studied optimal pricing probabilities $\{x_i\}_{i=1}^n$ for given price levels $\cV=\{v_i\}_{i=1}^n$, assuming that the possible support of the buyer's willingness to pay is limited to the predetermined price levels $\{v_i\}_{i=1}^n$. This means the adversary can only generate worst-case valuation distributions whose support is within $\{v_i\}_{i=1}^n$, and the corresponding hindsight optimal price is also restricted to $\{v_i\}_{i=1}^n$. 
    Under this assumption, the problem can be solved by a finite linear programming problem, as both potential valuations and possible hindsight optimal pricing mechanisms are finite. In contrast, our model does not confine the buyer's willingness to pay to the predefined set of prices $\cV$. Although our selling mechanism involves a finite number of price levels, the buyer's valuation can range across the entire interval $[\lv,\uv]$. Consequently, the adversary can choose any distribution within this continuous range, making the hindsight optimal pricing mechanism potentially any price within $[\lv,\uv]$. Hence, our problem is essentially an infinite linear programming problem since the adversary's decision is infinite-dimensional. 
    As a result, our \Cref{lemma:fixv-support} yields different competitive ratios and pricing probabilities from those in Theorem 2.1 in \cite{eren2010monopoly}. Specifically, when the buyer's potential valuations are confined to a finite set $\{v_i\}_{i=1}^n$, the competitive ratio is larger, given by $\bigof{n-\sum_{i=2}^n \frac{v_{i-1}}{v_i}}^{-1}$ \citep{eren2010monopoly}. However, when valuations can span the entire interval $[\lv,\uv]$, the competitive ratio is reduced to $\big( 1 -n+\sum_{i=1}^n \frac{v_{i+1}}{v_{i}} \big)^{-1}$, as illustrated in \Cref{lemma:fixv-support}.
\end{remark}

\begin{remark}
    The robust screening framework can also be adapted as a bid-price control policy in the single-leg revenue management problem considered in \cite{ball2009toward}. In particular, the fare classes correspond to the price levels $\cV=\{v_i\}_{i=1}^n$, and the protection level for each fare class $v_i$ is defined as $\sum_{i'=i}^n x_{i'}$.  \cite{ball2009toward} optimize over the bid control policy to maximize the competitive ratio against the optimal policy, assuming that the fare classes $\cV=\{v_i\}_{i=1}^n$ are \emph{predetermined} and \emph{fixed}. However, in our framework, the adversary has the superiority to \textit{optimize} the fare classes $\cV$, in addition to the bid control policy and the worst-case demand sequence. This means that for any seller-defined bid control policy determined by $(\cV, \bx)$, nature is empowered to choose the most adversarial fare classes along with the bid control policy $(\cV^*, \bx^*)$ to minimize the competitive ratio. This capability of the adversary to adjust both the fare classes and the control policy adds complexity to the seller's decision-making process. 
    \end{remark}
    
\begin{figure}[htbp]
    \centering
       \caption{Competitive Ratios in \cite{ball2009toward}, \cite{eren2010monopoly} (dash-dot blue line with circle markers) and in \Cref{thm:support} (solid red line with star markers)}        \includegraphics[width=0.5\textwidth]{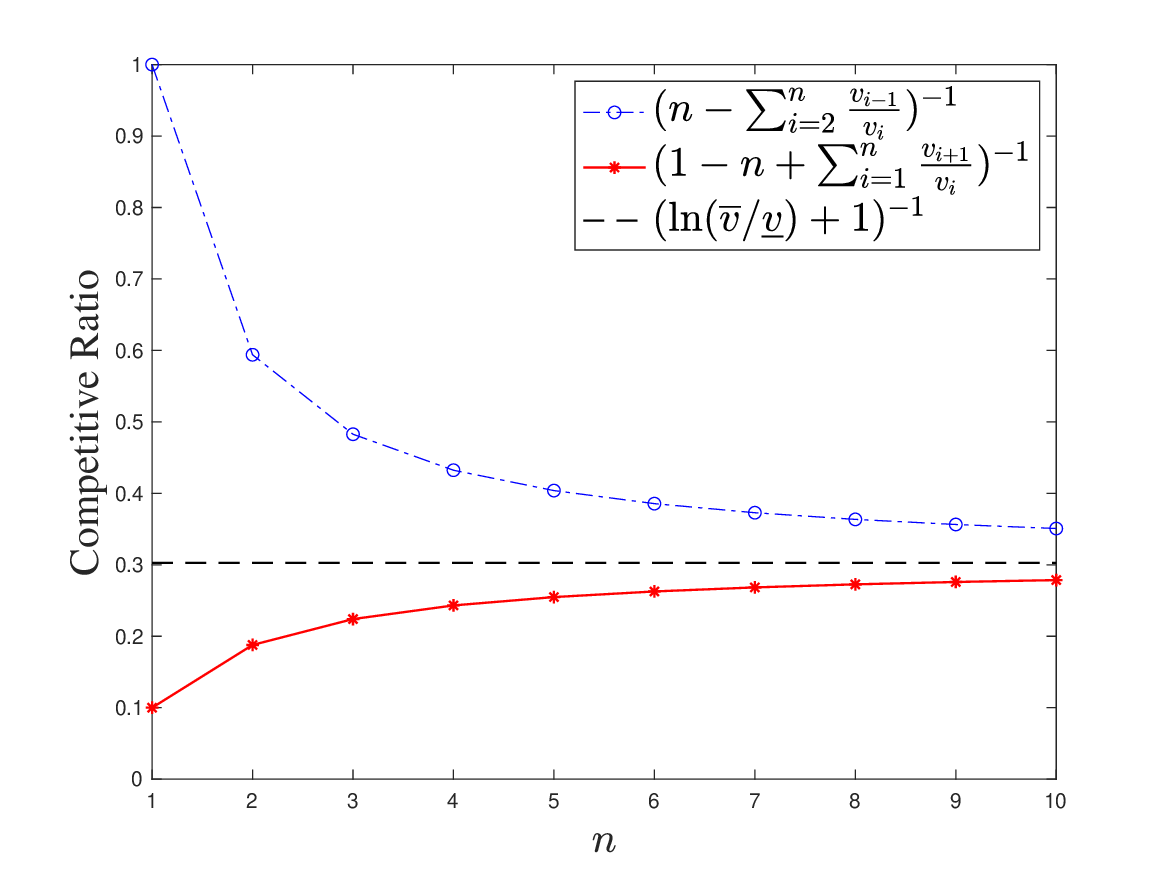}
        \label{fig:compareball}
\end{figure}

    The competitive ratio $\bigof{n-\sum_{i=2}^n \frac{v_{i-1}}{v_i}}^{-1}$, achieved in the revenue management problem in \cite{ball2009toward} aligns with the competitive ratio in \cite{eren2010monopoly}, because, in both models, $\cV$ determines both the seller's pricing options and the underlying fare classes (for the revenue management model) or willingness to pay (for the pricing model) that nature can choose from. In contrast, in our model, the potential fare classes or willingness to pay that nature can select can be arbitrary values in $[\lv,\uv]$.
    For given price levels $\cV$, let $\cR^\ddag = \bigof{n-\sum_{i=2}^n \frac{v_{i-1}}{v_i}}^{-1}$ denote the competitive ratio achieved by \cite{ball2009toward} and \cite{eren2010monopoly}, and let $\cR^\dag = \big( 1 -n+\sum_{i=1}^n \frac{v_{i+1}}{v_{i}} \big)^{-1}$ denote the competitive ratio achieved in \Cref{thm:support}. We have
{\small
\begin{align*}
\cR^\ddag &= \max_{\bpi\in \cP(\cV)} \inf_{\F\in \cP(\cV)}\frac{\Rev(\bpi,\F)}{\Rev(\OPT,\F)} \ge \max_{\bpi\in \cP([0,\uv])} \inf_{\F\in \cP([0,\uv])}\frac{\Rev(\bpi,\F)}{\Rev(\OPT,\F)}\ge \max_{\bpi\in \cP(\cV)} \inf_{\F\in \cP([0,\uv])}\frac{\Rev(\bpi,\F)}{\Rev(\OPT,\F)}  = \cR^\dag.
\end{align*}
}
To illustrate this distinction, \Cref{fig:compareball} presents a comparison between $\cR^\ddag$ and $\cR^\dag$ under the same choice of price levels $
v_i=\lv^{\frac{n+1-i}{n}}\cdot \uv^{\frac{i-1}{n}}, \, i=1,\dots,n
$, as defined in \Cref{thm:support} for $\uv/\lv = 10$ and $n=1,2,\dots,10$.
As the number of price levels $n$ increases, the competitive ratio $\cR^\dag$ for the $n$-level mechanism (solid red line with star markers) increases and approaches $\max_{\bpi\in \cP([0,\uv])} \inf_{\F\in \cP([0,\uv])}\frac{\Rev(\bpi,\F)}{\Rev(\OPT,\F)} = \big(\ln(\uv/\lv)+1\big)^{-1}$. On the contrary, $\cR^\ddag$ (dash-dot blue line with circle markers) from \cite{ball2009toward} and \cite{eren2010monopoly} decreases in $n$, converging to $\big(\ln(\uv/\lv)+1\big)^{-1}$ as $n\to \infty$.
\section{Moment Ambiguity Set}
\label{sec:mean}
We mainly focus on the mean ambiguity set in this section since it satisfies \Cref{assume:phi}, which enables us to deduce the closed-form optimal solutions for $n=1,2$ employing the small-size linear programming reformulation presented in \Cref{thm:finitelp0}. In \Cref{sec:meansupport}, we obtain the optimal selling mechanisms and competitive ratio for 1-level, and 2-level pricing and investigate the effectiveness of 2-level pricing when the mean and support of the willingness to pay are known. Then in \Cref{sec:mean-var}, we provide a closed-form approximation for 2-level pricing under the mean-variance ambiguity set. Both subsections underscore the significant advantage of adopting the 2-level pricing compared with 1-level pricing, particularly as the variability of willingness to pay increases (as indicated by either a larger support or a higher variance).
\subsection{Mean Ambiguity Set}
\label{sec:meansupport}
Suppose the seller knows that the buyer's valuation has a mean of at least $\mu$ and support of $[0,\uv]$. In this case, the ambiguity set of the buyer's valuation distribution is 
$
\cF=\left\{\F\in \cP([0,\uv]): \,\int_{0}^{\uv} v\,d\F(v) \ge \mu\right\}.
$
Here $\phi(v)=v-\mu$, which is a continuous function with no breakpoints on $[0,\uv]$, so for any given price level set $\cV=\{v_1,\dots, v_n\}$, we only need to specify $\lphi_{j} =\lim_{v\rightarrow v_j-} \phi(v) = v_j-\mu$ for $ j=1,\dots,n+1$. Then Problem \eqref{primal-finite0} becomes
\small \begin{equation}
\label{eq:meansupport}
\ba
\displaystyle
\sup\limits_{\lmb\in \R_+^{n},x_1,\dots,x_n, r\in \R^+} & r \\
 \mathrm{s.t.}\quad  &  (v_{j}-\mu)\lm_i\le \sum_{l=1}^{j-1} v_lx_l \quad \forall i \in [n],  j=1,\dots,i-1 \\
 & r\cdot v_i+(v_{j}-\mu)\lm_i\le \sum_{l=1}^{j-1}v_lx_l \quad \forall i \in [n], \  j=i,\dots,n+1 \\
 & r\cdot \uv\le \sum_{l=1}^n v_lx_l  \\
 & \sum_{i=1}^n x_i = 1
\ea
\end{equation}\normalsize
Next, we will evaluate the competitive ratios achieved by different price levels.
\subsubsection*{1-Level Pricing}
Now we investigate the performance of 1-level pricing by illustrating the optimal solution to Problem \eqref{eq:meansupport}. When $n=1$, the seller adopts a posted-price selling scheme, so we must have $x_1=1$. Hence, for any price level $v_1$, Problem \eqref{eq:meansupport} is simplified as follows:
\small \begin{equation}
\label{eq:meansupport-1}
\ba
\displaystyle
\cR(\{v_1\})=\sup\limits_{\lm\in \R_+, r\in \R^+} & r \\
 \mathrm{s.t.}\quad  & r\cdot v_1 + (v_{1}-\mu)\lm_1\le 0 \quad  \\
&  r\cdot v_1 + (\uv-\mu)\lm_1\le v_1 \quad  \\
 & r\cdot \uv\le v_1
\ea
\end{equation}\normalsize
\Cref{thm:meansupport-1} characterizes the optimal competitive ratio and posted price solved by Problem \eqref{eq:meansupport-1}.
\begin{proposition}
    \label{thm:meansupport-1}
Under the mean and support ambiguity set, the optimal competitive ratio achieved by a posted price policy is   $\cR_1^* = 1-\sqrt{1-\mu/\uv}$  and the optimal price is  $v_1=\uv-\sqrt{\uv^2-\mu\uv}$.
\end{proposition}

Let us illuminate \Cref{thm:meansupport-1} with a deeper intuitive exploration. For any fixed posted price $v_1$, nature selects a two-point valuation distribution: $v_1-$ with probability $\frac{\uv-\mu}{\uv-v_1}$ and $\uv$ with probability $\frac{\mu-v_1}{\uv-v_1}$. For optimal pricing, nature can adopt one of two possible paths.  When $v_1$ is small, the hindsight optimal price is $\uv$; when  $v_1$ is large, it is $v_1-$. The corresponding competitive ratios are $\frac{v_1}{\uv}$ and $\frac{\mu-v_1}{\uv-v_1}$, respectively. 
Regardless of the seller's choice of $v_1$, nature will select the price that yields the lower competitive ratio. Hence, $\cR(\{v_1\})=\min\{\frac{v_1}{\uv}, \frac{\mu-v_1}{\uv-v_1}\}$. The seller then optimizes $v_1$ to make nature indifferent between the two strategies.
In view of \Cref{thm:meansupport-1}, as demand dispersion increases, the competitive ratio decreases.
Moreover, the optimal posted price $v_1$ increases with $\mu$ and decreases with $\uv$.
This suggests that as the seller is more optimistic about the market (higher $\mu$), they set a higher price $v_1$; while greater valuation variability (higher $\uv$) leads to a lower price to hedge against demand uncertainty, resulting in a more conservative pricing strategy.

Regardless of the value of $v_1$ chosen by the seller, the worst-case distribution chosen by nature always assigns some probability at $\uv$. Thus, the seller can extract a revenue of merely $v_1$ from a buyer with valuation $\uv$, thereby limiting the performance of 1-level pricing. Next, we will show that introducing a price between $v_1$ and $\uv$ could significantly improve the performance ratio.

\subsubsection*{2-Level Pricing}
\label{sec:meansupport-level2}
We evaluate the performance when the seller randomizes between two prices $v_1,v_2$ with probability $x_1,x_2$, respectively. Similar to the analysis for $n=1$, to ensure a positive competitive ratio, we have $v_1< \mu$. Hence, for any given price level $v_1,v_2$, Problem \eqref{eq:meansupport} reduces to 
\small \begin{subequations}
\begin{align}
\displaystyle
\addtocounter{equation}{-1}
\displaystyle
\cR(\{v_1,v_2\})=\sup\limits_{\lmb\in \R_+^{ 2},x_1,x_2, r\in \R^+} & r  \label{eq:meansupport-2}\\
 \mathrm{s.t.}\quad 
 & r\cdot v_1+(v_1-\mu)\lm_1\le 0 \label{primal-meansupport-level2-11}\\
   & r\cdot v_1+(v_2-\mu)\lm_1\le v_1 x_1 \label{primal-meansupport-level2-12}\\
   & r\cdot v_1+(\uv-\mu)\lm_1\le v_1x_1+v_2x_2  \label{primal-meansupport-level2-13}\\
    & r\cdot v_2+(v_2-\mu)\lm_2\le v_1 x_1\label{primal-meansupport-level2-14}\\
    & r\cdot v_2+(\uv-\mu)\lm_2\le v_1x_1+v_2x_2  \label{primal-meansupport-level2-15}\\
&  r\cdot \uv\le v_1x_1+v_2x_2 \label{primal-meansupport-level2-16} \\
 & x_1+x_2 = 1\label{primal-meansupport-level2-7}
\end{align}
\end{subequations} \normalsize
In \Cref{thm:meansupport-2}, we characterize the optimal 2-level pricing mechanism and its competitive ratio.
\begin{proposition}
    \label{thm:meansupport-2}
In the optimal $2$-level pricing mechanism, the lower price is equal to the optimal price in $1$-level pricing, i.e., $v_1=\uv-\sqrt{\uv^2-\mu\uv}$, and the other parameters are defined as follows:
\begin{enumerate}
     \item  If $\mu/\uv\le 0.49$, $v_2=\sqrt{v_1\uv},\ x_1 = \frac{1}{2-\sqrt{v_1/\uv}}, \ x_2 = 1-x_1, \
    \cR_2^* = \frac{1}{2\sqrt{\overline{v}/v_1}-1}$.
\item  If $\mu/\uv>0.49$, $v_2=\frac{v_{1}+\sqrt{v_{1}^2\,+8v_1\uv}}{4},$ $x_1=\frac{{v_2(2v_2\uv - v_2^2 -  \mu\uv)}}{{(v_2-v_1)(2v_2\uv - v_2^2 -  \mu\uv) + v_1\uv( \uv-\mu)}}$,  $\ x_2 = 1-x_1, \ $ $\cR_2^* = \frac{{v_1v_2(\uv-\mu)}}{{(v_2-v_1)(2v_2\uv - v_2^2 -  \mu\uv) + v_1\uv( \uv-\mu)}}$.  
 \end{enumerate}
\end{proposition}
The 2-level pricing mitigates the conservativeness of $1$-level pricing and improves the competitive ratio from $\mathcal{O}(\mu/\uv)$ to $\mathcal{O}(\sqrt{\mu/\uv})$ when $\mu$ is small.
We further analyze nature's strategy and provide an intuitive explanation of \Cref{thm:meansupport-2}.  For any price level $v_1\le \mu\le v_2$, nature has four potential strategies to choose from.
\begin{enumerate}
    \item The worst-case distribution of valuation is a two-point distribution with probability $\frac{\mu-v_1}{\uv-v_1}$  at $\uv$ and $\frac{\uv-\mu}{\uv-v_1}$ at $v_1-$.
    \begin{enumerate}
        \item The hindsight optimal price is $\uv$ and the performance ratio is $\frac{v_1x_1+v_2x_2}{\uv}$. 
        \item The hindsight optimal price is $v_1-$ and the performance ratio is $\frac{(\mu-v_1)(v_1x_1+v_2x_2)}{v_1(\uv-v_1)}$. 
    \end{enumerate}
    \item The worst-case distribution of valuation is a two-point distribution with probability $\frac{\mu-v_1}{v_2-v_1}$ at $v_2-$ and $\frac{v_2-\mu}{v_2-v_1}$ at $v_1-$.
    \begin{enumerate}
        \item The hindsight optimal price is $v_2-$, and the performance ratio is $\frac{v_1x_1}{v_2}$. 
        \item The hindsight optimal price is $v_1-$ and the performance ratio is $\frac{x_1(\mu-v_1)}{v_2-v_1}.$
    \end{enumerate}
\end{enumerate}
For any pricing policy $(v_1,v_2,x_1,x_2)$ chosen by the seller, nature will select one of the four strategies above that yields the lowest competitive ratio. Hence, the competitive ratio when  $v_1\le \mu\le v_2$ is given by 
$r=\min\bigofff{\frac{v_1x_1+v_2x_2}{\uv}, \ \frac{(\mu-v_1)(v_1x_1+v_2x_2)}{v_1(\uv-v_1)}, \,  \frac{v_1x_1}{v_2},\,\frac{x_1(\mu-v_1)}{v_2-v_1}}.$
On the other hand, when $v_1\le v_2\le \mu$, nature's strategy falls within three different scenarios: the first two are the same as the first two scenarios when $v_1\le \mu\le v_2$, and the third one has a two-point worst-case distribution with probability $\frac{\uv-\mu}{\uv-v_2}$ at $v_2-$ and with probability $\frac{\mu-v_2}{\uv-v_2}$ at $\uv$, with a hindsight optimal price of $v_2-$. The performance ratio under the third scenario is $\frac{\mu-v_2}{\uv-v_2}x_2+\frac{v_1x_1}{v_2}$. Therefore, when $v_2\le \mu$, the competitive ratio is given by $r = \min\bigofff{\frac{v_1x_1+v_2x_2}{\uv},\, \frac{(\mu-v_1)(v_1x_1+v_2x_2)}{v_1(\uv-v_1)},\, \frac{\mu-v_2}{\uv-v_2}x_2+\frac{v_1x_1}{v_2}
}.$
In each case, the seller chooses $(v_1,v_2,x_1,x_2)$ to maximize the worst-case competitive ratio. After optimizing $x_1,x_2,v_1,v_2$ in both cases, the optimal selling mechanism is presented in \Cref{thm:meansupport-2}.
\Cref{thm:meansupport-2} reveals that the lower price $v_1$ equals the robust posted price in $1$-level pricing. Following the proof of \Cref{thm:meansupport-2}, \Cref{cor:mean-v2} characterizes the monotonicity and the range of $v_2$ with respect to varying values of $\mu$.
\begin{corollary}
    The higher price $v_2$  in 2-level pricing is increasing in $\mu$ when $\mu/\uv\le 0.49$ or when $\mu/\uv> 0.49$. Besides, it is greater than $\mu$ if $\mu/\uv\le 0.49$ and less than $\mu$ if $\mu/\uv> 0.49$.
    \label{cor:mean-v2}
\end{corollary}

\subsubsection*{3-Level Approximation}
\begin{figure}[htbp]
    \centering
    \caption{Competitive Ratios by $n$-Level Pricing for $n=1,2,\infty$ and Approximation for $n=3$ under the Mean Ambiguity Set}
    \label{fig:leveln-mean}
        \includegraphics[width=0.6\linewidth]{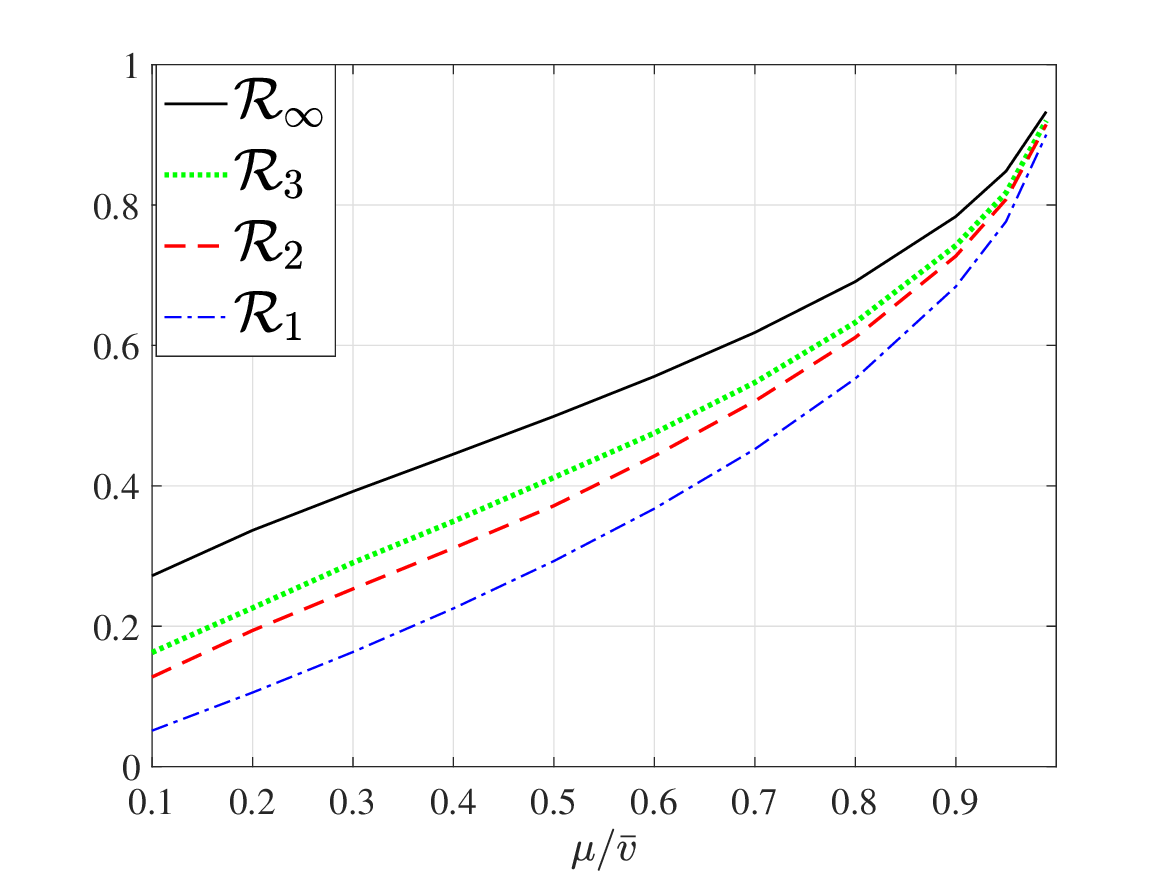}
\end{figure}

When the seller takes more price levels, deriving the closed-form solution to Problem \eqref{eq:meansupport} is more complicated. Therefore, we can not directly solve for the optimal pricing policy. Nevertheless, since it is easy and efficient to solve the simple small-size linear programming problem \eqref{eq:meansupport}, it is feasible to employ a grid search across potential price points to approximate the optimal pricing policy. Next, we conduct an exhaustive search over a discretized price space for each of the three prices ranging from 0 to $\uv$, with a granularity of $0.01\uv$. For every distinct pricing triplet, we solve  Problem \eqref{eq:meansupport} to determine the optimal pricing probabilities. After evaluating the competitive ratios for each price combination, we identify and select the one with the maximal competitive ratio.
\Cref{fig:leveln-mean} depicts the comparison between competitive ratios generated by the 1-level pricing in \Cref{thm:meansupport-1} (denoted by $\cR_1$), 2-level pricing in \Cref{thm:meansupport-2} (denoted by $\cR_2$) and the numerical approximation for 3-level pricing (denoted by $\cR_3$) against the optimal robust mechanism derived by \cite{wang2024minimax} (denoted by $\cR_{\infty}$). 
It shows that the improvement of the competitive ratio from 1-level pricing to 2-level pricing is significant. Introducing a second price cuts the shortfall toward continuous pricing approximately in half, and incorporating a third price can reduce the gap to optimal robust screening by more than half for $\mu/\uv\in [0.1,1]$. This demonstrates the substantial benefits of incorporating a small menu size into robust pricing.
\subsection{Discussions on Mean-Variance Ambiguity Set}
\label{sec:mean-var}
Having established the effectiveness of 2-level pricing under the mean-support ambiguity set, we now examine its performance under the mean-variance ambiguity set, where the buyer's valuation has a mean of $\mu$ and a variance of no more than $\sigma^2$. Following Theorem 1 in \cite{wang2024minimax}, the $n$-level pricing problem under the mean-variance ambiguity set can be formulated below.
{
\small
\begin{equation}
\label{eq:mean-var}
\ba
\displaystyle
\sup\limits_{\substack{\bv\in \R^n_+, \bx\in \R^n_+,\  r\in \R_+\\ \lmb_1 \in \R^{[0,\infty)}, \lmb_2 \in \R^{[0,\infty)}_- }} & r \\
 \mathrm{s.t.}\quad  &  (v-\mu)\lm_{1}(p) + (v^2-\mu^2-\sigma^2)\lm_2(p)\le \sum_{l=1}^n v_lx_l\cdot \mathbbm{1}[v\ge v_l], \, \ \forall p\in[0,\infty), \ \forall v\in[0,p)  \\
 & r\cdot p+ (v-\mu)\lm_{1}(p) + (v^2-\mu^2-\sigma^2)\lm_2(p)\le \sum_{l=1}^n v_lx_l\cdot \mathbbm{1}[v\ge v_l], \, \forall p\in[0,\infty), \ \forall v\in[p, \infty)
\ea
\end{equation}
}
Since the mean-variance ambiguity set does not satisfy \Cref{assume:phi}, we cannot apply the same finite LP approach in \Cref{thm:finitelp0}. Hence, it is challenging to obtain the closed-form optimal solution for the infinite-dimensional linear programming problem under the mean-variance ambiguity set. Nevertheless, we could provide a feasible 2-level pricing mechanism and show that this approximation already obtains much improvement compared to deterministic pricing.

\begin{proposition}
\label{thm:mean-var}
    The following selling mechanism and competitive ratio are feasible for Problem \eqref{eq:mean-var} with $n=2$, i.e. the 2-level pricing problem under the mean-variance ambiguity set:
 $v_2 = \sqrt{v_1\bigof{\mu +\frac{\sigma ^2}{\mu -v_{1}}}}$, $x_1 = \frac{\sqrt{\sigma ^2+\mu \,\left(\mu -v_{1}\right)}}{2\,\sqrt{\sigma ^2+\mu \,\left(\mu -v_{1}\right)}-\sqrt{v_{1}\,\left(\mu -v_{1}\right)}}$, $x_2=1-x_1$, 
 $r=\bigof{2\,\sqrt{\frac{1}{v_{1}}\bigof{\mu +\frac{\sigma ^2}{\mu -v_{1}}}}-1}^{-1}$, where
    \begin{itemize}
        \item for $\sigma \le \sqrt{\sqrt{5}-2}\mu$, $v_1\in [0,\mu)$ is the solution to $(\mu-v_1)^3 = (2v_1-\mu)\sigma^2$
        \item for $\sigma > \sqrt{\sqrt{5}-2} \mu$, $v_1\in [0,\mu)$ is the solution to $(\mu-v_1)^3 = \frac{2}{9+\sqrt{5}}(7 v_1-3\mu)\sigma^2$.
    \end{itemize}
\end{proposition}
The two-level pricing mechanism proposed in \Cref{thm:mean-var}, while not theoretically optimal, is straightforward to calculate and execute.  In this approximation, the lower price $v_1$ is slightly lower than $\mu$ when $\sigma$ is small, and it approaches $\frac{3}{7}\mu$ from above as $\sigma\to\infty$. 
The higher price $v_2$ is close to $\sqrt{v_1\mu}$ when $\sigma$ is small, since $v_2 = \sqrt{v_1\bigof{\mu +\frac{\sigma ^2}{\mu -v_{1}}}} = \sqrt{v_1\bigof{\mu +\frac{(\mu-v_1) ^2}{2v_{1}-\mu}}} = v_1 \sqrt{\frac{v_1}{2v_1-\mu}} = v_1 \sqrt{\frac{1}{1-(\mu/v_1-1)}} \to v_1\sqrt{\mu/v_1}=\sqrt{v_1\cdot \mu}$, where the second equality is according to the definition of $v_1$ for $\sigma<\sqrt{\sqrt{5}-2}\mu$. When $\sigma$ is large, the higher price $v_2$ goes to $\sqrt{\frac{3}{7}\mu^2+\frac{3}{4}\sigma^2}$, which approximately increases linearly in $\sigma$. When $\sigma$ is close to zero, the pricing probability at the lower price is close to 1, since $x_1 = \frac{\sqrt{\sigma ^2+\mu \,\left(\mu -v_{1}\right)}}{2\,\sqrt{\sigma ^2+\mu \,\left(\mu -v_{1}\right)}-\sqrt{v_{1}\,\left(\mu -v_{1}\right)}}=\frac{\sqrt{\frac{(\mu-v_1)^3}{2v_1-\mu}+\mu \,\left(\mu -v_{1}\right)}}{2\,\sqrt{\frac{(\mu-v_1)^3}{2v_1-\mu}+\mu \,\left(\mu -v_{1}\right)}-\sqrt{v_{1}\,\left(\mu -v_{1}\right)}} =\frac{\sqrt{v_1}}{2\sqrt{v_1}-\sqrt{2v_1-\mu}} \to 1$, where the second equality is due to definition of $v_1$ and the last approximation is because $v_1\to \mu$ when $\sigma$ is small. When $\sigma$ is large, the pricing probability at both price levels goes to $\frac{1}{2}$, since  $x_1 = \frac{\sqrt{\sigma ^2+\mu \,\left(\mu -v_{1}\right)}}{2\,\sqrt{\sigma ^2+\mu \,\left(\mu -v_{1}\right)}-\sqrt{v_{1}\,\left(\mu -v_{1}\right)}}=\frac{1}{2-\sqrt{\frac{v_{1}\,\left(\mu -v_{1}\right)}{\sigma ^2+\mu \,\left(\mu -v_{1}\right)}}}\to \frac{1}{2}$. 
A brief characterization of the 2-level price approximation is as follows: when the valuation has a low variability, both two price levels are lower than and close to $\mu$; and when the valuation has a large variability, the seller tends to assign equal probability to two price levels where the lower price is lower than $\mu$ and bounded from below by $\frac{3}{7}\mu$ and the higher price increases linearly in $\sigma$. 

\Cref{tab:mean-var} compares the approximation ratio defined by \Cref{thm:mean-var} with the competitive ratio of 1-level pricing in \cite{chen2022distribution} and $\infty$-level pricing in \cite{wang2024minimax}. \Cref{tab:mean-var} shows that when $\sigma/\mu$ is large, the improvement in competitive ratio from 1-level pricing to 2-level pricing, i.e. $\frac{\cR_2}{\cR_1}$ is greater than that from 2-level pricing to $\infty$-level pricing, i.e. $\frac{\cR_{\infty}}{\cR_2}$. For example, when $\sigma/\mu=10$, the competitive ratio increases tenfold from $0.25\%$ for 1-level pricing to $2.53\%$ for 2-level pricing, while the improvement from 2-level to $\infty$-level pricing is fivefold, from $2.53\%$ to $14.93\%$.
\begin{table}[htbp]
  \centering
  \caption{Competitive Ratios Based on the 1-level pricing (Deterministic Pricing), the 2-Level Approximation Defined in \Cref{thm:mean-var}, and $\infty$-level pricing (Robust Screening)}
  \small
    \begin{tabular}{lccccccccccc}
    \toprule
    $\sigma/\mu$      & 0.5   & 1     & 2     & 3     & 4     & 5     & 6     & 7     & 8     & 9     & 10 \\
    \hline
    $\cR_1$ & 37.28\% & 17.05\% & 5.57\% & 2.63\% & 1.52\% & 0.98\% & 0.68\% & 0.51\% & 0.39\% & 0.31\% & 0.25\% \\
     $\cR_2$ & 43.93\% & 25.82\% & 13.25\% & 8.74\% & 6.49\% & 5.15\% & 4.27\% & 3.65\% & 3.18\% & 2.82\% & 2.53\% \\
    $\cR_\infty$   & 56.08\% & 40.28\% & 28.04\% & 23.17\% & 20.52\% & 18.82\% & 17.62\% & 16.72\% & 16.00\% & 15.42\% & 14.93\% \\
    \bottomrule
    \end{tabular}%
  \label{tab:mean-var}%
\end{table}%

\begin{figure}[htbp]
    \centering
    \caption{$\cR_1/\cR_{\infty}$ and $\cR_2/\cR_{\infty}$ under the Mean-Variance Ambiguity Set}
    \label{fig:mean-var}
      \includegraphics[width=0.5\linewidth]{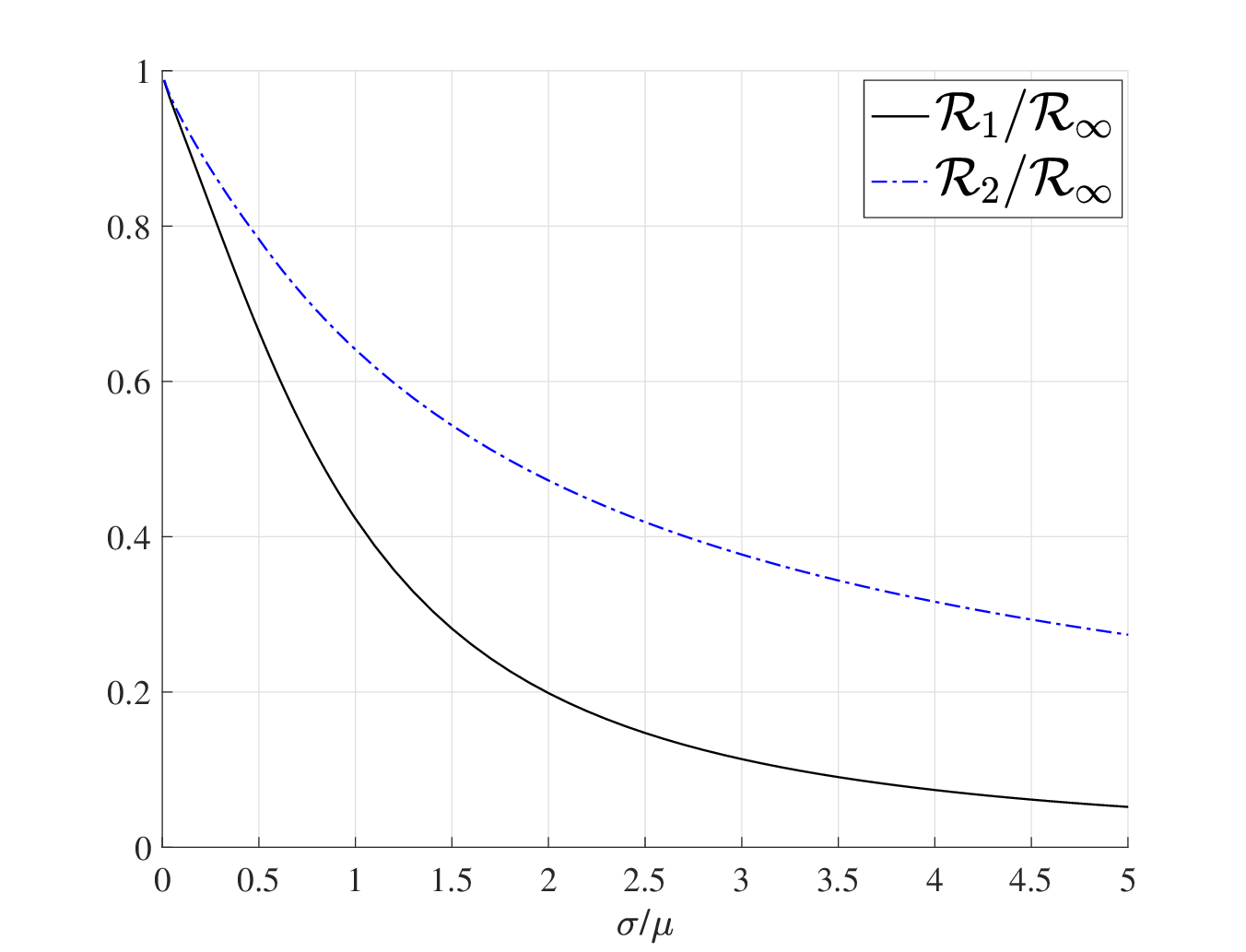}
\end{figure}

Moreover, \Cref{tab:mean-var} indicates that the benefit of adding price levels increases as valuation variability increases.
For example, when $\sigma/\mu=1$, the competitive ratio improves from $17.05\%$ with deterministic pricing to $25.82\%$ with 2-level pricing; whereas when $\sigma/\mu=5$, it increases fivefold from $0.98\%$ to $5.15\%$.
This is because when valuations are close to $\mu$, a deterministic price slightly lower than $\mu$ can already secure most of the optimal revenue. 
Thus, in our 2-level pricing approximation, the lower price $v_1$ is exactly equal to optimal deterministic pricing, and the pricing probability is higher for the lower price, so 2-level pricing performs similarly to 1-level pricing. 
Nevertheless, as valuation variability increases, the robust deterministic price decreases to hedge against low valuations, reducing its efficacy in capturing higher valuations and thus lowering the competitive ratio.  
In contrast, the performance decline of 2-level pricing is considerably slower than that of 1-level pricing. 
The 2-level pricing mechanism sets a low price $v_1$ to secure revenue from the low valuation and simultaneously a higher price $v_2$, which scales with $\sigma$, to exploit the high valuation variability. Hence, the advantage of 2-level pricing is considerably substantial when valuation has high variability. For a detailed comparison, \Cref{fig:mean-var} depicts $\cR_1/\cR_{\infty}$ and  $\cR_2/\cR_{\infty}$ with the coefficient of variation $\sigma/\mu$  ranging from 0 to 5, where $\cR_1, \cR_2, \cR_{\infty}$ stand for the competitive ratios achieved by 1-level, 2-level, and $\infty$-level pricing models, respectively. It shows that under a moderate variability of valuation, $\cR_2$ improves significantly from $\cR_1$. As $\sigma/\mu$ increases, the decreasing rate of the performance ratio for the 2-level pricing is slower than that of the 1-level pricing.
\normalsize
Given the slower decreasing trend in competitive ratio for 2-level pricing observed in \Cref{tab:mean-var} and \Cref{fig:mean-var}, we aim to analyze how the competitive ratio diminishes with $\sigma/\mu$.
To obtain a more straightforward and intuitive characterization of the relationship between the approximation ratio and $\sigma/\mu$,
\Cref{cor:mean-var} provides an effective and simple lower bound of the approximation ratio in \Cref{thm:mean-var}.

\begin{corollary}
\label{cor:mean-var}
The competitive ratio of the 2-level pricing approximation proposed in \Cref{thm:mean-var} is bounded from below by  $\bigof{\frac{7\sqrt{3}}{3}\sqrt{\frac{4}{7}+\bigof{\frac{\sigma}{\mu}}^2}}^{-1}$.
\end{corollary} 

\Cref{cor:mean-var} shows that the competitive ratio of the 2-level pricing is at least in the order of $\Theta(\frac{\mu}{\sigma})$, a notable improvement from the competitive ratio of deterministic pricing, which is $\Theta\of{(\frac{\mu}{\sigma})^2}$. At the same time, we also notice that for very large $\sigma/\mu$, there is still a large gap between the 2-level approximation and the optimal robust screening, whose competitive ratio is in the order of $\Theta\bigof{\frac{1}{2\ln\left(\sigma/\mu\right)}}$ according to \cite{giannakopoulos2023robust,wang2024minimax}.
\section{Quantile Ambiguity Set}
\label{sec:quantile}
In this section, we consider the ambiguity set $
\cF=\left\{\F\in \cP([0,\uv]): \,\int_{0}^{\uv} \mathbbm{1}[v\ge \omega] \,d\F(v) \ge \xi \right\}
$. This implies that based on historical sales data or market surveys, if the seller adopts a posted price at $\omega$, then the probability that the buyer is willing to purchase the product is at least $\xi$. 
Under the quantile ambiguity set, we have that $\phi(v) =\mathbbm{1}[v\ge \omega]-\xi$ and $\lphi_j=\lim_{v\rightarrow v_j-}\mathbbm{1}[v\ge \omega]-\xi=\mathbbm{1}[v_j>\omega]-\xi$. Hence, for any given price level set $\cV=\{v_1,\dots,v_n\}$, Problem \eqref{primal-finite0} is reduced to
\small \begin{equation}
\label{eq:quantile}
\ba
\displaystyle
\sup\limits_{\lmb \in \R^{n+1}_+ \bx\in\R_+^n, r\in \R^+} & r \\
 \mathrm{s.t.}\quad  &   (\mathbbm{1}[v_j > \omega]-\xi)\lm_{i}\le \sum_{l=1}^{j-1} v_lx_l, \ \forall \  i\in [n+1], \, j\in[i-1],  \\
 & r\cdot v_i+ (\mathbbm{1}[v_j > \omega]-\xi)\lm_{i}\le \sum_{l=1}^{j-1} v_lx_l, \ \forall \  i\in [n+1], \, j=i,\dots,n+1 \\
 & \sum_{i=1}^n x_i = 1
\ea
\end{equation}
\normalsize
 In the following, we evaluate the performance of robust mechanisms with different price levels. 
 
\subsubsection*{1-Level Pricing.}
\label{sec:quantile-1}
Suppose the seller posts a deterministic price at $v_1$. If $v_1> \omega$, then the worst-case distribution reduces to a single-point distribution at $\omega$, which satisfies the quantile constraint that the probability that the valuation is greater than $\omega$ is no less than $\xi\le 1$. In this case, the revenue achieved by posting price $v_1$ is zero, and thus the competitive ratio is zero. Hence, we consider only price $v_1$ up to $\omega$, and then Problem \eqref{eq:quantile} reduces to the following problem:
\small \begin{equation}
\label{eq:quantile-1}
\ba
\displaystyle
\sup\limits_{\lmb \in \R^{2}_+, r\in \R^+} & r \\
 \mathrm{s.t.}\quad  
  & r\cdot v_1 + (-\xi)\lm_{1} \le 0 \\
  & r\cdot v_1 + (1-\xi)\lm_{1} \le v_1 \\
  & r\cdot \uv+ (1-\xi)\lm_{2} \le v_1. \\
\ea
\end{equation}\normalsize
\Cref{thm:quantile-1} characterizes the optimal competitive ratio achieved by 1-level pricing.
\begin{proposition}
Under quantile ambiguity set $
\cF=\bigofff{\F\in \cP([0,\uv]): \,\int_{0}^{\uv} \mathbbm{1}[v\ge \omega] \,d\F(v) \ge \xi }
$, the optimal posted price is $v_1=\min\{\omega,\xi\uv\}$, and the corresponding 
competitive ratio of 1-level pricing is  $\cR_1 =\min\left\{
\xi,  \omega/\uv\right\}$.
\label{thm:quantile-1}
\end{proposition}
For any fixed posted price $v_1\le \omega$, nature can employ one of two strategies. In one scenario, the worst-case distribution of valuation is degenerate at $\uv$ and the corresponding hindsight optimal price is also $\uv$. In the second scenario, nature adopts a two-point discrete distribution: with a probability of $1-\xi$ at $v_1-$ and with a probability of $\xi$ at $\omega$, and the corresponding hindsight optimal price is $v_1-$. The competitive ratio in the first case is $v_1/\uv$ and in the second case, it is $\xi$. Regardless of the posted price $v_1$ set by the seller, nature selects the strategy that results in a lower competitive ratio. Therefore, the competitive ratio for a single fixed price ${v_1}$ is $\min\{v_1/\uv, \xi\}$. Based on nature's strategy, the seller should choose the maximum $v_1$ that does not exceed $\omega$, so resulting competitive ratio becomes $\cR_1=\min\{\omega/\uv, \xi\}$. 
Moreover, when $\xi\uv\le \omega$, pricing between $\xi\uv$ and $\omega$ results in the same competitive ratio. 
Notice that the worst-case distribution selected by nature always places some probability mass above $\omega$, so in the following, we will show that incorporating an additional price level between $v_1$ and $\uv$ could substantially enhance the competitive ratio. 

\subsubsection*{2-Level Pricing.}
Suppose the seller can randomize between two prices. We aim to jointly optimize the prices, $v_1$ and $v_2$, as well as their corresponding probabilities, $x_1$ and $x_2$. The optimal pricing strategy is presented in \Cref{thm:quantile-2}.
\begin{proposition}
    \label{thm:quantile-2}
Under quantile ambiguity set $
\cF=\Bigofff{\F\in \cP([0,\uv]): \,\int_{0}^{\uv} \mathbbm{1}[v\ge \omega] \,d\F(v) \ge \xi }$, the optimal 2-level pricing strategy and the corresponding competitive ratio are as follows:
   \begin{enumerate}
    \item If $\xi\uv\le \omega$, then $v_1=\xi\uv$, $v_2=\min\{\omega,\sqrt{\xi}\uv\}$, $x_1=\frac{v_2-\xi\uv}{\frac{\xi\uv^2}{v_2}-2\xi \uv+v_2}$, $x_2=1-x_1$, and $\cR_2^*=\frac{1-\xi}{\frac{\uv}{v_2}+\frac{v_2}{\xi\uv}-2}$
    \item If $\xi\uv> \omega$, then $v_1=\omega$, $v_2=\max\{\frac{\omega}{\xi},\sqrt{\omega\uv}\}$, $x_1 = \frac{v_2^{2}}{v_2^{2}-\omega v_{2}+\omega\overline{v}}$, $x_2=1-x_1$, and $\cR_2^*=\frac{1}{\frac{v_2}{\omega}+\frac{\overline{v}}{v_2}-1}$. 
\end{enumerate}
\end{proposition}
The optimal 2-level pricing satisfies that the lower price level $v_1=\min\{\omega,\xi\uv\}$ which is equal to the optimal price in 1-level pricing. We now evaluate the improvement in competitive ratio by incorporating 2-level pricing.
For a given incumbent price $\omega$, when the conversion rate $\xi$ is low, satisfying $\xi\le \of{\omega/\uv}^2$, then $\cR_1^* =\xi$, which approaches zero linearly as $\xi\to 0$. In contrast, $\cR_2^*=\frac{1-\xi}{\frac{\uv}{v_2}+\frac{v_2}{\xi\uv}-2} = \frac{1-\xi}{\frac{\uv}{\sqrt{\xi}\uv}+\frac{\sqrt{\xi}\uv}{\xi\uv}-2} = \frac{1-\xi}{2\of{\frac{1}{\sqrt{\xi}}-1}} = \frac{\sqrt{\xi}+\xi}{2}$, and $\cR_2^*/\cR_1^*=\frac{1}{2}(1+\frac{1}{\sqrt{\xi}})$, which goes to infinity as $\xi\to 0$. Hence, when $\xi$ is small, the improvement from 1-level pricing to 2-level pricing is more significant.

\subsubsection*{$\infty$-Level Pricing.}
Next, we obtain the benchmark where the seller can adopt the optimal robust mechanism by solving Problem \eqref{primal} with feasible mechanism set $\bPi= \{ \bpi:[0,\uv]\rightarrow \R^+\, \mid \, \int_0^{\uv} \pi(u)du = 1
\}$. Under the quantile ambiguity set, Problem \eqref{primal} becomes the following infinite linear programming problem:
\small \begin{equation}
\label{eq:quantile-continuous}
\ba
\displaystyle
\sup\limits_{\lmb \in \R^{[0,\uv]}_+,\ \bpi\in\bPi, \ r\in \R^+} & r \\
 \mathrm{s.t.}\quad  &   \mathbbm{1}[v\ge \omega] \cdot \lm(p)-\xi\cdot \lm(p)\le \int_0^v \pi(u)u\, du\quad \forall p\in[0,\uv], \ \forall v\in[0,p),  \\
 & r\cdot p+  \mathbbm{1}[v\ge \omega] \cdot \lm(p)-\xi\cdot \lm(p)\le \int_0^v \pi(u)u\, du\quad \forall p\in[0,\uv], \ \forall v\in[p, \uv], \\
 & \int_0^{\uv} \pi(v)\, dv = 1.
\ea
\end{equation}\normalsize
\Cref{thm:quantile-inf} characterizes the optimal mechanism and competitive ratio for $n=\infty$.
\begin{proposition}
    \label{thm:quantile-inf}
    Under quantile ambiguity set $
\cF=\bigofff{\F\in \cP([0,\uv]): \,\int_{0}^{\uv} \mathbbm{1}[v\ge \omega] \,d\F(v) \ge \xi}
$, the optimal price density function $\pi(v)$ and the corresponding payment function $t(v)$ are defined as:
   \small 
  \beqn \begin{cases}
\pi(v) = 
\begin{cases}
\frac{r}{(1-\xi)v} & \forall v\in[\xi\hp, \omega) \\
\frac{r}{v} & \forall v\in[\hp,\uv] \\
\end{cases} \\
\pi(v)\, \text{has a probability mass of } \frac{r(\hp-\omega)}{\omega(1-\xi)} \text{ at }  \omega
\end{cases}
\quad t(v)=
\int_0^v \pi(u)u\, du = 
\begin{cases}
0 & \forall v\in[0,\xi\hp) \\
\frac{r\cdot (v-\xi\hp)}{1-\xi} & \forall v\in[\xi\hp, \omega) \\
r\cdot \hp & \forall v\in[\omega,\hp) \\
r\cdot v & \forall v\in[\hp,\uv] \\
\end{cases}
   \eeqn \normalsize 
 where $\hp=\min\{(2-\xi)\omega,\uv\}$, and the corresponding competitive ratio $\cR_\infty$ is
$$r=
\begin{cases}\big(
\ln(\uv)-\ln(\omega(2-\xi))-\frac{\ln(\xi(2-\xi))}{1-\xi}+1
\big)^{-1} & \mbox{if } \ (2-\xi)\omega<\uv\\
\big(
\frac{\uv-\omega}{(1-\xi)\omega}+\frac{\ln(\omega)-\ln(\xi\uv)}{1-\xi}
\big)^{-1} & \mbox{if } \ (2-\xi)\omega\ge \uv.
\end{cases}
$$ 
\end{proposition}
\Cref{thm:quantile-inf} implies the allocation function (or the cumulative price distribution function) is defined as 
\small
$$q(v)=\int_0^v \pi(u)\, du = 
\begin{cases}
0 & \forall v\in[0,\xi\hp) \\
\frac{r}{1-\xi}\ln\frac{v}{\xi\hp} & \forall v\in[\xi\hp, \omega) \\
1-r\ln(\uv/\hp) & \forall v\in[\omega,\hp) \\
1-r\ln(\uv/v) & \forall v\in[\hp,\uv].
\end{cases}$$ \normalsize
Based on the characterization of $\pi(v)$ and $q(v)$, we show that as the conversion rate $\xi$ increases, the seller prices at higher valuations with greater probability.
\begin{corollary}
\label{cor:quantile-dominance}
   Given an incumbent price $\omega$, if $\xi^{\dag}\le\xi^{\ddag}$, the optimal pricing density function $\pi^{\dag}$ derived from conversion rate $\xi^{\dag}$ is stochastically dominated by the optimal pricing density $\pi^{\ddag}$ derived from $\xi^{\ddag}$. 
\end{corollary}
\Cref{cor:quantile-dominance} is intuitive because as more customers accept price $\omega$, the seller becomes more confident in the customers' valuation of the product. Hence, the pricing strategy derived from a higher conversion rate  $\xi^{\ddag}$ is less conservative than that derived from a lower conversion rate $\xi^{\dag}$.
\subsubsection*{3-Level Approximation and Comparison Across Price Levels.}
As it is complicated to derive the closed-form solution to Problem \eqref{eq:quantile} for higher price levels, we search over the price combinations to approximate the optimal pricing policy by solving the simple small-scale linear programming problem \eqref{eq:quantile} when $n=3$. Similar to the analysis in the mean ambiguity set, we conduct an exhaustive search over a discretized price space for each of the three prices, ranging from 0 to $\uv$ with a granularity of $0.01\uv$. For each price combination, we solve  Problem \eqref{eq:quantile} to determine the optimal pricing probabilities and then aggregate the competitive ratios garnered across all price combinations to identify the combination with the highest competitive ratio.
\begin{figure}[htbp]
\centering
\captionsetup{justification=centering}
\caption{Competitive Ratios by $n$-Level Pricing for $n=1,2,\infty$ and Approximation for $n=3$ Under the Quantile Ambiguity Set}
\label{fig:quantile}
\begin{subfigure}[t]{0.49\textwidth}
\includegraphics[width=\textwidth,keepaspectratio]{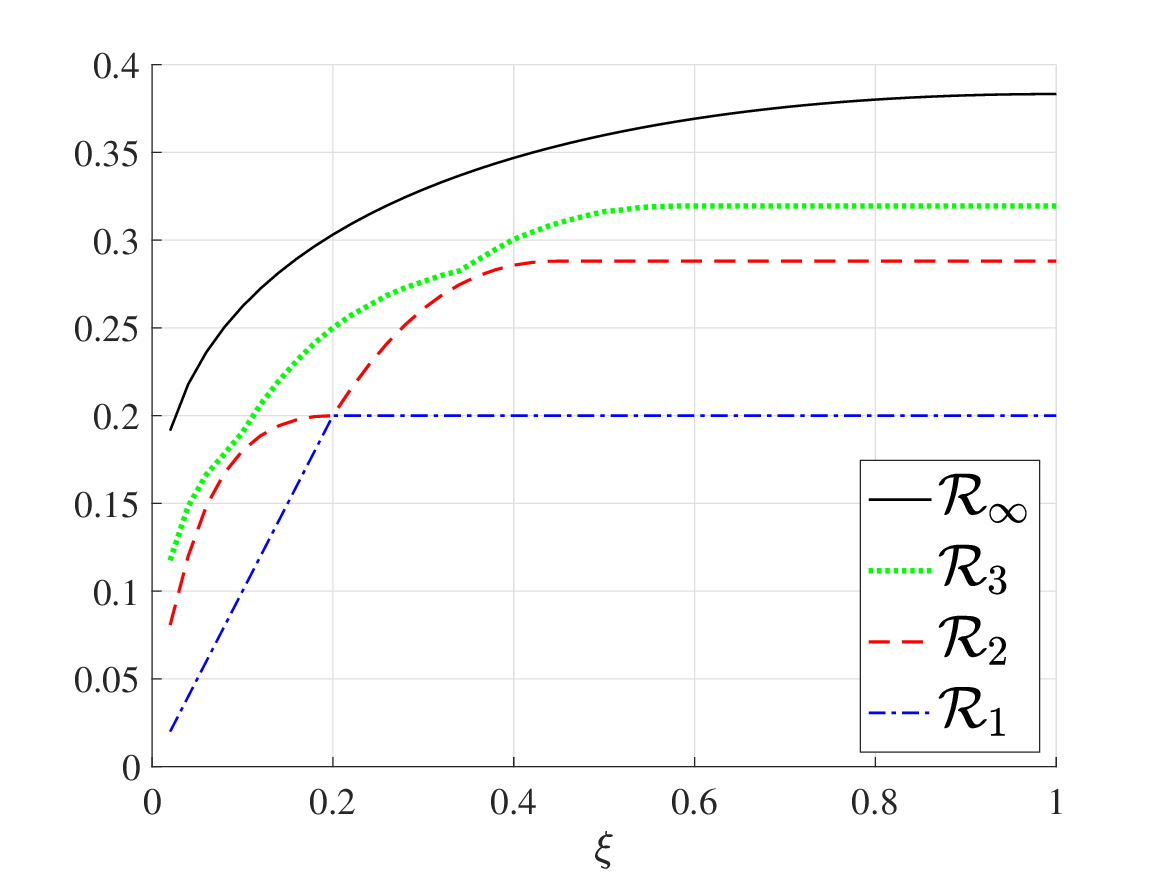}
\caption{$\omega/\uv=0.2$}
\label{fig:quantile-w02}
\end{subfigure}
\begin{subfigure}[t]{0.49\textwidth}
\includegraphics[width=\textwidth,keepaspectratio]{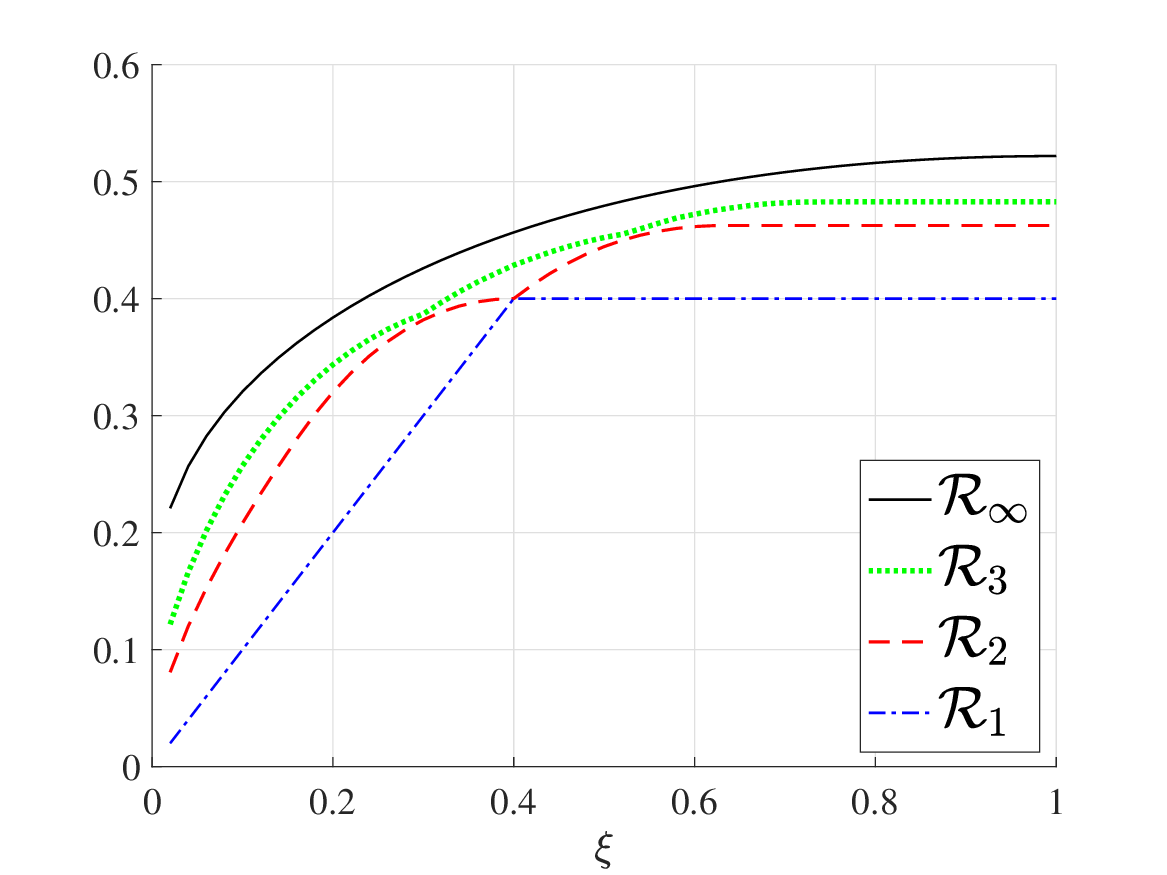}
\caption{$\omega/\uv=0.4$}
\label{fig:quantile-w04}
\end{subfigure}
\begin{subfigure}[t]{0.49\textwidth}
\includegraphics[width=\textwidth,keepaspectratio]{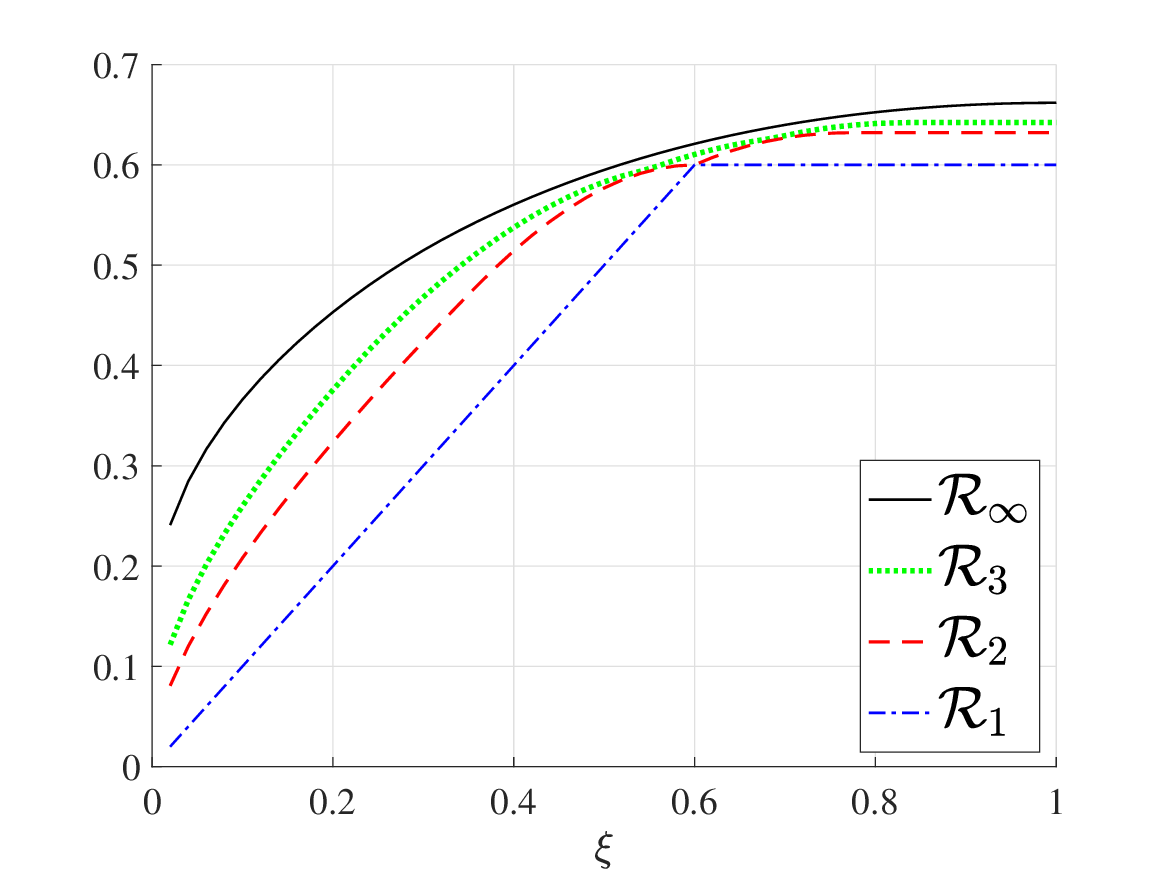}
\caption{$\omega/\uv=0.6$}
\label{fig:quantile-w06}
\end{subfigure}
\begin{subfigure}[t]{0.49\textwidth}
\includegraphics[width=\textwidth,keepaspectratio]{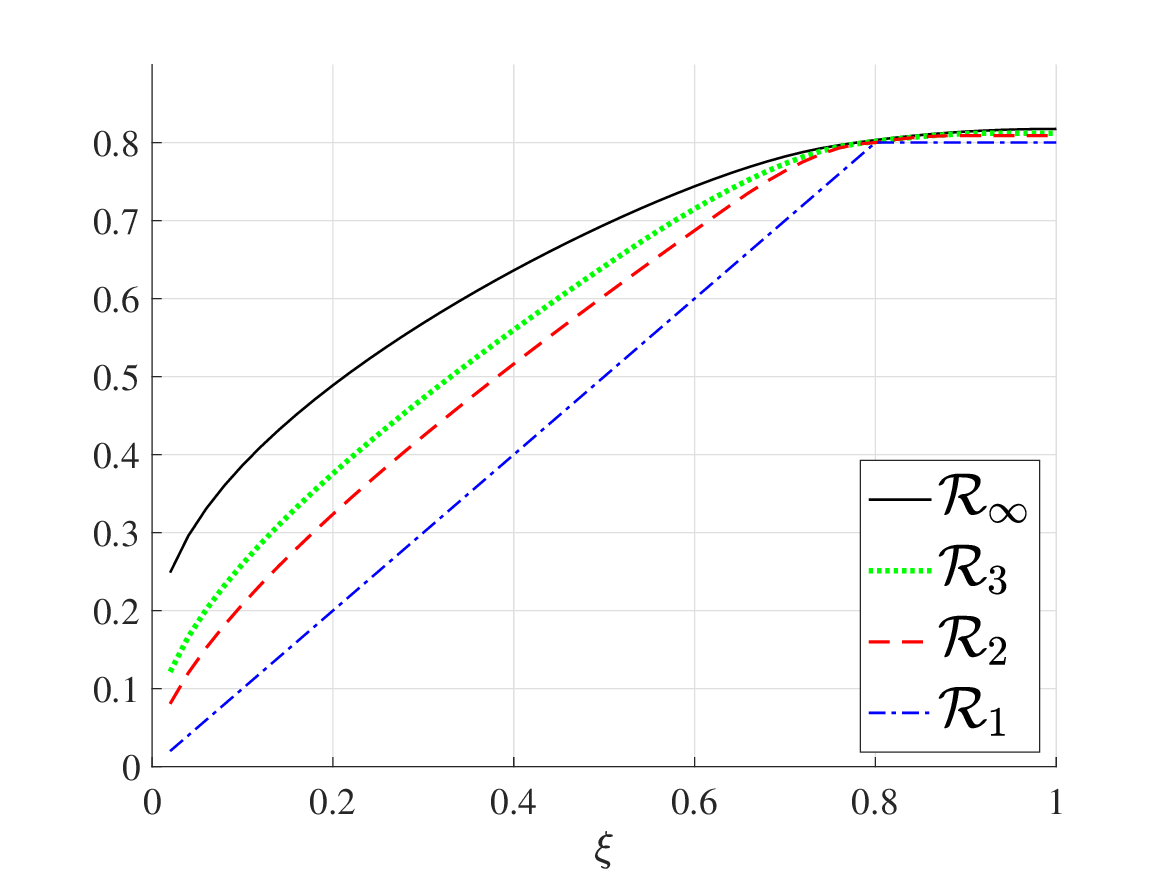}
\caption{$\omega/\uv=0.8$}
\label{fig:quantile-w08}
\end{subfigure}
\end{figure}
\Cref{fig:quantile} compares the competitive ratio achieved by 1-level pricing ($\cR_1$) as defined in \Cref{thm:quantile-1}, 2-level pricing ($\cR_2$) as defined in \Cref{thm:quantile-2}, the $\infty$-level pricing ($\cR_{\infty}$) as defined in \Cref{thm:quantile-inf}, alongside the numerical lower-bound approximation for 3-level pricing ($\cR_3$). 
We consider various quantile information with $\omega/\uv$ ranging from $0.2$ to $0.8$, and $\xi$ varying from 0 to 1. \Cref{fig:quantile} shows that 2-level pricing significantly improves upon 1-level pricing, especially as $\xi$ approaches zero, where it can outperform 1-level pricing by several folds.  Moreover, when $\omega/\uv$ is moderate or large, the competitive ratio achieved by the 2-level pricing is very close to that achieved by $\infty$-level pricing, particularly when $\xi\ge \omega/\uv$.
Though there is no improvement from 1-level pricing to 2-level pricing when $\xi = \omega/\uv$, the augmentation to 3-level pricing under this condition yields considerable benefits. This is attributable to the fact that, as per \Cref{thm:quantile-2}, 2-level pricing reduces to a single price point at $\omega$ when $\xi = \omega/\uv$, while 3-level pricing benefits from setting prices at three distinct levels: below, at, and above $\omega$ under this condition. The performance of the 3-level pricing approximation is very close to that of the $\infty$-level pricing across a variety of $\omega$ and $\xi$ values.

\begin{remark}
     \cite{allouah2023optimal} provides a finite linear programming formulation that effectively approximates the $\infty$-level pricing problem when $n\to \infty$ for regular (and MHR) valuation distributions with quantile information at one point. For any fixed price level set $\cV$, their finite LP provides a \emph{lower bound} of the competitive ratio. In contrast, the finite LP in \eqref{eq:quantile} determines the \emph{optimal} selling mechanism and competitive ratio for the given price level set $\cV$. Thus, while both LPs can approximate the $\infty$-level pricing as $n\to \infty$, our LP in \eqref{eq:quantile} can be adopted to perform a grid search over prices to approximate the optimal $n$-level selling mechanism for small $n$.
\end{remark}

\section{Alternative Robust Metrics}
\label{sec:alternative}
In this section, we apply our framework to general robust metrics. Consider a robust metric $\mathsf{OBJ}_\alpha:= \sup_{\bpi\in \bPi_n} \mathsf{OBJ}_\alpha(\bpi)$, where $\alpha\in [0,1]$ and 
{\small \begin{align} \mathsf{OBJ}_\alpha(\bpi):=\inf_{\F\in \cF}{\min_{p\in [0,\uv]} \ \int_{0}^{\uv}\bigof{\int_0^v \pi(u)u\, du}\, d\F(v) -\alpha \cdot  p(1-\F(p-))}
\label{general}
\end{align}}
When $\alpha=0$, Problem \eqref{general} recovers the maximin revenue objective and when $\alpha=1$, Problem \eqref{general} recovers the minimax absolute regret objective.
For the succinctness of the paper, we focus on the mean ambiguity set, and the same techniques apply to other ambiguity sets.
Applying similar ideas in \Cref{thm:finitelp0}, we can reduce Problem \eqref{general} into a finite linear programming problem.
\begin{proposition}
\label{prop:general-mean}
 Suppose the buyer's valuation has a mean of $\mu$ and an upper bound of $\uv$.
    For a given price level set $\cV$, $\mathsf{OBJ}_\alpha$ can be solved by the following linear programming problem.
\begin{align*}
    \sup_{\bx\in \R_+^n, \lmb\in \R^{n+1}} \  & \Delta \nonumber\\
 \mbox{s.t.}  \quad &   \Delta  \  \le   \sum_{l=1}^{j-1}\bigof{v_lx_l} - \lm_i \cdot (v_j-\mu) \quad \forall i=1,\dots,n+1, \  j=1,\dots,i-1\\
 &   \Delta  \  \le   \sum_{l=1}^{j-1}\bigof{v_lx_l} -\alpha v_i - \lm_i \cdot (v_j-\mu) \quad \forall i=1,\dots,n+1, \  j=i,\dots,n+1 \\
  &  \Delta  \  \le   \lm_i \mu \quad \forall i=1,\dots,n\ \\
    & \sum_{j=1}^n x_j =1
\end{align*}
\end{proposition}
where $v_{n+1} = \uv$. When $\alpha=0$, \Cref{prop:maximin revenue meansupport lp} shows that the problem above can be further simplified.
\begin{proposition}
 Given the price level set $\cV$, the optimal $n$-level mechanism under the maximin revenue metric and mean ambiguity set is solved by the following linear programming problem.
    \begin{align*}
\max_{\bx\in \R_+^n,\lambda_0,\lambda_1} & \ \lambda_0  \\
\mathrm{s.t.} \quad & 
\lambda_0+\lambda_1\cdot (v_j-\mu) \le \sum_{i=1}^{j-1} v_ix_i\quad  \forall \ j=1,\dots,n+1 \\
& \sum_{j=1}^n x_j = 1 
\end{align*}
\label{prop:maximin revenue meansupport lp}
\end{proposition}
This finite LP yields the optimal $n$-level mechanism under the maximin revenue metric.
\begin{proposition}
    For the mean ambiguity set, the optimal $n$-level pricing mechanism under the maximin revenue metric is $
    v_j = v_1\cdot \bigof{\uv/v_1}^{\frac{j-1}{n}}, \,  x_j = \frac{1}{n}, \, \forall j= 1,\dots, n, 
$
where $v_1$ is the unique solution to $ \mu/v + n \bigof{v/\uv}^{1/n}=n+1 $. The optimal robust revenue is $\lambda_0^* = (\mu-v_1)\cdot n^{-1}\cdot \bigof{ (\uv/v_1)^{1/n} -1 }^{-1}.$
\label{prop:maximin revenue meansuport closedform}
\end{proposition}
Based on \Cref{prop:maximin revenue meansuport closedform}, we can derive the closed-form optimal $n$-level mechanism for different $n$.
\begin{itemize}
    \item For $n=1$, the optimal deterministic pricing is $v_1=\uv-\sqrt{\uv^2-\mu\uv}$, and the optimal robust revenue is $(\sqrt{\uv}-\sqrt{\uv-\mu})^2$, which matches the result in \cite{chen2024screening}.
    \item For $n=2$, $v_1$ is the unique solution to $\mu-3v+2v^{3/2}/\sqrt{\uv}=0$. The optimal 2-level mechanism posts prices at $v_1$ and $\sqrt{v_1\uv}$ with equal probability $\frac{1}{2}$. The optimal robust revenue is $\frac{\mu-v_1}{2}\bigof{\sqrt{\uv/v_1}-1}^{-1}$.
    \item As $n\to \infty$, the optimal selling mechanism is a linear-payment mechanism as follows
\begin{align*}
    t(v) = \begin{cases}
        0 & v \in [0,v_1] \\
        \frac{v-v_1}{\ln\uv-\ln v_1}& v \in (v_1,\uv]
    \end{cases}  \quad    q(v) = \begin{cases}
        0 & v \in [0,v_1]\\
        \frac{\ln v-\ln v_1}{\ln\uv-\ln v_1}& v \in (v_1,\uv]
    \end{cases} 
\end{align*}
where $v_1$ is the solution to $\mu=v\cdot\bigof{1+\ln\uv-\ln v}$, matching the result in \cite{carrasco2018optimal,chen2024screening}. 
\end{itemize}

When $\alpha=1$, based on the finite LP in \Cref{prop:general-mean}, we provide the closed-form optimal 1-level mechanism and the corresponding regret for the minimax absolute regret metric in \Cref{prop:minimax regret meansupport1}.
\begin{proposition}
    The optimal 1-level selling mechanism under the mean ambiguity set for minimax absolute regret objective is to post a price at $v_1=\frac{\uv\,\left(\mu +\uv-\sqrt{\left(\uv-\mu\right)\,\left(3\,\mu +\uv\right)}\right)}{2\,\mu }$, and the corresponding optimal minimax regret is $\frac{\mu-\uv+\sqrt{\left(\uv-\mu\right)\,\left(3\,\mu +\uv\right)}}{2}$.
    \label{prop:minimax regret meansupport1}
\end{proposition}

\section{Summary and Future Directions}
In this paper, we introduce the $n$-level pricing problems that establish a favorable balance between theoretical robustness and practical implementation. We propose a unified framework that addresses a broad class of ambiguity sets including support, mean, and quantile information. Our framework enables us to derive closed-form optimal $n$-level pricing mechanisms and competitive ratios under the support ambiguity set, as well as for 1- and 2-level pricing under the mean and quantile ambiguity sets. Our closed-form competitive ratios highlight the effectiveness of 2-level pricing. 
Although we primarily focus on the support, mean and quantile ambiguity sets, our framework extends to more general ambiguity sets, and our main insights may also hold under general ambiguity sets.
For instance, we provide a closed-form approximation for 2-level pricing under the mean-variance ambiguity set. 
Our major implication is that merely randomizing between just two prices can significantly improve the performance of deterministic pricing. One future exploration would be to quantify the competitive ratio for general $n$-level pricing under different ambiguity sets, which could provide deeper insights into how the performance guarantee of a robust mechanism depends on the menu complexity. Another interesting direction is applying our framework to multi-agent robust mechanism design, where simple robust mechanisms could prove valuable.
\ACKNOWLEDGMENT{The author thanks the department editor, associate editor, and three anonymous referees for their valuable and constructive suggestions that significantly improved the exposition of this paper. This research is supported by the National Natural Science Foundation of China [Grant 72394395].}	

\bibliographystyle{ormsv080}
\bibliography{myref.bib}
\ECSwitch
\ECHead{Electronic Companion to ``The Power of Simple Menus in Robust Selling Mechanisms''}
\section{Empirical Performance of 1-,2-,$\infty$-level Pricing under Mean and Quantile Ambiguity Set}
In this section, we will investigate the empirical performance of 1-level, 2-level, and $\infty$-level pricing under different ambiguity sets. For each given distribution, we adopt only its partial information, i.e., mean or quantile information to design the robust 1-level, 2-level, and $\infty$-level pricing, and then we evaluate the empirical performance of the robust mechanisms under the true distribution, compared against the optimal pricing given the true distribution.
Our experiment results demonstrate the superiority of 2-level pricing across different distributions. 
\subsection{Mean Ambiguity Set}
Now that we have discussed the competitive ratio of different mechanisms against the adversary in \Cref{sec:meansupport}, we now evaluate the empirical performance of the 1-level pricing, 2-level pricing, and $\infty$-level pricing under given distributions.
We choose beta distribution since it exhibits a variety of shapes under different parameters. We summarize the performance of the three pricing mechanisms under different shapes of valuation density functions as follows.
\begin{itemize}
    \item For a unimodal valuation where the mode is non-zero, regardless of the valuation density being flat or concentrated, 2-level pricing exhibits superior performance. This outperformance stems from (1) the lower price in 2-level pricing effectively captures substantial revenue from the mode, as long as the mode is not significantly lower than the mean. (2) the higher price can capture the high valuations from the distribution's right tail.
    \item When the valuation distribution is unimodal with a mode at zero,  the 2-level and $\infty$-level pricing demonstrate comparable performance, notably surpassing the 1-level pricing. The reason for the advantage of 2-level pricing is similar to the previous scenario. Meanwhile, in this case, the $\infty$-level pricing is sometimes more effective in extracting revenue from the high valuation.
    \item The strengths of $\infty$-level pricing become apparent in a bimodal distribution with modes 0 and 1. In this scenario, 2-level outperforms 1-level pricing but is inferior to $\infty$-level pricing. The reason for the superiority of $\infty$-level pricing is that the dispersion of $\infty$-level pricing can capture the revenue at mode 1 effectively, yet the higher price in 2-level pricing is not as effective in extracting revenue at mode 1.
\end{itemize}

\begin{figure}[htb]
\centering
\captionsetup{justification=centering}
\caption{Performance Ratios for $n$-Level Pricing for $n=1,2,\infty$ Based on the Mean Ambiguity Set Under Beta Distributions}
\label{fig:mean_beta}
\begin{subfigure}[t]{0.45\textwidth}
\includegraphics[width=\textwidth,keepaspectratio]{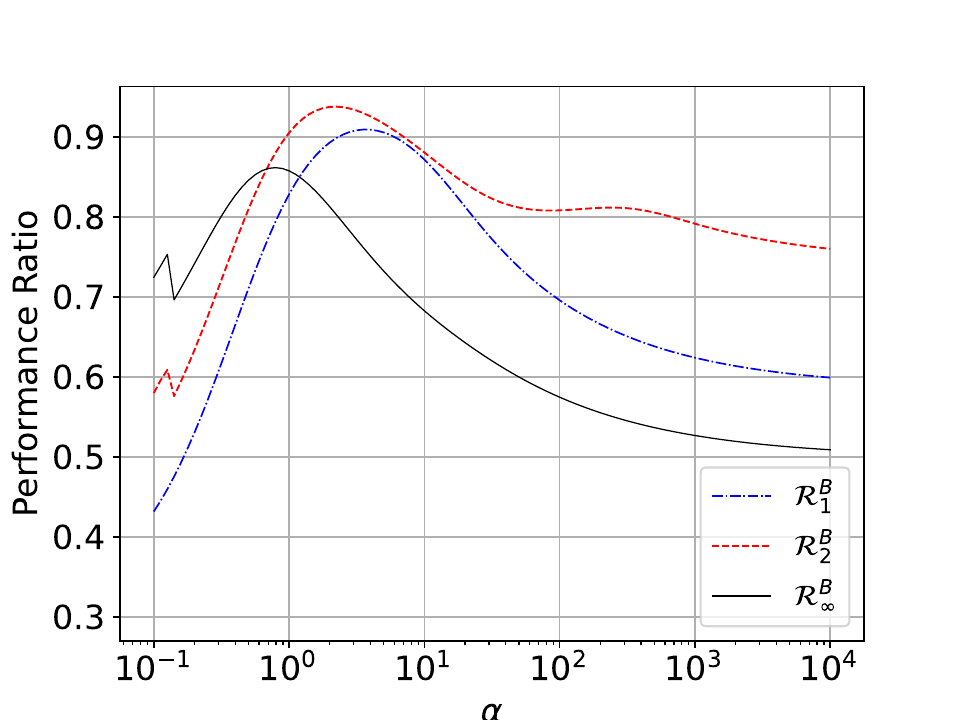}
\caption{$\alpha=\beta$}
\label{fig:mean_beta_equal}
\end{subfigure}
\begin{subfigure}[t]{0.45\textwidth}
\includegraphics[width=\textwidth,keepaspectratio]{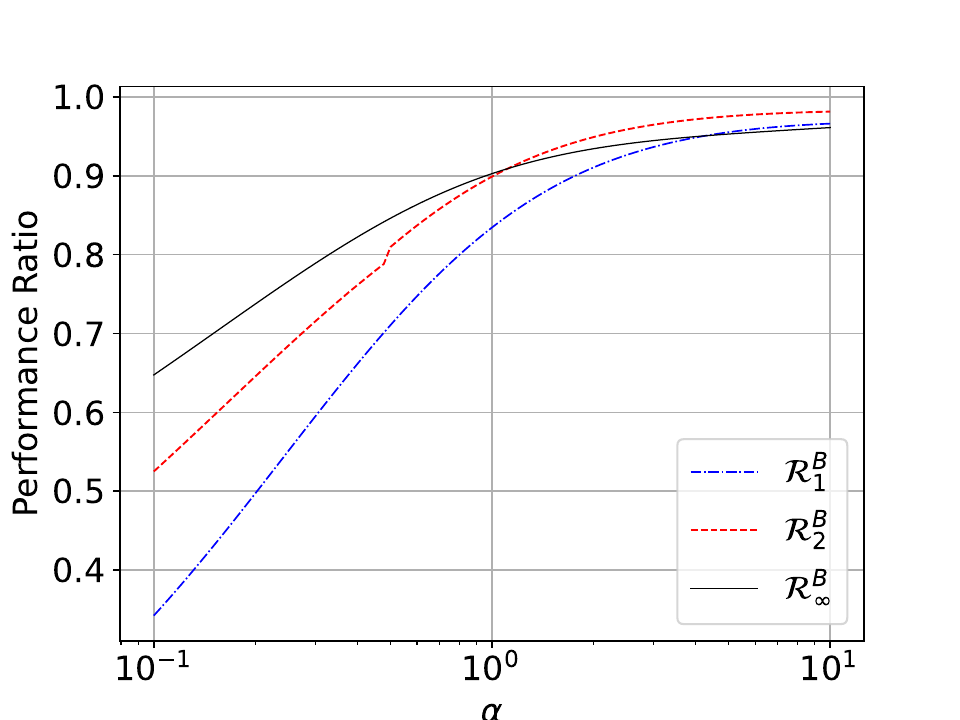}
\caption{$\beta = 0.5$}
\label{fig:mean_beta_05}
\end{subfigure}
\begin{subfigure}[t]{0.45\textwidth}
\includegraphics[width=\textwidth,keepaspectratio]{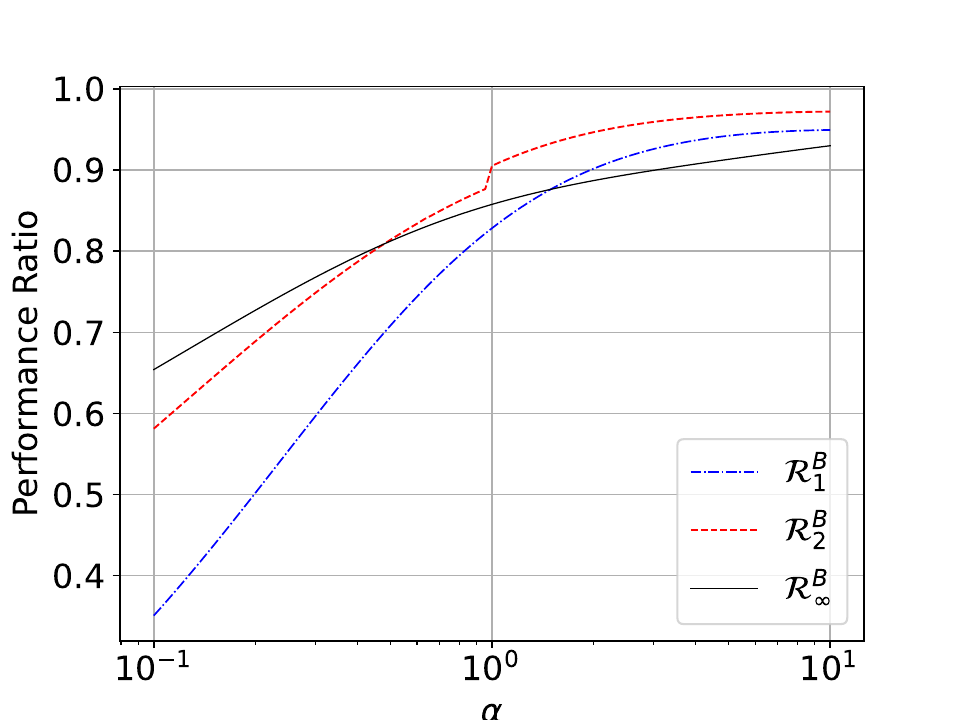}
\caption{$\beta = 1$}
\label{fig:mean_beta_1}
\end{subfigure}
\begin{subfigure}[t]{0.45\textwidth}
\includegraphics[width=\textwidth,keepaspectratio]{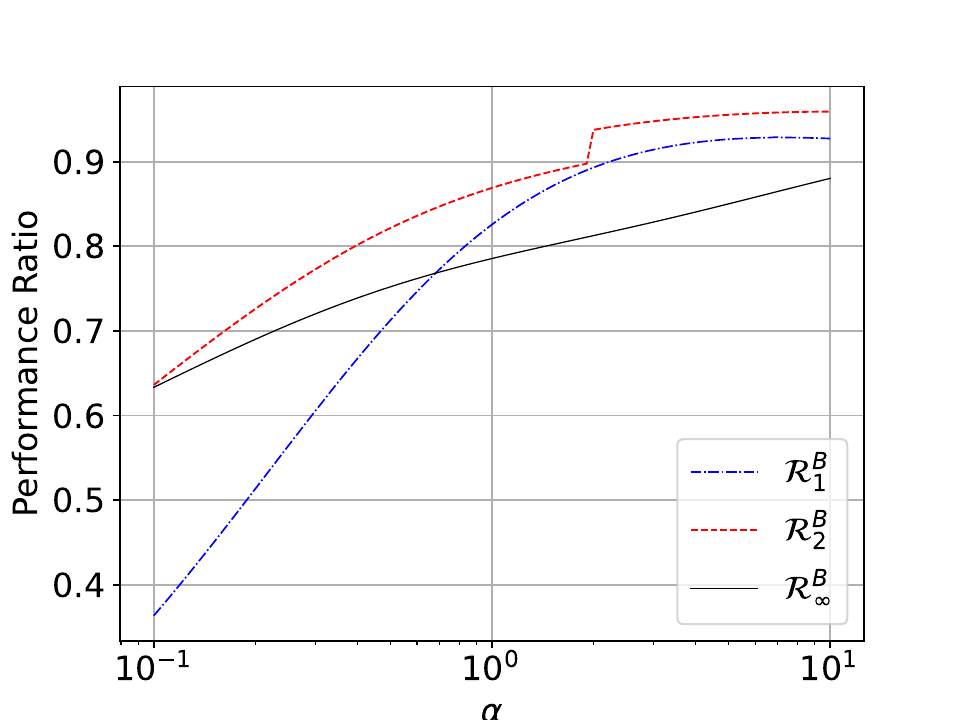}
\caption{$\beta = 2$}
\label{fig:mean_beta_2}
\end{subfigure}
\end{figure}
Now we delve into the detailed empirical performance of the three mechanisms. In \Cref{fig:mean_beta}, we present the performance ratio of 1-level pricing in blue dash-dot lines, 2-level pricing in red dashed line, and $\infty$-level pricing in black solid lines. When $\alpha=\beta$, in \Cref{fig:mean_beta_equal}, we let $\alpha=\beta$ and vary $\alpha$ from 0.1 to 10. When $\alpha=\beta<1$, the valuation density is a bimodal U-shape function with modes at 0 and 1. As $\alpha$ and $\beta$ become smaller, the valuation is more concentrated on 0 and 1, which approximates the 2-point Bernoulli distribution with an equal probability of $1/2$.
In this case, as shown in \Cref{fig:mean_beta_equal},  the $\infty$-level pricing outperforms 1-level and 2-level pricing. The performance ratio of 1-level and 2-level pricing, especially the 1-level pricing, drops significantly as the valuation distribution degenerates to the two-point distribution, while the $\infty$-level pricing is more robust to the decreasing of $\alpha$ and $\beta$. 
The intuition is that the 1-level pricing can only extract about $\frac{v_1}{2}$ under the two-point distribution, and the 2-level pricing yields $\frac{x_1v_1+(1-x_1)v_2}{2}$, while the $\infty$-level pricing spreads more evenly across the valuation, extracting more revenue from valuation 1.
On the other hand, when $\alpha=\beta>1$, the valuation is a symmetric unimodal function with a mode of $1/2$.
As $\alpha$ and $\beta$ become larger, the valuation is more concentrated in the mean, i.e. 0.5, and the distribution has a more similar shape to the normal distribution. \Cref{fig:mean_beta_equal} indicates that as the valuation becomes more concentrated to the mean, the performance of 2-level pricing is significantly better than the other two mechanisms, and it is more robust to the increase of $\alpha$. The intuition is that as the valuation becomes more concentrated at the mean, the $\infty$-level pricing loses some revenue due to the dispersion above $\mu$, and the regret from $1$-level pricing is the gap between $v_1$ and the optimal pricing which approaches $\mu$. In contrast, the 2-level pricing mechanism takes advantage of the randomization between $v_1$ and $v_2$ that is slightly lower than $\mu$.
Finally, when $\alpha=\beta=1$, the valuation follows a uniform distribution on $[0, 1]$. Under the uniform distribution, the 2-level pricing also displays a superior performance ratio compared to 1-level and $\infty$-level pricing. This simulation highlights that a robust mechanism with a higher competitive ratio does not necessarily yield superior empirical performance under a given distribution. Nevertheless, the 2-level pricing exhibits robustness in the face of changes to the shape of the valuation distribution.

\Cref{fig:mean_beta_05} depicts the performance of 1-level, 2-level and $\infty$-level pricing as $\alpha$ ranges from 0.1 to 10 with $\beta$ set at $0.5$. When $\alpha<1$, the valuation density is U-shaped with two modes at 0 and 1. As per the previous discussion, the $\infty$-level pricing scheme demonstrates superior performance over the other two schemes. As $\alpha$ becomes greater than 1, the density function becomes unimodal with mode 1, and it is convex and increasing. As the valuation becomes more concentrated at mode 1, the 2-level pricing has a slightly higher performance than the other two schemes.

\Cref{fig:mean_beta_1} presents the performance of the three pricing schemes as $\alpha$ ranges from 0.1 to 10 and $\beta=1$. In this case, the density function becomes  $\alpha x^{\alpha-1}$. For $\alpha<1$, the density has a mode 0 and is convex and decreasing. In this scenario, both the 2-level  and $\infty$-level pricing schemes significantly outperform the 1-level pricing scheme. The reason is that since the density function is positively skewed with a right tail, the mean of valuation is low. Thus, the 1-level pricing has a conservative posted price that can not capture the revenue achievable from the high valuation in the right tail. In contrast, the 2-level pricing can mitigate this conservativeness effectively. As $\alpha$ becomes close to 1, the density function becomes flatter ultimately converging to a uniform distribution, and the 2-level pricing becomes more effective than the $\infty$-level pricing. As $\alpha$ becomes greater than 1, the density function becomes negatively skewed and strictly increases. It is concave, linear, and convex when $1<\alpha<2$, $\alpha=2$, and $\alpha>2$, respectively. As shown in \Cref{fig:mean_beta_1}, the 2-level pricing scheme outperforms the other two selling mechanisms when $\alpha\ge 1$. As $\alpha$ increases, the valuation is more concentrated at 1, leading to all three mechanisms approaching a performance of 1.

\Cref{fig:mean_beta_2} displays the performance of the three mechanisms as $\alpha$ ranges from 0.1 to 10 and $\beta=2$. When $\alpha<1$, similar to the previous case, the density has a mode 0 and is convex and decreasing. In this scenario, the performance of 2-level pricing is more effective than the other two mechanisms. When $\alpha>1$, the density function is unimodal, and the performance of 1-level pricing increases significantly. Though it is still below the performance of 2-level pricing, it surpasses that of the $\infty$-level pricing for all $\alpha \ge 1$. 

Our numerical results demonstrate the effectiveness of 2-level pricing under different shapes of distributions. The empirical performance of 2-level pricing has a significant improvement from the 1-level pricing, and 2-level pricing is more robust to parameter variations. 
In addition, while the competitive ratio of 2-level pricing is not as high as that of $\infty$-level pricing, the empirical performance of 2-level pricing can be superior to that of $\infty$-level pricing across many different distributions. 
\subsection{Quantile Ambiguity Set}
Now that we have characterized the competitive ratio for different selling schemes in \Cref{sec:quantile}, we proceed to assess the empirical performance of these mechanisms. Our analysis employs similar distributions as presented in the previous subsection, specifically using  Beta distributions with $\beta$ values of 0.5, 1, and 2, while varying $\alpha$ between 0.1 and 10. For each Beta distribution, we generate $(1-\xi)$-quantiles for $\xi$ values of 0.3, 0.5, and 0.7, and then the robust selling mechanisms are designed based on these quantile values. The main insights from the empirical performance are outlined below, followed by detailed discussions.

\begin{itemize}
    \item When the valuation distribution does not have a mode at zero, then it is more beneficial to leverage the information of a moderate or large quantile level $\xi$, which yields high performance for all three robust mechanisms.
    \item 
     When the valuation distribution presents two modes at 0 and 1 and negative skew, 2-level pricing, and $\infty$-level pricing both have good performance, irrespective of whether $\xi$ is small, moderate, or large.
      \item 
    In cases where the valuation distribution has two modes at 0 and 1 with positive skew, it is advantageous to utilize information from a small or moderate quantile level $\xi$. While 2-level pricing and $\infty$-level pricing outperform the 1-level pricing, none of the robust mechanisms perform very well in this scenario.
     \item Across all types of distributions, the performance of 2-level pricing is robust to the choice of quantile-level information.
\end{itemize}
\begin{figure}[htpb]
\centering
\captionsetup{justification=centering}
\caption{Performance Ratio for $n$-Level Pricing for $n=1,2$ and Infinity Based the Quantile Ambiguity Set Under Beta Distributions}
\label{fig:quantile_beta}
\begin{subfigure}[t]{0.32\textwidth}
\includegraphics[width=\textwidth,keepaspectratio]{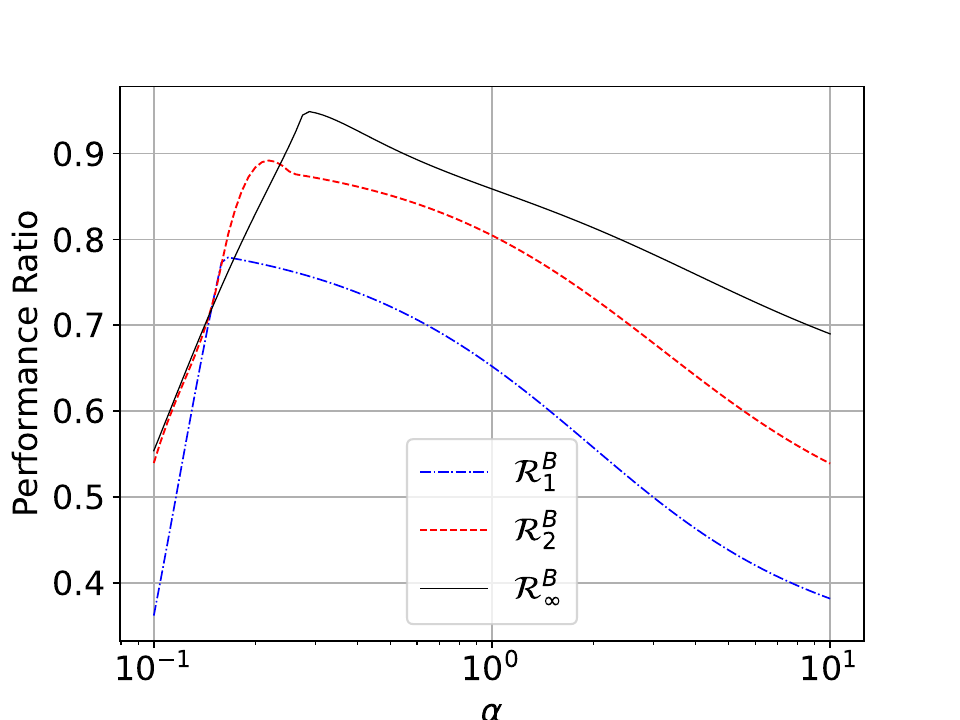}
\caption{$\beta=0.5, \xi=0.3$}
\label{fig:quantile-beta05-xi03}
\end{subfigure}
\begin{subfigure}[t]{0.32\textwidth}
\includegraphics[width=\textwidth,keepaspectratio]{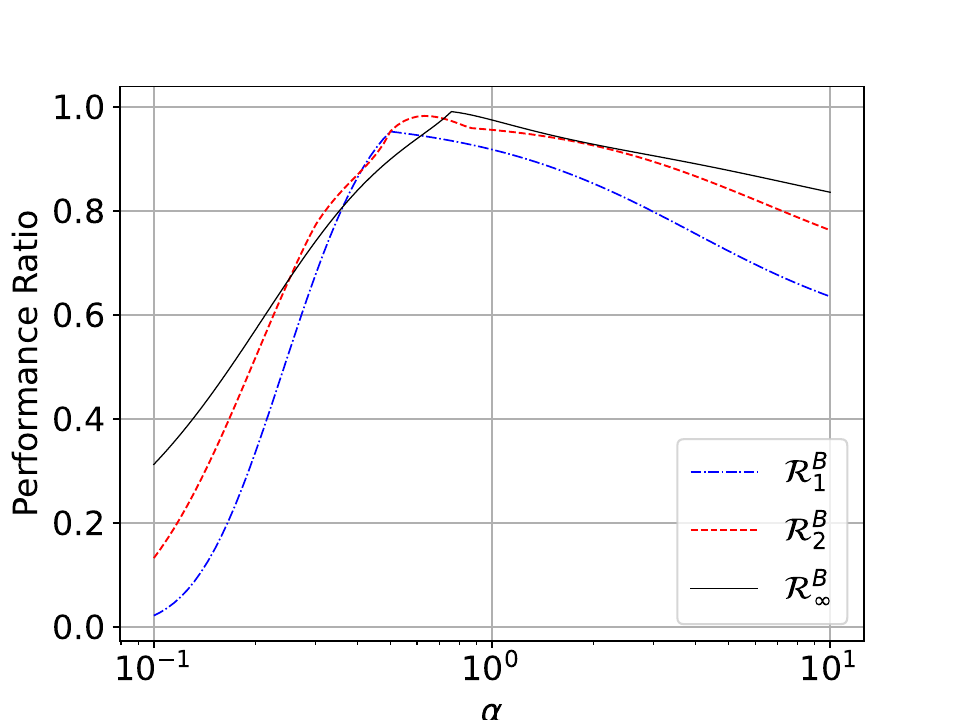}
\caption{$\beta=0.5, \xi=0.5$}
\label{fig:quantile-beta05-xi05}
\end{subfigure}
\begin{subfigure}[t]{0.32\textwidth}
\includegraphics[width=\textwidth,keepaspectratio]{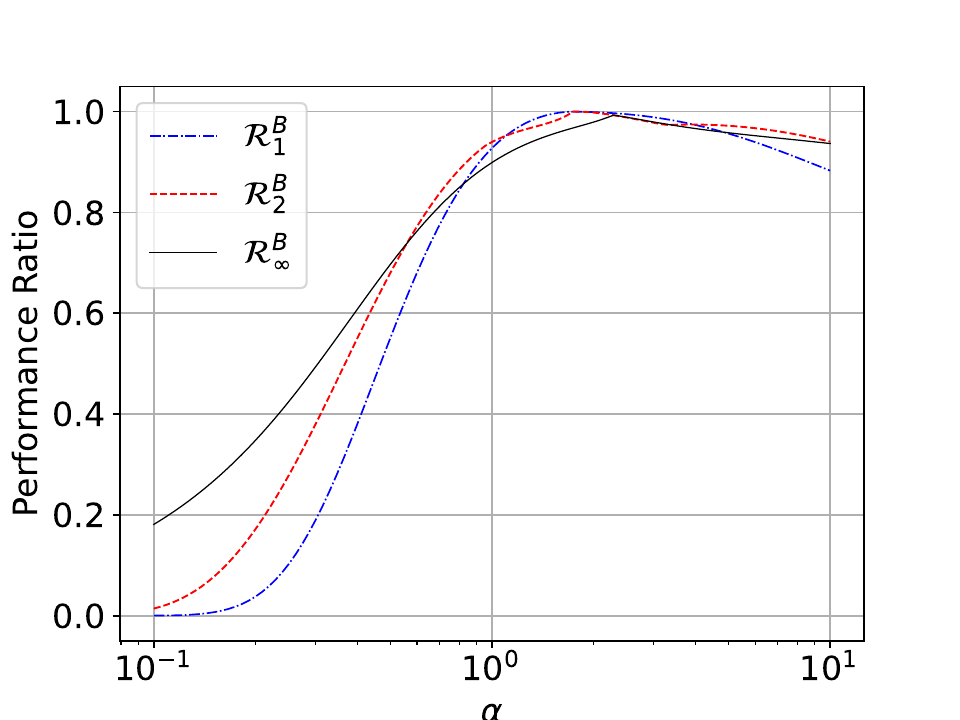}
\caption{$\beta=0.5, \xi=0.7$}
\label{fig:quantile-beta05-xi07}
\end{subfigure}
\begin{subfigure}[t]{0.32\textwidth}
\includegraphics[width=\textwidth,keepaspectratio]{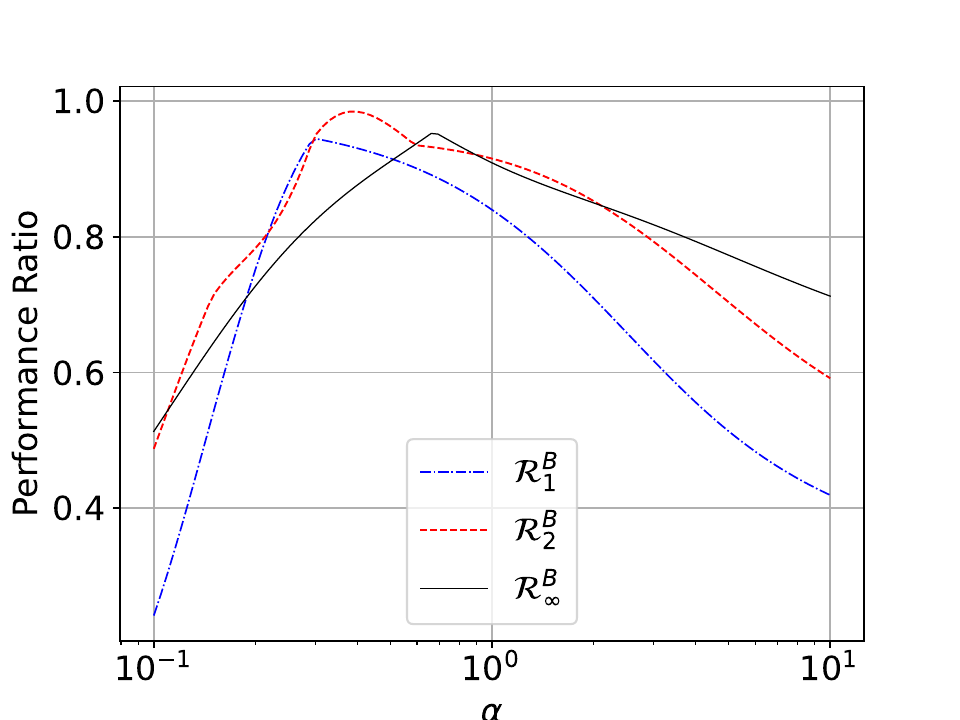}
\caption{$\beta=1, \xi=0.3$}
\label{fig:quantile-beta1-xi03}
\end{subfigure}
\begin{subfigure}[t]{0.32\textwidth}
\includegraphics[width=\textwidth,keepaspectratio]{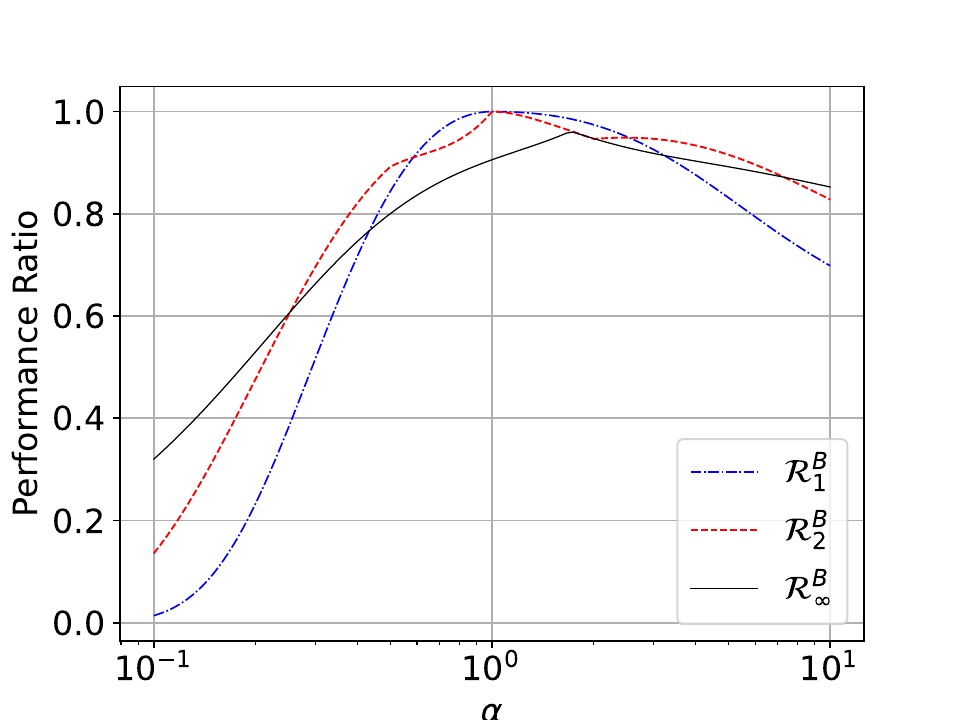}
\caption{$\beta=1, \xi=0.5$}
\label{fig:quantile-beta1-xi05}
\end{subfigure}
\begin{subfigure}[t]{0.32\textwidth}
\includegraphics[width=\textwidth,keepaspectratio]{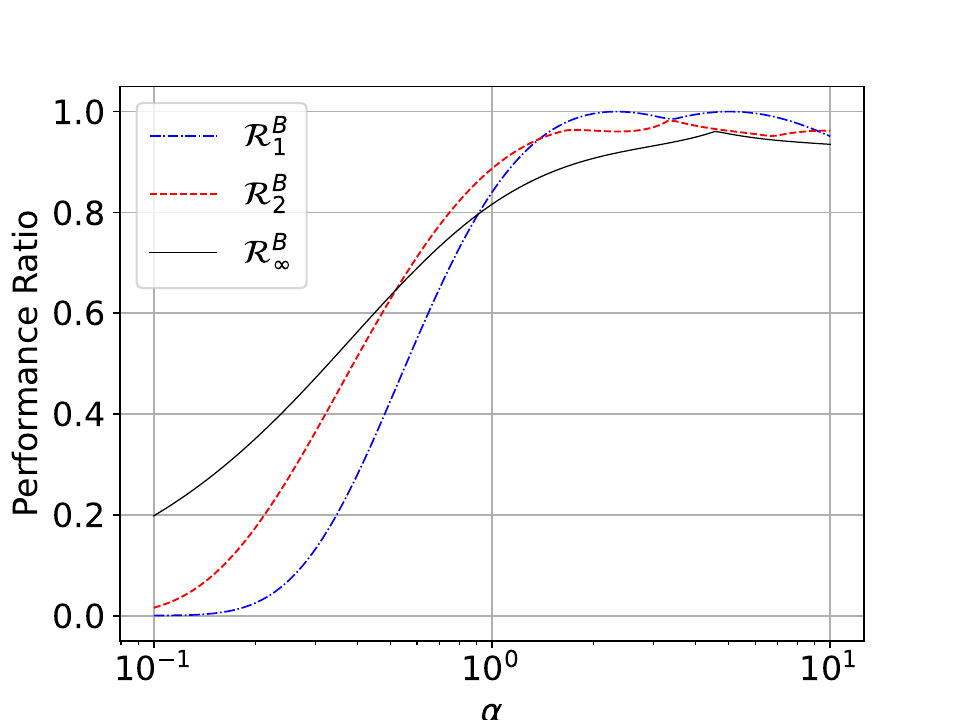}
\caption{$\beta=1, \xi=0.7$}
\label{fig:quantile-beta1-xi07}
\end{subfigure}
\begin{subfigure}[t]{0.32\textwidth}
\includegraphics[width=\textwidth,keepaspectratio]{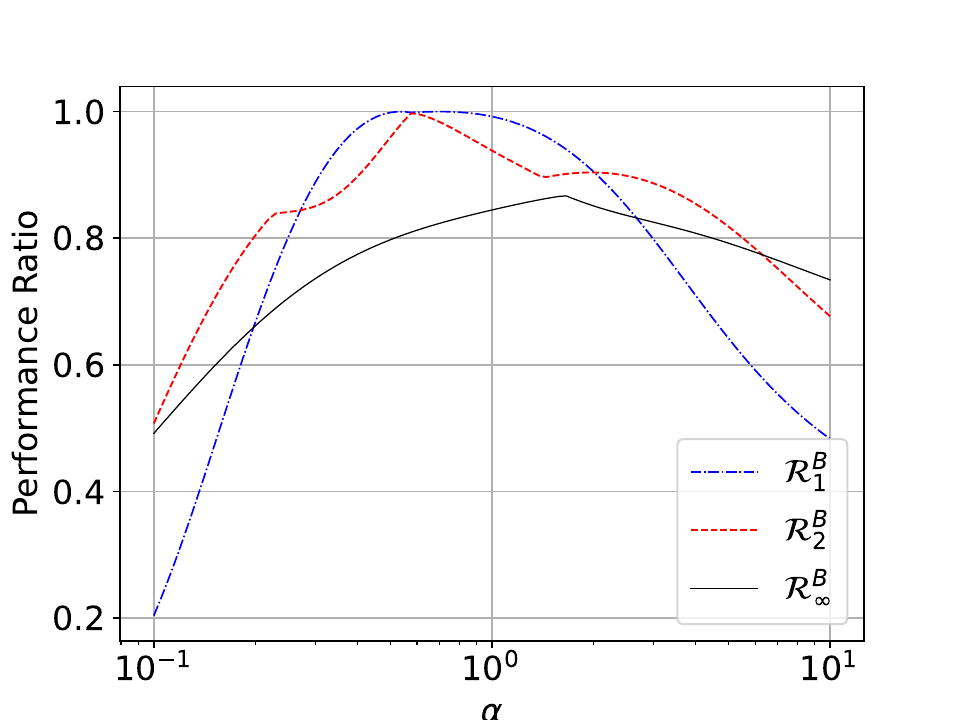}
\caption{$\beta=2, \xi=0.3$}
\label{fig:quantile-beta2-xi03}
\end{subfigure}
\begin{subfigure}[t]{0.32\textwidth}
\includegraphics[width=\textwidth,keepaspectratio]{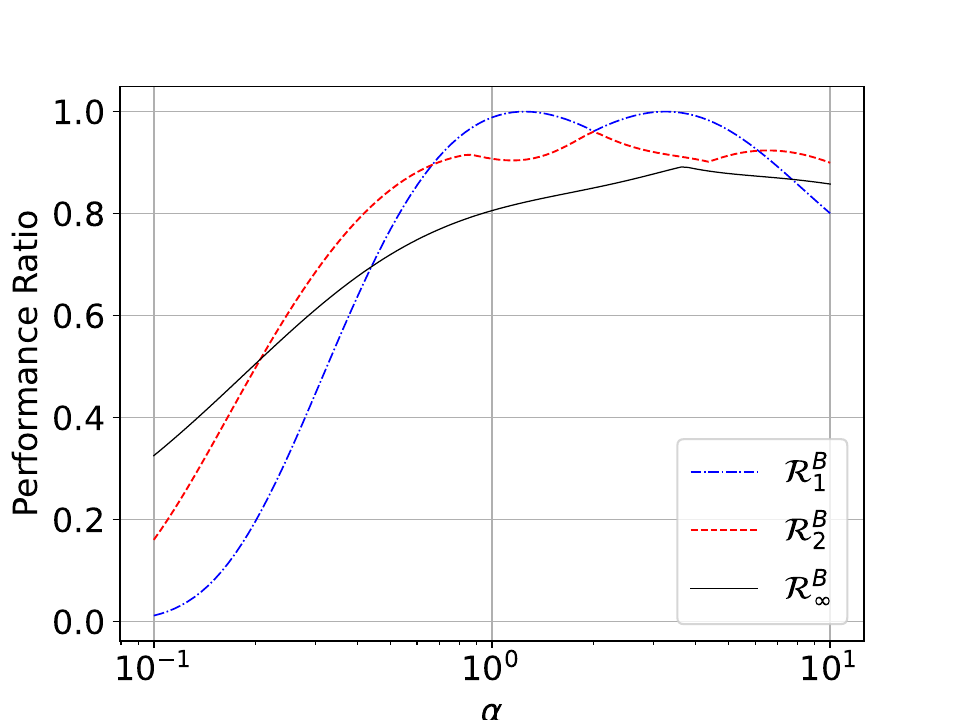}
\caption{$\beta=2, \xi=0.5$}
\label{fig:quantile-beta2-xi05}
\end{subfigure}
\begin{subfigure}[t]{0.32\textwidth}
\includegraphics[width=\textwidth,keepaspectratio]{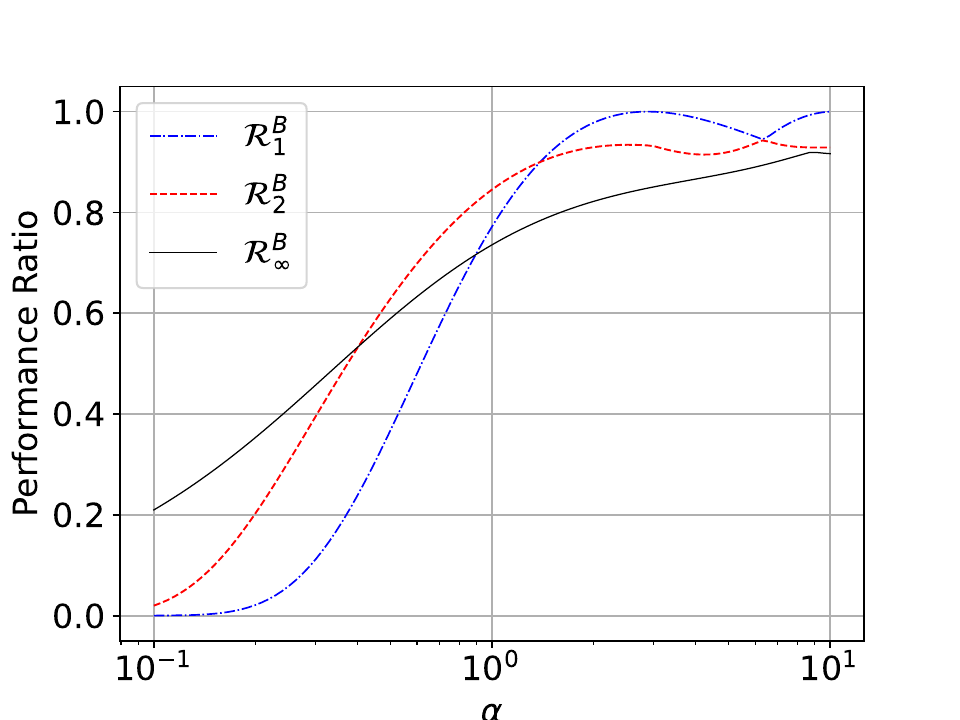}
\caption{$\beta=2, \xi=0.7$}
\label{fig:quantile-beta2-xi07}
\end{subfigure}
\end{figure}

In \Cref{fig:quantile_beta}, we observe that when $\alpha$ is small, the valuation distribution has a mode at 0 or two modes at 0 and 1. Given this configuration, the $(1-\xi)$-quantile tends to be small, meaning that a robust mechanism based on a near-zero quantile $\omega$ does not assure substantial revenue. Consequently, none of the three robust mechanisms has a good performance in this scenario. For $\beta = 0.5$ and $\alpha$ approximating or exceeding 1, the valuation distribution mode shifts to 1. In these cases, adopting a moderate or large quantile level $\xi$, such as $\xi = 0.5$ or $\xi=0.7$, consistently leads to good performance for all the robust selling mechanisms. For small $\xi=0.3$, the performance of the $\infty$-level pricing and 2-level pricing is more robust to the increase of $\alpha$. This is attributed to the fact that for small $\xi$, even if the $(1-\xi)$-quantile is large, the 1-level pricing posts price at $\xi$, which results in a revenue loss. In contrast, in 2-level pricing, even though the lower price is also at $\xi$, the higher price set at $\min\{\omega,\sqrt{\xi}\}$ can extract the revenue at valuation $\sqrt{\xi}$ efficiently.

When $\beta=1$ or $\beta=2$, and $\alpha$ is not very close to 0.1, the performance of 1-level pricing and 2-level pricing outperform $\infty$-level pricing under moderate $\alpha$ for any choice of quantile information. On the other hand, $\infty$-level pricing outperforms 1-level pricing and 2-level pricing for large $\alpha$ and small $\xi$. The reason that 1-level and 2-level pricing have superior performance under moderate $\alpha$ and $\beta$ is that in these scenarios, the valuation distribution is a single mode so adopting a moderate or large $\xi$ can secure a large amount of revenue captured by the mode in the distribution. Similarly, the performance is inferior if the valuation distribution has a mode at 0. In light of the numerical results, we can see that the 2-level pricing is very robust to the choice of quantile level $\xi$ across different distributions.
\section{Proofs}
\subsection{Proofs in \Cref{sec:basic formulation}}
\begin{proof}{Proof of \Cref{prop:lpformulation_inf}.}
Problem \eqref{original} can be represented as $\sup_{\bpi\in \bPi_n} \CR(\bpi)=
\sup\limits_{\bpi\in \bPi_n}\inf\limits_{\F\in \cF}\min\limits_{p\in [0,\uv]}\frac{\int_{0}^{\uv}(\int_0^v \pi(u)u\, du)\, d\F(v)}{ p(1-\F(p-))}$, so we can exchange the order of minimization. Hence, Problem \eqref{original} is equivalent to 
\begin{equation*}
\sup_{\bpi\in \bPi_n} \CR(\bpi)=
\sup_{\bpi\in \bPi_n}\min_{p\in [0,\uv]}\inf_{\F\in \cF} \ \frac{\int_{0}^{\uv}(\int_0^v \pi(u)u\, du)\, d\F(v)}{ p(1-\F(p-))}
\end{equation*}
Denoting $r=\CR(\bpi)$, we have the following equivalent formulation to the above problem.
\beq
\sup_{r\in \R_+, \bpi\in \bPi_n} & & r \nonumber\\
\mbox{s.t.} & & r\cdot p\ \le \ \inf_{\F\in \cF} \ \frac{\int_{0}^{\uv}(\int_0^v \pi(u)u\, du)\, d\F(v)}{ (1-\F(p-))}, \quad \forall p\in [0,\uv] \label{inner}
\eeq 
Incorporating the definition of $\cF$ defined in \eqref{worstf}, the infimum problem on the right-hand side of constraint \eqref{inner} for each $p\in[0,\uv]$ can be written in the following problem:
\beqn
\begin{aligned}
\displaystyle
\inf_{d\F\ge 0}\quad & \frac{\int_{0}^{\uv}(\int_0^v \pi(u)u\, du)\, d\,\F(v)}{ 1-\F(p-)}
 \\
 \mathrm{s.t.} \quad & \int_{v\in[0,\uv]}\,d\,\F(v) = 1\\
& \int_{v\in[0,\uv]}\phi_k(v)\,d\,\F(v) \ge 0, \ 
\forall k=1,2\ldots,K.
\end{aligned}
\eeqn
Then we introduce new decision variables $H_p(v) =\frac{\F(v)}{ 1-\F(p-)}$, in order to get rid of the nonlinear objective, and the problem above is equivalent to a linear programming problem: 
\beqn
\begin{aligned}
\displaystyle
\inf_{dH_p\ge 0}\quad & \int_{0}^{\uv}(\int_0^v \pi(u)u\, du)\,d\,H_p(v)
 \\
 \mathrm{s.t.} \quad & \int_{v\in[p,\uv]}\,d\,H_p(v) = 1\\
& \int_{v\in[0,\uv]}\phi_k(v)\,d\,H_p(v) \ge 0, \
\forall \ k=1,2\ldots,K.
\end{aligned}
\eeqn
Let $\lmb_p=(\lm_0(p),\lm_1(p),\dots,\lm_K(p))\in \R \times \R^{K}_+$ be the Lagrange multipliers associated with the constraints above. Let us consider the Lagrangian function 
\begin{align*}
\cL(H_p(\cdot), \lmb_p) &= \int_{0}^{\uv}(\int_0^v \pi(u)u\, du)\,d\,H_p(v) + \lm_0(p)\cdot \bigof{1- \int_{v\in[p,\uv]}\,d\,H_p(v)} - \sum_{k=1}^K \lm_k(p)\cdot \int_{v\in[0,\uv]}\phi_k(v)\,d\,H_p(v) \\
& = \lm_0(p)+\int_{0}^{\uv}\bigof{\int_0^v \pi(u)u \, du - \lm_0(p) \mathbbm{1}[v\ge p] -\sum_{k=1}^K \lm_k(p)\phi_k(v) }
 \,d\,H_p(v) 
\end{align*}
Since $\cL$ is bilinear in $H_p$ and $\lmb_p$, it satisfies that $\cL(\cdot, \lmb_p)$ is continuous and quasi-convex in $H_p$ for any $\lmb_p$ and $\cL(H_p, \cdot)$ continuous and quasi-concave in $\lmb_p$ for any $H_p$. 
Moreover, the feasible sets of $H_p(\cdot)$ and $\lmb_p$ are both convex subsets within their respective linear topological spaces. By Sion's theorem \citep{sion1958general}, we have that 
\begin{align*}
\inf_{dH_p\ge 0} \sup_{\lmb_p\in \R \times \R^{K}_+}  \cL(H_p(\cdot), \lmb_p) = \sup_{\lmb_p\in \R \times \R^{K}_+}\inf_{dH_p\ge 0}   \cL(H_p(\cdot), \lmb_p) 
\end{align*}
Hence, the optimal objective value for the above problem is equal to that of its dual problem, which can be formulated as follows. 
\beqn
\ba
\displaystyle
\sup_{\lmb_p\in \R \times \R^{K}_+} \quad & \lm_0(p)
 \\
 \mathrm{s.t.} 
& \sum_{k=1}^{K}\phi_k(v)\lm_{k}(p)\le \int_0^v \pi(u)u\, du \quad\forall v\in[0,p) \\
& \lm_0(p)+\sum_{k=1}^{K}\phi_k(v)\lm_{k}(p)\le \int_0^v \pi(u)u\, du \quad\forall v\in[p, \uv] 
\ea
\eeqn

Substituting the right-hand side of constraint \eqref{inner} with the above maximization problem, we have that
\beqn
\sup_{\lmb\in \R^{(K+1)\times [0,\uv]}_+, r\in \R_+, \bpi\in \bPi_n} & & r \\
\mbox{s.t.} & & r\cdot p\ \le \ \lm_0(p), \quad \forall p\in [0,\uv]  \\
&& \sum_{k=1}^{K}\phi_k(v)\lm_{k}(p)\le \int_0^v \pi(u)u\, du \quad\forall p\in [0,\uv], 
 \, v\in[0,p) \\
&& \lm_0(p)+\sum_{k=1}^{K}\phi_k(v)\lm_{k}(p)\le \int_0^v \pi(u)u\, du \quad\forall p\in [0,\uv],
 \,  v\in[p, \uv] 
\eeqn 
Notice that $\lm_0(p)$ can be eliminated by letting $\lm_0(p)=r\cdot p$, and then the problem above becomes Problem \eqref{primal}, which completes our proof of \Cref{prop:lpformulation_inf}.\QED
\end{proof}

\proof{Proof of \Cref{thm:finitelp0}.}
Problem \eqref{primal} has an infinite number of constraints characterized by an infinite sequence of $p\in[0,\uv]$ and $v\in[0,\uv]$. To reduce the number of constraints, we first prove the redundancy of most constraints indexed by $v$. We demonstrate that based on the optimal choice of $\lmb\in \R^{K\times [0,\uv]}_+$, the number of potentially active constraints is finite.
To reduce the number of decision variables indexed by $p\in[0,\uv]$, we prove the existence of an optimal solution to Problem \eqref{primal}, wherein $\lambda_k(p)$ is a piecewise constant function of $p$. In particular, the set of distinct values that $\lambda_k(p)$ can take over the domain $p\in[0,\uv]$ is finite.
In this way, we are able to reduce both the dimension of decision variables and that of constraints to finite. 

Given the price level set $\cV=\{v_1,v_2,\dots, v_n\}$, we denote $x_i$ the probability that the seller posts a price at $v_i$ for all $i=1,\dots, n$, and then the payment function can be represented by a step function of $v$. Specifically, the right-hand side of the constraints in Problem \eqref{primal} satisfies $\int_0^v \pi(u)u\, du=\sum_{l=1}^j x_lv_l$, if $v\in[v_j,v_{j+1})$ for all $j=0,1,\dots,n+1$. Embedding this relationship into the right-hand side of the constraints in \eqref{primal}, we have an equivalent formulation of \eqref{primal} as follows. Here we denote $v_0=0, v_{n+1}=\uv$, and $j(v) = \argmax_{j=0,1,\dots,n} v_j\le v$.
 \begin{equation*}
\ba
\displaystyle
\sup\limits_{\lmb \in \R^{K\times [0,\uv]}_+,\ \bx\in\R_+^n, \ r\in \R^+} & r \\
 \mathrm{s.t.}\quad  &   \sum_{k=1}^{K}\phi_k(v)\lm_{k}(p)\le \sum_{l=1}^{j(v)} x_lv_l\quad \forall  p\in[0,\uv], \  v\in[0,p) \\
 & r\cdot p+ \sum_{k=1}^{K}\phi_k(v)\lm_{k}(p)\le \sum_{l=1}^{j(v)}  x_lv_l\quad \forall p\in[0,\uv], \ \forall v\in[p, \uv]\\
& \sum_{i=1}^n x_i=1
\ea
\end{equation*} 
Due to \Cref{assume:phi}, since each $\phi_k(v)$ is piecewise continuous with $\cB$ as the joint set of breakpoints, we have that for all $\lmb$, $\sum_{k=1}^{K}\phi_k(v)\lm_{k}(p)$ is also piecewise continuous with breakpoints taking values from $\cB$. 
At the same time, since $\phi_k(\cdot)$ is nondecreasing within each segment, for all $\lmb\ge \allzero$, $\sum_{k=1}^{K}\phi_k(v)\lm_{k}(p)$ is also nondecreasing within each segment. Notice that the right-hand side $\sum_{l=1}^{j(v)}  x_lv_l$ is a piecewise constant function of $v$, so the difference between them  $\sum_{k=1}^{K}\phi_k(v)\lm_{k}(p)-\sum_{l=1}^{j(v)}  x_lv_l$ is a piecewise continuous function that is nondecreasing within each segment. Besides, the function added by a constant term $r\cdot p$ also satisfies this property, which corresponds to the second set of constraints $r\cdot p+ \sum_{k=1}^{K}\phi_k(v)\lm_{k}(p)- \sum_{l=1}^{j(v)}  x_lv_l$. 
Though the constraints require the negativity of an infinite number of piecewise increasing functions indexed by $p\in[0,\uv]$, each piecewise increasing function achieves its maximum at the right endpoint in each segment. Therefore, the only active constraints in the above optimization problem are indexed by $v\rightarrow u_{j}-$ for some $u_j\in \cV\cup \cB \cup \{\uv\}$. Hence, we reduce the number of constraints and obtain an equivalent formulation as follows:
\begin{equation*}
\ba
\displaystyle
\sup\limits_{\lmb \in \R^{K\times [0,\uv]}_+,\ \bx\in\R_+^n, \ r\in \R^+} & r \\
 \mathrm{s.t.}\quad  &   \sum_{k=1}^{K}\lphi_{k,j}\lm_{k}(p)\le \sum_{l\in S_{j}}x_lv_l\quad \forall  p\in[0,\uv], \  j=1,\dots, j^{\dag}(p)\\
 & r\cdot p+ \sum_{k=1}^{K}\lphi_{k,j}\lm_{k}(p)\le \sum_{l\in S_{j}} x_lv_l\quad \forall p\in[0,\uv], \ j=j^{\dag}(p)+1,\dots,n+m+1\\
& \sum_{i=1}^n x_i=1 . 
\ea
\end{equation*} 
where  $S_{j}=\{l =1,\dots,n\mid v_l<u_j\}$ for $j=1,\dots,n+m+1$ and $j^{\dag}(p) = \argmax_j  u_j\le p$. Next, we aim to reduce the dimension of decision variables $\lmb \in \R^{K\times [0,\uv]}_+$. For any $i=0,1,\dots,n+m$, the constraints corresponding to $\lmb(p)$ for $p\in[u_{i},u_{i+1})$ are represented as
\begin{subequations}
    \begin{align}
 &   \sum_{k=1}^{K}\lphi_{k,j}\lm_{k}(p)\le \sum_{l\in S_{j}}x_lv_l\quad \forall  \  j\in [i] \label{consp1}\\
 & r\cdot p+ \sum_{k=1}^{K}\lphi_{k,j}\lm_{k}(p)\le \sum_{l\in S_{j}} x_lv_l\quad \forall \ j=i+1,\dots,n+m+1. \label{consp2}
\end{align}
\end{subequations} 
Notice that for each constraint with the same index $j$, the right-hand side $\sum_{l\in S_{j}}x_lv_l$ is the same regardless of $p$. At the same time, the coefficient for $\lambda_k(p)$ on the left-hand side, $\lphi_{k,j}$ is the same for all $p\in[u_{i},u_{i+1})$. Thus, for the constraints with the same indexed by $j$, the only difference in the constraints across different $p$ lies in the term $r\cdot p$. 
Since $r\cdot p$ is increasing in $p$, any feasible variable $\{\lm_k(u_{i+1}-)\}_{k=1}^K$ satisfying the following constraints
\begin{align*}
 &   \sum_{k=1}^{K}\lphi_{k,j}\lm_{k}(u_{i+1}-)\le \sum_{l\in S_{j}}x_lv_l\quad \forall  \  j\in [i]\\
 & r\cdot u_{i+1}+ \sum_{k=1}^{K}\lphi_{k,j}\lm_{k}(u_{i+1}-)\le \sum_{l\in S_{j}} x_lv_l\quad \forall \ j=i+1,\dots,n+m+1
\end{align*}
is also feasible for all constraints in \eqref{consp1} and \eqref{consp2} for any $p\in[u_{i},u_{i+1})$. Thus, we could let $\lm_k(p)=\lm_k(u_{i+1}-)$ for all $p\in[u_{i},u_{i+1})$, and satisfying the constraints corresponding to $p=u_{i+1}-$ implies the feasibility of all other constraints indexed by $p\in[u_{i},u_{i+1})$. Hence, it is sufficient to determine the decision variables $\lmb_k(\cdot)$ indexed by the right endpoints of all segments partitioned by the valuations in $\cV\cup \cB$.  The above analysis indicates that any feasible solution $\lmb_k(\cdot)$ can be represented as a piecewise constant function of $p$, whose range contains at most $n+m+1$ values. Therefore, we can reduce the dimension of decision variables $\lmb$ to $K\times (n+m+1)$. Hence, incorporating all constraints for $i\in[n+m+1]$ and slightly reindexing the constraints provides the equivalent formulation in problem \eqref{primal-finite0}. \QED\endproof
\subsection{Proofs in \Cref{sec:support}}
\proof{Proof of \Cref{lemma:fixv-support}.}
As discussed above, if $v_1>\lv $, then $r=0$, so we must have $v_1=\lv$. Hence, constraint \eqref{eq:supportcon3} becomes $ r\cdot v_1 \le x_1\cdot v_1$, which is implied by the first constraint in \eqref{eq:supportcon1}, i.e., $ r\cdot v_2 \le x_1\cdot v_1$. Therefore, the optimal competitive ratio $r$ solved by Problem \eqref{eq:supportobj} is equivalent to
\beqn
r=\min_{i=2,\dots,n+1}\left\{
\frac{\sum_{j=1}^{i-1} x_j\cdot v_j}{v_i}
\right\}
\eeqn 
For fixed $x_1,x_2,\dots, x_{t-1}$, let us define 
\begin{equation*}
\begin{aligned}
f_t(y\mid x_1,x_2,\dots,x_{t-1})=\max_{\{x_{t},\dots, x_{n}\mid \sum_{j=t}^n x_j\le y\}} \min_{i=t+1,\dots,n+1}\left\{
\frac{\sum_{j=1}^{i-1} x_j\cdot v_j}{v_i}\right\}
\end{aligned}
\end{equation*}
Then we have that 
\begin{equation*}
\begin{aligned}
f_t(y\mid x_1,x_2,\dots,x_{t-1})=\max_{0\le x_{t}\le y} \min\left\{
\frac{\sum_{j=1}^{t} x_j\cdot v_j}{v_{t+1}}, \ \,  f_{t+1}(y-x_t\mid x_1,\dots, x_{t-1},x_t)\right\}
\end{aligned}
\end{equation*}
Next we show that the second term is decreasing in $x_t$. For any $x_t$, suppose the optimal solution to problem $f_{t+1}(y-x_t\mid x_1,x_2,\dots,x_{t-1},x_t)$ is $x_{t+1},\dots, x_n$. Then consider solution $x_t'=x_t-\epsilon$,  $x_{t+1}'=x_{t+1}+\epsilon, x_{i}'=x_{i},\, \forall i\ge t+2$. Since
$\frac{\sum_{j=1}^{i-1} x_j'\cdot v_j}{v_i}= \frac{\sum\limits_{j<i,j\neq t,t+1} x_j'\cdot v_j+(x_{t}-\epsilon)v_t+(x_{t+1}+\epsilon)v_{t+1} }{v_i}=\frac{\sum\limits_{j<i} x_j\cdot v_j+\epsilon(v_{t+1}-v_t) }{v_i}>\frac{\sum_{j=1}^{i-1} x_j\cdot v_j}{v_i}$, for any $i\ge t+2$, we have that 
$\min\limits_{i=t+2,\dots,n+1}\left\{
\frac{\sum_{j=1}^{i-1} x_j'\cdot v_j}{v_i}\right\}\ge \min\limits_{i=t+2,\dots,n+1}\left\{
\frac{\sum_{j=1}^{i-1} x_j\cdot v_j}{v_i}\right\}$. It follows that 
\beqn 
f_{t+1}(y-x_t'\mid x_1,\dots, x_{t-1},x_t')&= &\max_{\{x_{t+1},\dots, x_{n}\mid \sum_{j=t+1}^n x_j\le y-x_t'\}} \min_{i=t+2,\dots,n+1}\left\{
\frac{\sum_{j=1}^{i-1} x_j\cdot v_j}{v_i}\right\}\\
&\ge & \min_{i=t+2,\dots,n+1}\left\{
\frac{\sum_{j=1}^{i-1} x_j'\cdot v_j}{v_i}\right\}\\
&\ge & \min_{i=t+2,\dots,n+1}\left\{
\frac{\sum_{j=1}^{i-1} x_j\cdot v_j}{v_i}\right\}\\
& =& f_{t+1}(y-x_t\mid x_1,\dots, x_{t-1},x_t)
\eeqn 
which implies that $ f_{t+1}(y-x_t\mid x_1,\dots, x_{t-1},x_t)$ is decreasing in $x_t$.
Now that $\frac{\sum_{j=1}^{t} x_j\cdot v_j}{v_{t+1}}$ is increasing in $x_t$ and  $ f_{t+1}(y-x_t\mid x_1,\dots, x_{t-1},x_t)$ is decreasing in $x_t$, the optimal $x_t$ equates the two terms in order to maximize the minimum of the two terms. Therefore, we have proved that $\frac{\sum_{j=1}^{t} x_j\cdot v_j}{v_{t+1}}=\frac{\sum_{j=1}^{t+1} x_j\cdot v_j}{v_{t+2}}$ for any $t$ by backward induction, so we have that 
\beqn 
r =  \frac{x_1\cdot v_1}{v_{2}}= \dots=\frac{\sum_{j=1}^{i}x_j\cdot v_j}{v_{i+1}}=\dots=\frac{\sum_{j=1}^{n}x_j\cdot v_j}{\uv}
\eeqn 
Hence, the optimal pricing probabilities are given by $x_1=r \cdot \frac{v_2}{\lv} ,\,  x_i = r\cdot\frac{v_{i+1}-v_{i}}{v_i},\, \forall \ i=2,\dots,n$. Taking into account $\sum_{i=1}^n x_i=1$, we have that $r= \big( \frac{v_2}{v_1} +\sum_{i=2}^n \frac{v_{i+1}-v_{i}}{v_{i}} \big)^{-1}$.
\endproof
\proof{Proof of \Cref{thm:support}.}
The optimal competitive ratio can be represented as
$
r =\big( \frac{v_2}{v_1} +\sum_{i=2}^n \frac{v_{i+1}-v_{i}}{v_{i}} \big)^{-1}
= \big(\sum_{i=1}^n \frac{v_{i+1}}{v_{i}} - (n-1) \big)^{-1}
$.
Given that $v_1=\lv, v_{n+1}=\uv$ are fixed, we now consider how $r$ changes with $v_i$ for $i=2,\dots,n$. 
The derivative of $\sum_{i=1}^{n} \frac{v_{i+1}}{v_{i}} $ with respect to $v_i$ is $-\frac{v_{i+1}}{v_i^2}+\frac{1}{v_{i-1}}$, which implies that $\sum_{i=1}^{n} \frac{v_{i+1}}{v_{i}} $ is decreasing in $v_i$ if $v_i\le \sqrt{v_{i-1}{v_{i+1}}}$ and increasing in $v_i$ if $v_i\ge \sqrt{v_{i-1}{v_{i+1}}}$.
Hence, the optimal $v_i$ satisfies $v_i=\sqrt{v_{i-1}{v_{i+1}}}$ for all $i=2,\dots,n$. It follows that $v_i=\lv\cdot (\uv/\lv)^{\frac{i-1}{n}}=\lv^{\frac{n+1-i}{n}}\cdot \uv^{\frac{i-1}{n}}$, for all $i=1,\dots, n$. Then by \Cref{lemma:fixv-support}, we obtain the optimal pricing probabilities for different price levels. \QED
\endproof 
\subsection{Proofs in \Cref{sec:mean}}
\proof{Proof of \Cref{thm:meansupport-1}.}
Notice that when $v_1> \mu$, since $\lm_1\ge 0$, the first constraint implies $r\cdot v_1\le 0$, suggesting $r\le 0$. Besides, $r=0, \lm_1=0$ forms a feasible solution to the above problem, so when $v_1>\mu$, the optimal competitive ratio achieved by a deterministic posted price $v_1$ is $0$. Subsequently, our analysis will focus on characterizing the non-trivial optimal solution for the scenario $v_1\le \mu$.

In the case where $v_1\le \mu$, the first constraint in \eqref{eq:meansupport-1} can be rewritten as $\lambda_1\ge \frac{r\cdot v_1}{\mu-v_1}$ and the second constraint can be rewritten as $\lambda_1\le \frac{v_1-r\cdot v_1}{\uv-\mu}$. For a feasible solution $\lm_1$ to exist, the intersection of these two conditions results in $r\le \frac{\mu-v_1}{\uv-v_1}$. Taking into account the third constraint $r\le \frac{v_1}{\uv}$, we have that the optimal objective value $r$ in Problem \eqref{eq:meansupport-1} is 
\beqn 
\cR(\{v_1\})=\min\{\frac{v_1}{\uv}, \frac{\mu-v_1}{\uv-v_1}\}
\eeqn 
Consequently, the optimal posted price selling scheme is deduced by
$
v_1^*=\argmax_{v_1}\{\min\{\frac{v_1}{\uv}, \frac{\mu-v_1}{\uv-v_1}\}\}
$.
Since the derivative of $\frac{\mu-v_1}{\uv-v_1}$ with respective to $v_1$ is $\frac{\mu-\uv}{(\uv-v_1)^2}\le 0$, the second term $\frac{\mu-v_1}{\uv-v_1}$ is decreasing in $v_1$. Moreover, since the first term is $\frac{v_1}{\uv}$ is increasing in $v_1$, the optimal $v_1$ is obtained at $\frac{v_1}{\uv}=\frac{\mu-v_1}{\uv-v_1}$. Hence, the optimal 1-level pricing scheme is to post a deterministic price at $v_1^*=\uv-\sqrt{\uv^2-\mu\uv}$. Consequently, the optimal competitive ratio achieved by the 1-level pricing scheme is 
$$
\cR_1^* = \max_{v_1}\cR(\{v_1\}) = \frac{v_1^*}{\uv} = 1-\sqrt{1-\mu/\uv}
$$ \QED
 \endproof

\proof{Proof of \Cref{thm:meansupport-2}.}
If $v_1\ge \mu$, then constraint\eqref{primal-meansupport-level2-11} implies that $r\le 0$, so we focus only on the case where $v_1<\mu$. 
In the following analysis, we further simplify Problem \eqref{eq:meansupport-2} by discussing the value of $v_2$.
\begin{enumerate}
    \item When $v_2\ge \mu$, since the coefficient of $\lm_2$ in the constraints \eqref{primal-meansupport-level2-14} and \eqref{primal-meansupport-level2-15} is nonnegative, the optimal $\lm_2$ is $0$. Hence, the constraints \eqref{primal-meansupport-level2-14} and \eqref{primal-meansupport-level2-15} are equivalent to $r\cdot v_2\le v_1 x_1$ and $r\cdot v_2\le v_1x_1+v_2x_2 $, respectively. Based on constraints
    \eqref{primal-meansupport-level2-11} to \eqref{primal-meansupport-level2-13}, the existence of a feasible $\lm_1$ is equivalent to $r\le \frac{x_1(\mu-v_1)}{v_2-v_1}$ and $r\le \frac{(\mu-v_1)(v_1x_1+v_2x_2)}{v_1(\uv-v_1)}$. Hence, the optimal competitive ratio $r$ is solved by 
    \beqn 
r = \min\left\{\frac{v_1x_1+v_2x_2}{\uv}, \ \frac{v_1x_1}{v_2},\,\frac{(\mu-v_1)(v_1x_1+v_2x_2)}{v_1(\uv-v_1)}, \, \frac{x_1(\mu-v_1)}{v_2-v_1}
\right\}
    \eeqn 
    Now we optimize $x_1$. The second and the fourth terms are increasing in $x_1$ and the first and the third terms are decreasing in $x_1$. After jointly optimizing $x_1,x_2, v_1,v_2$, we have that 
    \beqn
    v_1 = \uv-\sqrt{\uv^2-\mu\uv},\quad v_2=\sqrt{v_1\uv},\quad x_1 = \frac{1}{2-\sqrt{v_1/\uv}}, \quad
    r = \frac{1}{2\sqrt{\frac{\overline{v}}{v_1}}-1}
    \eeqn 
    \item When $v_2\le \mu$, \eqref{primal-meansupport-level2-14} is equivalent to $\lm_2\ge \frac{rv_2-v_1x_1}{\mu-v_2}$ and \eqref{primal-meansupport-level2-15} is equivalent to $\lm_2\le \frac{v_1x_1+v_2x_2-rv_2}{\uv-\mu}$. In order that $\lm_2$ has a nonnegative solution, the two constraints hold if and only if $r$ satisfies $r\le \frac{\mu-v_2}{\uv-v_2}x_2+\frac{v_1x_1}{v_2}$.
    Based on constraints
    \eqref{primal-meansupport-level2-11}-\eqref{primal-meansupport-level2-13}, the existence of a feasible $\lm_1$ is equivalent to $r\le \frac{(\mu-v_1)(v_1x_1+v_2x_2)}{v_1(\uv-v_1)}$ and 
    $r\le \frac{v_2x_2(\mu-v_2)}{v_1(\uv-v_2)}+x_1$. Notice that $\frac{v_2x_2(\mu-v_2)}{v_1(\uv-v_2)}+x_1 =\frac{v_2}{v_1}\cdot \big(\frac{\mu-v_2}{\uv-v_2}x_2+\frac{v_1x_1}{v_2}\big)\ge\frac{\mu-v_2}{\uv-v_2}x_2+\frac{v_1x_1}{v_2}$, so constraint $r\le \frac{v_2x_2(\mu-v_2)}{v_1(\uv-v_2)}+x_1$ is implied by the previous constraint  $r\le \frac{\mu-v_2}{\uv-v_2}x_2+\frac{v_1x_1}{v_2}$ derived by 
    \eqref{primal-meansupport-level2-14}-\eqref{primal-meansupport-level2-15}.
    Hence, the optimal competitive ratio $r$ is solved by 
    \beqn 
r = \min\left\{\frac{v_1x_1+v_2x_2}{\uv},\, \frac{(\mu-v_1)(v_1x_1+v_2x_2)}{v_1(\uv-v_1)},\, \frac{\mu-v_2}{\uv-v_2}x_2+\frac{v_1x_1}{v_2}
\right\}
    \eeqn 
    After jointly optimizing $x_1,x_2, v_1,v_2$, we have that 
   { \small
    \beqn
    v_1 & =& \uv-\sqrt{\uv^2-\mu\uv}, \,
v_2=\frac{v_{1}+\sqrt{v_{1}^2\,+8v_1\uv}}{4},\, x_1  =\frac{{v_2(2v_2\uv - v_2^2 -  \mu\uv)}}{{(v_2-v_1)(2v_2\uv - v_2^2 -  \mu\uv) + v_1\uv( \uv-\mu)}},\\
r &= &\frac{{v_1v_2(\uv-\mu)}}{{(v_2-v_1)(2v_2\uv - v_2^2 -  \mu\uv) + v_1\uv( \uv-\mu)}}.
    \eeqn
    }
\end{enumerate}
After verifying the feasible region of $v_2$, the optimal competitive ratio solved in Problem \eqref{eq:meansupport-2} is 
\beqn 
\cR_2^* =\max\left\{
\frac{1}{2\sqrt{\frac{\overline{v}}{v_1}}-1},\quad \frac{{v_1v_2(\uv-\mu)}}{{(v_2-v_1)(2v_2\uv - v_2^2 -  \mu\uv) + v_1\uv( \uv-\mu)}}
\right\}
\eeqn 
where $v_1 = \uv-\sqrt{\uv^2-\mu\uv}$ and $v_2=\frac{v_{1}+\sqrt{v_{1}^2\,+8v_1\uv}}{4}$. Moreover, we observe that $\frac{1}{2\sqrt{\frac{\overline{v}}{v_1}}-1}\ge \frac{{v_1v_2(\uv-\mu)}}{{(v_2-v_1)(2v_2\uv - v_2^2 -  \mu\uv) + v_1\uv( \uv-\mu)}}$ when $\mu\le 0.49\uv$ and $\frac{1}{2\sqrt{\frac{\overline{v}}{v_1}}-1}\le \frac{{v_1v_2(\uv-\mu)}}{{(v_2-v_1)(2v_2\uv - v_2^2 -  \mu\uv) + v_1\uv( \uv-\mu)}}$ when $\mu\ge 0.49\uv$, which directly yields the competitive ratio as defined in \Cref{thm:meansupport-2}.
\QED
\endproof

\proof{Proof of \Cref{thm:mean-var}.}
The proof is by constructing a feasible solution to problem \eqref{eq:mean-var}.
We consider the following $\lmb$.  
\begin{align*}
 & \lm_1(p) = \lm_1(v_1-), \lm_2(p) = \lm_2(v_1-), \forall p\in [0,v_1) \\
 &   \lm_1(p) = \lm_2(p) = 0, \forall p\in [v_1,\hat{v}). \\
  &  \lm_1(p) = \lm_1(\hat{v}), \lm_2(p) = \lm_2(\hat{v}), \forall p\in [\hat{v}, \infty)
\end{align*}
where $\hat{v} = \frac{v_1x_1+v_2 x_2}{r},   r\cdot v_2 = v_1x_1.$
Hence, we only need to determine the $\lm_1(p),\lm_2(p)$ for $p=v_1, \hat{v}$.
\begin{enumerate}
    \item For $p\in [v_1,\hat{v}]$, when $v<p$, the left-hand side (LFS) of constraints \eqref{eq:mean-var} is 0 which satisfies that LFS is less than or equal to the right-hand side (RHS). When $v\ge p$, the LFS is equal to $r\cdot p$, also less than or equal to the RHS which is $t(v) = x_1v_1 \ge r\cdot v$ for $v\in [v_1,v_2]$ and $t(v) = x_1v_1+x_2v_2 \ge r\cdot v$ for $v\in [v_2,\hat{v}]$. 
\item 
For $p=\hat{v}$, consider the following set of variables:
\begin{align*}
\lm_2(\hat{v}) &= -\frac{(\mu-v_1)r}{(v_1-\mu)^2 +\sigma^2} \\
\lm_1(\hat{v}) & = \frac{v_1^2-\mu^2-\sigma^2}{\mu-v_1} \lm_2(\hat{v})
\end{align*}
We can verify that $\lm_1,\lm_2$ defined above satisfy the following conditions.
\begin{align*}
&  \lm_1(\hat{v}) (v_1-\mu) + \lm_2(\hat{v})(v_1^2-\mu^2-\sigma^2) = 0\\
&  \lm_1(\hat{v}) (\hat{v}-\mu) + \lm_2(\hat{v})(\hat{v}^2-\mu^2-\sigma^2) = 0\\
&  2\lm_2(\hat{v}) v_1+\lm_1(\hat{v}) = r\\
& \lm_2(\hat{v}) < 0.
\end{align*}
In the following, we will show that the set of equalities and inequality above is a sufficient condition that all constraints for $p\ge \hat{v}$ hold. 
The first and second constraint ensures that the left-hand side of constraints \eqref{eq:mean-var} is 0 at $v_1$ and $\hat{v}$. The third constraint ensures that the gradient of the left-hand side is $r$ at $v_1$. Now for $v\in[0,v_1]$, the left-hand side of constraints in  \eqref{eq:mean-var} is negative, so it is less than the right-hand side, which is 0 for $v\in[0, v_1]$. 
Next for $v\in [v_1, \hat{v}]$, since the right-hand side $t(v) = v_1x_1 \mathbbm{1}[v\ge v_1] +v_2x_2 \mathbbm{1}[v\ge v_2]$ is always above $t(v) = rv$ for all $v\in [v_1, \hat{v}]$, the left-hand side is always below the right-hand side as long as its gradient is smaller than $r$ between $[v_1,\hat{v}]$ and its value is 0 at $v_1$. 
Finally, for $v\ge \hat{v}$, since the derivative of function $(v-\mu)\lm_{1}(\hat{v}) + (v^2-\mu^2-\sigma^2)\lm_2(\hat{v})$ with respect to $v$ is less than or equal to $-r$, the left-hand side of constraint is no greater than $r\cdot v+ (v-\mu)\lm_{1}(\hat{v}) + (v^2-\mu^2-\sigma^2)\lm_2(\hat{v})\le (r-r)\cdot(v-\hat{v}) +r\hat{v} =r\hat{v} $, where $r\hat{v}$  is the right-hand side of the constraint. Therefore, we have proved that for $p\ge \hat{v}$, $ \lm_2(p) = \lm_2(\hat{v}) = -\frac{(\mu-v_1)r}{(v_1-\mu)^2 +\sigma^2},  \lm_1(p) = \lm_1(\hat{v}) = \frac{v_1^2-\mu^2-\sigma^2}{\mu-v_1} \lm_2(\hat{v}) $ is feasible for all the constraints where $p\ge \hat{v}, v\in [0,\infty)$.
\item Consider $p\le v_1$.
It is sufficient if we can find $\lm_1(v_1),\lm_2(v_1)$ satisfying the following constraints.
\begin{align}
   & rv_1+\lm_1(v_1)\cdot (v_1-\mu) +\lm_2(v_1) \cdot (v_1^2-\mu^2-\sigma^2) \le 0 \label{eq:v1-1}\\
  &  rv_1+\lm_1(v_1)\cdot (v_2-\mu) +\lm_2(v_1)\cdot (v_2^2-\mu^2-\sigma^2) \le v_1x_1\label{eq:v1-2}\\
  &  rv_1-\lm_1(v_1)\cdot \mu -\lm_2(v_1) (\mu^2+\sigma^2) - \frac{\lm_1(v_1)^2}{4\lm_2(v_1)} \le r \frac{\mu^2+\sigma^2 - v_1\mu}{\mu-v_1} \label{eq:v1-3} \\
&    \lm_2(v_1) <0  \label{eq:v1-4}\\
 &   \frac{\lm_1(v_1)}{-2\lm_2(v_1)} \ge  v_2 \label{eq:v1-5}
\end{align}
\eqref{eq:v1-1} is equivalent to 
\begin{align*}
\lm_{1}(v_1)\ge \frac{r v_{1}-\lambda _{2}(v_1)\,\left(\mu ^2+\sigma ^2-{v_{1}}^2\right)}{\mu -v_{1}} 
\end{align*}
\eqref{eq:v1-3} is equivalent to 
\begin{align*}
  \lm_{1}(v_1)\in  \Bigoff{ -2\,\lambda _{2}(v_1)\,\Bigof{\mu -\sqrt{-\frac{r(\mu-v_1)^2+r\sigma ^2}{\lambda _{2}(v_1)\,\left(\mu -v_1\right)}-\sigma^2}},\, -2\,\lambda _{2}(v_1)\,\Bigof{\mu +\sqrt{-\frac{r(\mu-v_1)^2+r \sigma ^2}{\lambda _{2}\,\left(\mu -v_1\right)}-\sigma^2}}}
\end{align*}

Solving $\lm_{1}(v_1) =  \frac{rv_{1}-\lambda _{2}(v_1)\,\left(\mu ^2+\sigma ^2-{v_{1}}^2\right)}{\mu -v_{1}} $ and $\lm_1(v_1) = -2\,\lambda _{2}(v_1)\,\Bigof{\mu +\sqrt{-\frac{r(\mu-v_1)^2+r\sigma ^2}{\lambda _{2}\,\left(\mu -v_1\right)}-\sigma^2}}$, we have the following solution to $\lm_1(v_1)$ and $\lm_2(v_1)$
{\scriptsize
\begin{align}
\lm_2(v_1) &= r\frac{2\,\sqrt{\left(\mu -v_{1}\right)\,\left(\mu ^2-v_{1}\,\mu +\sigma ^2\right)\,\left((\mu - v_1)^3 + (\mu - 2v_1)\sigma^2\right)}-2\,\mu \,\sigma ^2-4\,\mu \,{v_{1}}^2+5\,\mu ^2\,v_{1}+3\,\sigma ^2\,v_{1}-2\,\mu ^3+{v_{1}}^3}{{\left(\mu ^2-2\,\mu \,v_{1}+\sigma ^2+{v_{1}}^2\right)}^2}\label{eq:def lambda2}\\
\lm_1(v_1) & = \frac{r v_{1}-\lambda _{2}(v_1)\cdot\,\left(\mu ^2+\sigma ^2-{v_{1}}^2\right)}{\mu -v_{1}} \label{eq:def lambda1}
\end{align}}
Then we need to check whether the $\lm_1(v_1),\lm_2(v_1)$ are feasible for \eqref{eq:v1-2},\eqref{eq:v1-4}, and \eqref{eq:v1-5}. 
   \begin{itemize}
        \item For $\sigma \le \sqrt{\sqrt{5}-2}\,  \mu$, $v_1\in [0,\mu)$ is the solution to $(\mu-v_1)^3 = (2v_1-\mu)\sigma^2$. Then we have simplified formulation for $\lm_1(v_1), \lm_2(v_1)$ as follows, since $\left(\mu ^3-3\,\mu ^2\,v_{1}+\mu \,\sigma ^2+3\,\mu \,{v_{1}}^2-2\,\sigma ^2\,v_{1}-{v_{1}}^3\right)=0$. 
\begin{align*}
\lm_2(v_1) &=  r\cdot  \frac{-2\,\mu \,\sigma ^2-4\,\mu \,{v_{1}}^2+5\,\mu ^2\,v_{1}+3\,\sigma ^2\,v_{1}-2\,\mu ^3+{v_{1}}^3}{{\left(\mu ^2-2\,\mu \,v_{1}+\sigma ^2+{v_{1}}^2\right)}^2} \\
\lm_1(v_1) & = \frac{r v_{1}-\lambda _{2}\,\left(\mu ^2+\sigma ^2-{v_{1}}^2\right)}{\mu -v_{1}}
\end{align*}
  Embedding $(\mu-v_1)^3 = (2v_1-\mu)\sigma^2$, we have further simplified $\lm_1(v_1), \lm_2(v_1)$ :
\begin{align*}
\lm_2(v_1) &=  r\frac{\mu -2\,v_{1}}{{\left(\mu -v_{1}\right)}^2} \\
\lm_1(v_1) & = r\frac{2\,{v_{1}}^2}{{\left(\mu -v_{1}\right)}^2}
\end{align*}
Since the solution to $(\mu-v_1)^3 = (2v_1-\mu)\sigma^2$ is always greater than $\mu/2$, we have $\mu-2v_1<0$, so $\lm_2(v_1)<0$, and thus, \eqref{eq:v1-4} is satisfied. Besides, since $\frac{\lm_1(v_1)}{-2\lm_2(v_1)}=\frac{{v_{1}}^2}{2\,v_{1}-\mu }$, and  $v_2=v_{1}\,\sqrt{\frac{v_{1}}{2\,v_{1}-\mu}}$, we have that $\frac{\lm_1(v_1)}{-2\lm_2(v_1)}/v_2=\sqrt{\frac{v_{1}}{2\,v_{1}-\mu}}>1$, where the inequality is due to $v_1>2v_1-\mu$. Thus, \eqref{eq:v1-5} holds.

Hence, the only remaining constraint we need to verify is \eqref{eq:v1-2}. For $\sigma \le \sqrt{\sqrt{5}-2}\,  \mu$, $v_1\in [0,\mu)$ is the solution to $(\mu-v_1)^3 = (2v_1-\mu)\sigma^2$. After simplification, 
        $rv_1+\lm_1(v_1)\cdot (v_2-\mu) +\lm_2(v_1) (v_2^2-\mu^2-\sigma^2) - v_1x_1 = rv_{1}\cdot \bigof{- \sqrt{\frac{v_{1}}{ 2\,v_{1}-\mu}}-\frac{2\,v_{1}\,\left(\mu -v_{1}\,\sqrt{-\frac{v_{1}}{\mu -2\,v_{1}}}\right)}{{\left(\mu -v_{1}\right)}^2}+\frac{v_1}{\mu -v_{1}}} = r v_1\cdot \bigof{\frac{-v_1(v_1+\mu)}{{\left(\mu -v_{1}\right)}^2}
        + \frac{v_1^2-\mu^2+2\mu v_1}{{\left(\mu -v_{1}\right)}^2} \sqrt{\frac{v_{1}}{ 2\,v_{1}-\mu}}} $.
        This is less or equal to zero if $\sqrt{\frac{v_{1}}{ 2\,v_{1}-\mu}} \le  \frac{v_1(v_1+\mu)}{v_1^2-\mu^2+2\mu v_1}$ which is equivalent to $v_{1}\,{\left(\mu +v_{1}\right)}^2\,\left(2\,v_{1}-\mu \right) -{\left(-\mu ^2+2\,\mu \,v_{1}+{v_{1}}^2\right)}^2\ge 0 \Longleftrightarrow {\left(\mu -v_{1}\right)}^2\,\left(-\mu ^2+\mu \,v_{1}+{v_{1}}^2\right)\ge 0 \Longleftrightarrow v_1 \ge \frac{\sqrt{5}-1}{2}\mu$.           Since $v_1$ is solved by $(\mu-v_1)^3 = (2v_1-\mu)\sigma^2$, when $\sigma \le \sqrt{\sqrt{5}-2}\,  \mu$, $v_1$ satisfies $v_1 \ge \frac{\sqrt{5}-1}{2}\mu$. Hence, constraint \eqref{eq:v1-2} is also satisfied. Therefore, we have proved that the $\lm_1(v_1),\lm_2(v_1)$ we provide in \eqref{eq:def lambda2} and \eqref{eq:def lambda1} are feasible for all the constraints when $\sigma \le \sqrt{\sqrt{5}-2}\,  \mu$.
  
\item For $\sigma \ge \sqrt{\sqrt{5}-2}\,  \mu$, $v_1\in [0,\mu)$ is the solution to $\frac{9+\sqrt{5}}{6}(\mu-v_1)^3 = (\frac{7}{3}v_1-\mu)\sigma^2$. Since
{\footnotesize
\begin{align*}
\frac{\lm_1(v_1)}{-2\lm_2(v_1)} =\frac{rv_1}{-2 \lm_2(v_1)(\mu-v_1)} + \frac{\mu ^2+\sigma ^2-{v_{1}}^2}{2(\mu -v_{1})} >  \frac{\mu ^2+\sigma ^2-{v_{1}}^2}{2(\mu -v_{1})} = \frac{1}{2}\cdot \bigof{\mu+v_1 +\frac{\sigma^2}{\mu -v_{1}}} \ge \sqrt{v_1\cdot \bigof{\mu+\frac{\sigma^2}{\mu -v_{1}}}} = v_2 ,
\end{align*}}
we have that \eqref{eq:v1-5} holds.
For \eqref{eq:v1-4}, let us consider the numerator in \eqref{eq:def lambda2}. We first analyze the sum of all entries outside the square root symbol. Since  $3\,\sigma ^2\,v_{1}-2\,\mu \,\sigma ^2= (2v_1-\mu)\sigma^2 + (v_1-\mu)\sigma^2 < (\mu-v_1)^3 +  (v_1-\mu)\sigma^2 < (\mu-v_1)^3$, where the first inequality is due to $(\mu-v_1)^3 = (\frac{7}{3}v_1-\mu)\sigma^2>(2v_1-\mu)\sigma^2$, we have $-2\,\mu \,\sigma ^2-4\,\mu \,{v_{1}}^2+5\,\mu ^2\,v_{1}+3\,\sigma ^2\,v_{1}-2\,\mu ^3+{v_{1}}^3 < (\mu-v_1)^3 -4\,\mu \,{v_{1}}^2+5\,\mu ^2\,v_{1}-2\,\mu ^3+{v_{1}}^3 = -\mu(\mu-v_1)^2< 0$. 
Hence, the sum of entries outside the square root is negative.
Thus, the numerator is negative if and only if the square of the sum of all entries outside the square root symbol is greater than the sum of the entries inside the square root symbol, which is proved as follows: $\left(-2\mu^3 + 5\mu^2 v_1 + v_1^3 + 3v_1 \sigma^2 - 2\mu (2v_1^2 + \sigma^2)\right)^2 - 4(\mu - v_1)(\mu^2 - \mu v_1 + \sigma^2)\left((\mu - v_1)^3 + (\mu - 2v_1)\sigma^2\right)
= v_1^2\cdot ((\mu-v_1)^2+\sigma^2)^2 >0$. Thus, the numerator is negative, so $\lm_2(v_1)<0$ and \eqref{eq:v1-4} is satisfied.

Then the only remaining constraint we need to check is \eqref{eq:v1-2}. $\lm_2(v_1)$ and $\lm_1(v_1)$ are determined by $\mu,\sigma$ since  $v_1$ is uniquely determined by $\frac{9+\sqrt{5}}{6}(\mu-v_1)^3 = (\frac{7}{3}v_1-\mu)\sigma^2$. 
Due to its complicated closed-form, we numerically check that $rv_1+ \lm_1(v_1)\cdot (v_2-\mu) +\lm_2(v_1) (v_2^2-\mu^2-\sigma^2) \le v_1x_1 $ for all $\sigma/\mu\ge \sqrt{\sqrt{5}-2}$, where $v_2,x_1,r$ are uniquely determined by $v_1,\mu,\sigma$ as $v_2 = \sqrt{v_1\bigof{\mu +\frac{\sigma ^2}{\mu -v_{1}}}}$, $x_1 = \frac{\sqrt{\sigma ^2+\mu \,\left(\mu -v_{1}\right)}}{2\,\sqrt{\sigma ^2+\mu \,\left(\mu -v_{1}\right)}-\sqrt{v_{1}\,\left(\mu -v_{1}\right)}}$, and  $r=\bigof{2\,\sqrt{\frac{1}{v_{1}}\bigof{\mu +\frac{\sigma ^2}{\mu -v_{1}}}}-1}^{-1}$. Hence, $\lm_2(v), \lm_1(v_1)$
determined in  \eqref{eq:def lambda2} and \eqref{eq:def lambda1} are feasible for our constraints.
    \end{itemize}

\end{enumerate}
\QED
\endproof

\begin{proof}{Proof of \Cref{cor:mean-var}.}
We first take a look at the monotonicity of $v_1$ with respect to $\sigma$.
By definition of $v_1$ in \Cref{thm:mean-var}, we have when $\sigma<\sqrt{\sqrt{5}-2}$, $\sigma^2=\frac{{\left(\mu -v_{1}\right)}^3}{2\,v_{1}-\mu}$, and the derivative of the right-hand side with respect to $v_1$ is
$\frac{{\left(\mu -v_{1}\right)}^2\,\left(\mu -4\,v_{1}\right)}{{\left(\mu -2\,v_{1}\right)}^2}<0$. 
Notice that due to the definition of $v_1$ in this case, $v_1>\frac{\mu}{2}$ always holds. This implies that $v_1$ is decreasing in $\sigma$ when $\sigma<\sqrt{\sqrt{5}-2}$. 
On the other hand, when $\sigma>\sqrt{\sqrt{5}-2}$, $\sigma^2=\frac{{(9+\sqrt{5})\left(\mu -v_{1}\right)}^3}{2\,(7v_{1}-3\mu)}$, and the derivative of the right-hand side with respect to $v_1$ is $\frac{\left(9+\sqrt{5}\right)(\mu-7v_1)(\mu-v_1)^2}{(3\mu-7v_1)^2}<0$. This implies that $v_1$ is decreasing in $\sigma$ when $\sigma\ge \sqrt{\sqrt{5}-2}$. Since when $\sigma=\sqrt{\sqrt{5}-2}$, the two formulations result in the same $v_1$, we have that $v_1$ is decreasing in $\sigma$.

Since $r=\bigof{2\,\sqrt{\frac{1}{v_{1}}\bigof{\mu +\frac{\sigma ^2}{\mu -v_{1}}}}-1}^{-1}$, we consider $g_1(v_1) = \frac{1}{v_{1}}\bigof{\mu +\frac{\sigma ^2}{\mu -v_{1}}}$. The derivative of $g_1(v_1)$ with respect to $v_1$ is 
    $\frac{\partial g_1(v_1)}{\partial v_1} = -\frac{\mu +\frac{\sigma ^2\,\left(\mu -2\,v_{1}\right)}{{\left(\mu -v_{1}\right)}^2}}{{v_{1}}^2}$. Consider the sign of the numerator.
    When $v_1\le \frac{\mu}{2}$, then $\frac{\partial g_1(v_1)}{\partial v_1} <0$, which implies that $g_1(v_1)$ is decreasing in $v_1$.
    When $v_1> \frac{\mu}{2}$, we denote $g_2(v_1) = 
    -\mu -\frac{\sigma ^2\,\left(\mu -2\,v_{1}\right)}{{\left(\mu -v_{1}\right)}^2}
    $.
    If $v_1\ge \frac{\sqrt{5}-1}{2}\mu $, which implies that $\sigma^2\ge (\sqrt{5}-2)\mu^2$, then in this case we have that $\sigma^2=\frac{{\left(\mu -v_{1}\right)}^3}{2\,v_{1}-\mu}$, and $g_2(v_1) = -\mu+(\mu-v_1)=-v_1<0$. Thus, $g_1(v_1)$ is also decreasing in $v_1$ when $v_1\ge \frac{\sqrt{5}-1}{2}\mu$. If $v_1\in (\frac{1}{2}\mu, \frac{\sqrt{5}-1}{2}\mu)$, then $g_2(v_1) = 
    -\mu -\frac{\sigma ^2\,\left(\mu -2\,v_{1}\right)}{{\left(\mu -v_{1}\right)}^2} = \frac{-\left(\left(3+\sqrt{5}\right)\mu^2\right)+\left(13+3\sqrt{5}\right)\mu v_1-2\left(9+\sqrt{5}\right)v_1^2 }{14 v_1 -6\mu}<0
    $, so $g_1(v_1)$ is also decreasing in $v_1$ for $v_1\in (\frac{1}{2}\mu, \frac{\sqrt{5}-1}{2}\mu)$. Besides, since $g_1$ is continuous in $v_1$ for $v_1<\mu$, we have that $g_1(v_1)$ is decreasing in $v_1$ for all $v_1\in (0,\mu)$. Moreover, since $r$ is decreasing in $g_1$, we have that $r$ is increasing in $v_1$. Since $v_1>\frac{3}{7}\mu$ for all $\sigma$, we have that $r\ge r(v_1=\frac{3}{7}\mu) = \bigof{2\,\sqrt{\frac{7}{3\mu}\bigof{\mu +\frac{\sigma ^2}{\frac{4}{7}\mu}}}-1}^{-1} = \bigof{\frac{7\sqrt{3}}{3}\sqrt{\frac{4}{7}+\bigof{\frac{\sigma}{\mu}}^2}}^{-1}$.
\QED
\end{proof}

\subsection{Proofs in \Cref{sec:quantile}}
\proof{Proof of \Cref{thm:quantile-1}.}
Since we hope to maximize $r$, the optimal $\lm_2$ should be zero. 
Hence, the competitive ratio satisfies
\beqn 
r(v) = \min\left\{
\frac{\xi\lm_1}{v_1}, \  1-\frac{(1-\xi)\lm_1}{v_1}, \, \frac{v_1}{\uv}
\right\}
\eeqn 
Now let us optimize $\lm_1$ and $\xi$ that can maximize $ \min\left\{
\frac{\xi\lm_1}{v_1}, \  1-\frac{(1-\xi)\lm_1}{v_1}\right\}$. Since $0<\xi\le 1$ and $\lm_1\ge 0$, the first term $\frac{\xi\lm_1}{v_1}$ is increasing in $\lm_1$ and the second term $1-\frac{(1-\xi)\lm_1}{v_1}$ is decreasing in $\lm_1$. Hence, $ \min\left\{
\frac{\xi\lm_1}{v_1}, \  1-\frac{(1-\xi)\lm_1}{v_1}\right\} =\frac{\xi\lm_1}{v_1} $ when $\lm_1\le v_1$ and $ \min\left\{
\frac{\xi\lm_1}{v_1}, \  1-\frac{(1-\xi)\lm_1}{v_1}\right\} =1-\frac{(1-\xi)\lm_1}{v_1} $ when $\lm_1> v_1$. Thus, the optimal $\lm_1=v_1$ and $\min\left\{
\frac{\xi\lm_1}{v_1}, \  1-\frac{(1-\xi)\lm_1}{v_1}\right\} =\xi $. It follows that \beqn 
r(v_1) = \min\left\{
\xi, \ \frac{v_1}{\uv}
\right\}
\eeqn 
Based on the optimal competitive ratio, we are able to characterize the optimal posted price $v_1$ by
\beqn 
v^*=\argmax_{v_1\le \omega} r(v_1) = \begin{cases}
\omega & \text{ if } \omega \le \xi\uv \\
\text{any } v \text{ s.t.} v\in[\xi\uv, \omega] & \text{ if } \omega > \xi\uv 
\end{cases}
\eeqn 
and the corresponding optimal competitive ratio is \beqn 
\cR_1 =\max_{v_1\le \omega} r(v_1) =   \min\left\{
\xi, \ \frac{\omega}{\uv}\right\}
\eeqn 
\QED
\endproof

\proof{Proof of \Cref{thm:quantile-2}.}
Similar to the analysis in the 1-Level pricing, in order for $r$ to be nonnegative, we must have $ v_1\le \omega$. Then we partition our analysis based on two scenarios: when $v_2\ge \omega$ and when $v_2<\omega$.
\begin{enumerate}
    \item 
When $v_2\ge \omega$, Problem \eqref{eq:quantile} is reduced to
\begin{subequations}
\addtocounter{equation}{-1}
\begin{align}
\sup\limits_{\lmb \in \R^{3}_+ \bx\in\R_+^2, r\in \R^+} & r \label{eq:quantile-2-1} \\
 \mathrm{s.t.}\quad  &  r\cdot v_1 -\xi\lm_1 \le 0\label{eq:quantile-2-1-1}\\
 & r\cdot v_1 +(1-\xi)\lm_1\le v_1x_1\label{eq:quantile-2-1-2}\\
 &  r \cdot v_2 +(1-\xi) \lm_2 \le v_1x_1\label{eq:quantile-2-1-4}\\
  &  (1-\xi) \lm_3 \le v_1x_1\label{eq:quantile-2-1-5}\\
   &  r \cdot \uv +(1-\xi) \lm_3 \le v_1x_1+v_2x_2\label{eq:quantile-2-1-6}\\
 & x_1+x_2 = 1
 \end{align}
\end{subequations}
Since $1-\xi$ is nonnegative, the optimal choice of $\lm_2, \lm_3$ are zero by  constraints \eqref{eq:quantile-2-1-4}, \eqref{eq:quantile-2-1-5}, \eqref{eq:quantile-2-1-6}. Hence, the problem is simplified to 
\begin{subequations}
\addtocounter{equation}{-1}
\begin{align}
\sup\limits_{\lm_1 \in \R_+ \bx\in\R_+^2, r\in \R^+} & r \label{eq:quantile-2-1-s} \\
 \mathrm{s.t.}\quad  &  r\cdot v_1 -\xi\lm_1 \le 0\label{eq:quantile-2-1-1-s}\\
 & r\cdot v_1 +(1-\xi)\lm_1\le v_1x_1\label{eq:quantile-2-1-2-s}\\
  &  r \cdot v_2 \le v_1x_1 \\
   &  r \cdot \uv  \le v_1x_1+v_2x_2\label{eq:quantile-2-1-6-s}\\
 & x_1+x_2 = 1
 \end{align}
\end{subequations}
Therefore, $r$ is solved by 
\beqn 
r = \min\{\frac{\xi\lm_1}{v_1},\ \frac{v_1x_1-(1-\xi)\lm_1}{v_1}, \ \frac{v_1x_1}{v_2}, \ \frac{v_1x_1+v_2x_2}{\uv} \}
\eeqn 
Then we jointly optimize $x_1,v_1,v_2$, we have that 
\begin{enumerate}
    \item  When $\omega\leq \xi^2 \uv$, we have $v_1=\omega$, $v_2=\sqrt{\omega \uv}$, $r=\frac{1}{2\sqrt{\frac{\uv}{\omega}}-1}$.
   \item When $\xi^2 \uv<\omega \leq \xi \uv$, $v_1=\omega$, $v_2=\frac{\omega}{\xi}$, $r= \frac{1}{\frac{1}{\xi}+\frac{\xi \uv}{\omega}-1}$.
   \item When  $\omega>\xi \uv$, $r=\xi$.
\end{enumerate}

\item
When $v_2<\omega$, Problem \eqref{eq:quantile} is reduced to
\begin{subequations}
\addtocounter{equation}{-1}
\begin{align}
\displaystyle
\sup\limits_{\lmb \in \R^{3}_+ \bx\in\R_+^2, r\in \R^+} & r \label{eq:quantile-2-2}\\
 \mathrm{s.t.}\quad  &  r\cdot v_1 -\xi\lm_1 \le 0\label{eq:quantile-2-2-2}\\
 & r\cdot v_1 +(1-\xi)\lm_1\le v_1x_1+v_2x_2\label{eq:quantile-2-2-3}\\
&  r\cdot v_2 -\xi \lm_2 \le v_1x_1\label{eq:quantile-2-2-4}\\
 &  r \cdot v_2 +(1-\xi) \lm_2 \le v_1x_1+x_2v_2\label{eq:quantile-2-2-5}\\
   &  r\cdot \uv +(1-\xi) \lm_3 \le v_1x_1+v_2x_2\label{eq:quantile-2-2-7}\\
 & x_1+x_2 = 1 \label{eq:quantile-2-2-8}
 \end{align}
\end{subequations}
Since $1-\xi$ is nonnegative, the optimal choice of $\lm_3$ is zero in order to maximize $r$ by  constraint \eqref{eq:quantile-2-2-7}. Considering constraint \eqref{eq:quantile-2-2-2}, we have that $\lm_1\ge \frac{rv_1}{\xi}$. Embedding this to constraint \eqref{eq:quantile-2-2-3}, we have that $ r\cdot v_1 +(1-\xi)\frac{rv_1}{\xi}\le v_1x_1+v_2x_2$, which is equivalent to $r\le \frac{\xi(v_1x_1+v_2x_2)}{v_1}$. Moreover, \eqref{eq:quantile-2-2-4} is equivalent to $r\le \frac{v_1x_1+\xi\lm_2}{v_2}$ and  \eqref{eq:quantile-2-2-5} 
is equivalent to $r\le \frac{v_1x_1+x_2v_2-(1-\xi)\lm_2}{v_2}$. Thus, the optimal $\lm_2$ is to make $\frac{v_1x_1+\xi\lm_2}{v_2}= \frac{v_1x_1+x_2v_2-(1-\xi)\lm_2}{v_2}$, i.e., $\lm_2=x_2v_2$. Incorporating $x_1=1-x_2$, the problem is simplified to 
\begin{subequations}
\addtocounter{equation}{-1}
\begin{align}
\displaystyle
\sup\limits_{\bx\in\R_+^2, r\in \R^+} & r \label{eq:quantile-2-2-s}\\
 \mathrm{s.t.}\quad  &  r\le \frac{\xi(v_2+(v_1-v_2)x_1)}{v_1}\label{eq:quantile-2-2-2-s}\\
&  r \le \frac{v_1x_1+\xi v_2(1-x_1)}{v_2}\label{eq:quantile-2-2-3-s}\\
   &  r   \le \frac{v_2+(v_1-v_2)x_1}{\uv}\label{eq:quantile-2-2-4-s}\\
 & x_1+x_2  = 1 \label{eq:quantile-2-2-8-s}
 \end{align}
\end{subequations}
After jointly optimization $v_1,v_2,x_1$, we have that
\begin{enumerate}
    \item If $\omega\ge \xi\uv$, $v_1=\xi \uv, v_2=\min\left\{\omega, \sqrt{\xi}\uv\right\}, r=\frac{\xi(1-\xi)}{\frac{v_1}{v_2}+\frac{\xi v_2}{v_1}-2\xi} $.
    \item If $\omega\le \xi\uv$, $  v_1=v_2=\omega,   r=\frac{\omega}{\uv}$
\end{enumerate}

\end{enumerate}
The proof of \Cref{thm:quantile-2} is completed by summarizing all the above cases and writing in a compact form. In the following, we provide the detailed procedure of finding the optimal solutions to \eqref{eq:quantile-2-1-s} and \eqref{eq:quantile-2-2-s}.
\begin{enumerate}
    \item 
When $v_2\ge \omega$, according to Problem \eqref{eq:quantile-2-1-s}, $r$ is solved by 
\beqn 
r = \min\{\frac{\xi\lm_1}{v_1},\ \frac{v_1x_1-(1-\xi)\lm_1}{v_1}, \ \frac{v_1x_1}{v_2}, \ \frac{v_1x_1+v_2x_2}{\uv} \}
\eeqn 
Next we search for the optimal $\lm_1$. Since the first term is increasing in $\lm_1$ and the second term is decreasing in $\lm_1$, the optimal $\lm_1$ is where the two terms are equal to each other. Hence, $\lm_1 = v_1x_1$. Embedding $x_2=1-x_1$, then $r$ is represented as 
\beqn 
r = \min\{\xi x_1, \ \frac{v_1x_1}{v_2}, \ \frac{(v_1-v_2)x_1 + v_2}{\uv} \}
\eeqn 
Since $\xi x_1$ and $\frac{v_1 x_1}{v_2}$ are both increasing in $x_1$, and $\frac{(v_1-v_2)x_1 + v_2}{\uv}$ is decreasing in $x_1$, $r$ is maximized when the smaller one within the first two terms intersects with the third term. 
\begin{enumerate}
    \item  If $\frac{v_1}{v_2}\leq \xi$, let $\frac{v_1 x_1}{v_2}=\frac{(v_1-v_2)x_1 + v_2}{\uv}$.
Then we have $x_1=\frac{v_2^{2}}{v_2^{2}-v_{1}v_{2}+v_{1}\uv}=\frac{1}{1-\frac{v_1}{v_2}(1-\frac{\uv}{v_2})}\leq 1$, and $r=\frac{v_1 v_2}{v_2^{2}-v_{1}v_{2}+v_{1}\uv}=\frac{1}{\frac{v_2}{v_1}+\frac{\uv}{v_2}-1}$. To maximize $r$, 
\begin{enumerate}
    \item when $\omega \leq \xi \uv$, the optimal $v_1=\omega$, $v_2=\max\left\{\frac{\omega}{\xi}, \sqrt{\omega \uv}\right\}$. Hence we have:
    \begin{enumerate}
        \item If $ \omega>\xi^2 \uv$, then $v_1=\omega, v_2=\frac{\omega}{\xi}, r=\frac{1}{\frac{1}{\xi}+\frac{\xi \uv}{\omega}-1}$;
        \item If $ \omega\leq \xi^2 \uv$, then $ v_1=\omega, v_2=\sqrt{\omega \uv}, r= \frac{1}{2\sqrt{\frac{\uv}{\omega}}-1}$.
    \end{enumerate}

\item when $\omega\ge \xi \uv$, then $\frac{\omega}{\xi}\ge \uv$, and we have $\sqrt{\omega \uv}>\sqrt{\xi}\uv$, so $v_2=\uv$, $v_1=\xi \uv$, $r=\xi$.
\end{enumerate}
\item When $\frac{v_1}{v_2}\ge\xi$, letting
$\xi x_1=\frac{(v_1-v_2)x_1 + v_2}{\uv}$,
we have $x_1=\frac{v_2}{v_2-v_1+\xi \uv}=\frac{1}{1+\frac{\xi \uv-v_1}{v_2}}$. If $v_1\leq \xi \uv$, then $x_1\leq 1$, $r=\xi x_1=\frac{\xi v_2}{v_2-v_1+\xi \uv}$, which is increasing $v_1$ and $v_2$; if $v_1\ge\xi \uv$, then let $x_1=1$, $r=\xi$. Hence, by choosing the optimal $v_1$ and $v_2$, we have that 
\begin{enumerate}
    \item if $\omega \leq \xi \uv$, $v_1=\omega$, $v_2=\frac{\omega}{\xi}$, $r=\frac{1}{\frac{1}{\xi}+\frac{\xi \uv}{\omega}-1}$.
    \item if $\omega>\xi \uv$, regardless of either $v_1\leq \xi \uv$ or $\xi \uv<v_1\le \omega$, it follows that $r= \xi$.
\end{enumerate}
\end{enumerate}
Combining the above cases, we have that when $v_2\geq \omega$, 
\begin{itemize}
   \item If $\omega\leq \xi^2 \overline{v}$, then $v_1=\omega$, $v_2=\sqrt{\omega \overline{v}}$, $r= \frac{1}{2\sqrt{\frac{\overline{v}}{\omega}}-1}$;
    \item If $\xi^2 \overline{v}<\omega \leq \xi \overline{v}$, we have that $v_1=\omega$, $v_2=\frac{\omega}{\xi}$, $r=\frac{1}{\frac{1}{\xi}+\frac{\xi \overline{v}}{\omega}-1}$.
    \item If $\omega>\xi \overline{v}$, then $v_1\le \omega$ and $r=\xi$.
\end{itemize}

\item
When $v_2<\omega$, by Problem \eqref{eq:quantile-2-2-s}, we consider the following two scenarios. 
\begin{enumerate}
    \item If $v_1\ge \xi\uv$, then constraint \eqref{eq:quantile-2-2-2-s} implies constraint \eqref{eq:quantile-2-2-4-s}. Thus, $r=\min\{
     \frac{\xi(v_2+(v_1-v_2)x_1)}{v_1}, \frac{v_1x_1+\xi v_2(1-x_1)}{v_2}
    \}$. The first term is decreasing in $x_1$ and since $v_1\ge\xi\uv>\xi v_2$, the second term is increasing in $x_1$. Thus, the optimal $x_1$ makes the two terms equal to each other, i.e. $x_1 = \frac{\xi(v_2-v_1)}{\frac{v_1^2}{v_2}-2\xi v_1+\xi v_2} = \frac{\xi(v_2-v_1)}{(\frac{v_1}{v_2}-\xi) v_1+\xi (v_2-v_1)}<1$. 
    Hence, $ r = \frac{\xi(1-\xi)}{\frac{v_1}{v_2}+\frac{\xi v_2}{v_1}-2\xi} $.
    Denoting $\theta=\frac{v_1}{v_2}$, we have that $r=\frac{\xi(1-\xi)}{\theta+\frac{\xi}{\theta} -2\xi}$. The minimum value of the denominator is achieved at $\theta =\sqrt{\xi}$. Hence, as long as $\sqrt{\xi}\uv < \omega$, there exists $v_1\ge \xi\uv$ and $v_2<\omega$ such that $\frac{v_1}{v_2}=\sqrt{\xi} $, so the optimal $r=\frac{\sqrt{\xi}+\xi}{2}$. 
    Otherwise, if $\sqrt{\xi}\uv\ge \omega\ge \xi\uv$, then the optimal $v_1=\xi\uv$, $v_2=\omega$, and the optimal  $r=\frac{1-\xi}{\frac{\uv}{\omega}+\frac{\omega}{\xi\uv}-2}$. To write the cases above in a more compact form, the optimal $v_1,v_2,r$ can be represented as follows.
\begin{enumerate}
    \item If $\xi\uv\le \omega,$ then
$v_1 = \xi\uv,\, v_2=\min\{\omega, \sqrt{\xi}\uv\}, \, r = \frac{1-\xi}{\frac{\uv}{v_2}+\frac{ v_2}{\xi\uv}-2}.$
\item If $\xi\uv> \omega,$ this case can not happen since there are no $v_1\le v_2$ satisfying $v_1\ge \xi\uv, v_2<\omega$.
\end{enumerate}

    \item If $v_1\le \xi\uv$, then constraint \eqref{eq:quantile-2-2-4-s} implies constraint \eqref{eq:quantile-2-2-2-s}. Thus, $r=\min\{
  \frac{v_2+(v_1-v_2)x_1}{\uv}, \frac{v_1x_1+\xi v_2(1-x_1)}{v_2}
    \}$. 
    \begin{enumerate}
        \item If $v_1\le \xi v_2$, then both terms are decreasing in $x_1$ so the optimal $x_1=0$, resulting in $r=\min\{\frac{v_2}{\uv},\xi\}\le \min\{\frac{\omega}{\uv},\xi\}$.
        \item     If $v_1\ge \xi v_2$, then $\frac{v_2+(v_1-v_2)x_1}{\uv}=\frac{v_1x_1+\xi v_2(1-x_1)}{v_2}
    $ leads to $x_1 = \frac{v_2-\xi\uv}{\frac{\uv v_1}{v_2}+v_2-\xi\uv-v_1}$. 
    \begin{enumerate}
        \item   First, consider the case when $\omega \leq \xi \uv$. Then $v_2\leq \omega\leq \xi \uv$. In this case, to maximize $r$, $x_1=0$,  $r=\min\{
  \frac{v_2}{\uv}, \frac{\xi v_2}{v_2} \} = \frac{v_2}{\uv}$. Therefore, when $v_2=\omega$, $r=\frac{\omega}{\uv}$ is maximized. 
  \item  Second,  consider the case when $\omega \ge \xi \uv$. In this case, if $v_2\le \xi\uv$, then we have the optimal $r=\xi$, achieved at $v_2=\xi\uv$. On the other hand, if $v_2\ge \xi\uv$, the optimal $r= \frac{1-\xi}{\frac{v_2}{v_1}+\frac{\uv}{v_2}-\frac{\xi\uv}{v_1}-1}$. Then the optimal $v_2=\min\{\omega,\sqrt{v_1\uv}\}$. In either case, the optimal $v_1=\xi\uv$, and thus $v_2=\min\{\omega,\sqrt{\xi }\uv\}$, $r = \frac{1-\xi}{\frac{\uv}{v_2}+\frac{ v_2}{\xi\uv}-2}$. 
    \end{enumerate}

    \end{enumerate}
 
\end{enumerate}
Combining the above cases, we have that when $v_2\le \omega$, 
  \begin{itemize}
    \item If $\xi\uv\le \omega,$ then
$v_1 = \xi\uv,\, v_2=\min\{\omega, \sqrt{\xi}\uv\}, \,x_1 = \frac{\xi(v_2-v_1)}{\frac{v_1^2}{v_2}-2\xi v_1+\xi v_2}, r = \frac{\xi(1-\xi)}{\frac{v_1}{v_2}+\frac{\xi v_2}{v_1}-2\xi}$.
\item If $\xi\uv \ge \omega,$ then $v_1=v_2=\omega, r=\frac{\omega}{\uv}$.
\end{itemize}
\end{enumerate}
Then we summarize the results above.
\begin{enumerate}
    \item When $\omega\ge\xi \overline{v}$, we have that
    \begin{enumerate}
        \item If $\xi \overline{v}\le \omega<\sqrt{\xi} \overline{v}$, then the optimal prices satisfy $v_1=\xi \overline{v}$,  
        $v_2=\omega$ and 
        $r= \frac{\xi(1-\xi)}{\frac{\xi \overline{v}}{\omega}+\frac{\omega}{\overline{v}}-2\xi}$.
        \item If $\omega\geq \sqrt{\xi} \overline{v}$, then the optimal $v_1=\xi \overline{v}$ and $r=\begin{cases}
       \frac{\sqrt{\xi}(1+\sqrt{\xi})}{2} & \text{if } v_2 = \sqrt{\xi}\uv\\
       \xi &\text{if } v_2=\uv.
        \end{cases} $
    \end{enumerate}
Since $\xi \leq 1$, $\frac{\sqrt{\xi}(1+\sqrt{\xi})}{2}\geq \xi$, so it follows that when $\omega\ge \sqrt{\xi}\uv$, the optimal $v_2=\sqrt{\xi}\uv$. 
Writing the results from (a) and (b) in a compact form, we have that when $\omega>\xi \overline{v}$, the optimal 2-level pricing selling mechanism and competitive ratio are represented as 
$$v_1=\xi \overline{v}, v_2=\min\left\{\omega, \sqrt{\xi}\uv\right\}, x_1=\frac{v_2-\xi\uv}{\frac{\xi\uv^2}{v_2}-2\xi \uv+v_2}, r=\frac{(1-\xi)}{\frac{\uv}{v_2}+\frac{v_2}{\xi\uv}-2}.$$

\item
When $\omega \leq \xi \overline{v}$, we have that
\begin{enumerate}
    \item If $\omega \leq \xi^2 \overline{v}$, then the optimal prices satisfy $v_1 = \omega$, and $r=\begin{cases}
     \frac{\omega}{\overline{v}} &  \text{ if } v_2=\omega   \\
      \frac{1}{2\sqrt{\frac{\overline{v}}{\omega}}-1} &  \text{ if } v_2=\sqrt{\omega \overline{v}}. 
    \end{cases}$
    \item If $\omega \geq \xi^2 \overline{v}$, then the optimal prices satisfy $v_1 = \omega$, and $r=\begin{cases}
     \frac{\omega}{\overline{v}} &  \text{ if } v_2=\omega   \\
     \frac{1}{\frac{1}{\xi}+\frac{\xi \overline{v}}{\omega}-1} &  \text{ if } v_2=\frac{\omega}{\xi}. 
    \end{cases}$
\end{enumerate}
Notice that for $\omega\le \uv$, we always have $\frac{\uv}{\omega}-(2\sqrt{\frac{\overline{v}}{\omega}}-1)=(\sqrt{\frac{\overline{v}}{\omega}}-1)^2 \ge 0$, so $ \frac{\omega}{\overline{v}}\le  \frac{1}{2\sqrt{\frac{\overline{v}}{\omega}}-1}$. Hence, in case (a), the optimal prices are $v_1=\omega$, $v_2=\sqrt{\omega \overline{v}}$, and $r= \frac{1}{2\sqrt{\frac{\overline{v}}{\omega}}-1}$. On the other hand, $\frac{\uv}{\omega}-(\frac{1}{\xi}+\frac{\xi \overline{v}}{\omega}-1)=(\frac{\uv}{\omega}-\frac{1}{\xi})(1-\xi) \ge 0$, so 
$\frac{\omega}{\overline{v}}\le
     \frac{1}{\frac{1}{\xi}+\frac{\xi \overline{v}}{\omega}-1} $. Hence, in case (b), the optimal prices are $v_1=\omega$, $v_2= \frac{\omega}{\xi}$ and $r= \frac{1}{\frac{1}{\xi}+\frac{\xi \overline{v}}{\omega}-1}$.
Therefore, when $\omega \leq \xi \overline{v}$,  the optimal 2-level pricing selling mechanism and competitive ratio are represented as $$v_1=\omega, v_2=\max\left\{\frac{\omega}{\xi}, \sqrt{\omega \overline{v}} \right\}, x_1=\frac{v_2^{2}}{v_2^{2}-\omega v_{2}+\omega\overline{v}}, r=\frac{1}{\frac{v_2}{\omega}+\frac{\overline{v}}{v_2}-1}.$$
\end{enumerate}
Summarizing the results in (I) and (II), we complete the proof of \Cref{thm:quantile-2}.
   \begin{enumerate}
    \item If $\xi\uv\le \omega$, then $v_1=\xi\uv$, $v_2=\min\{\omega,\sqrt{\xi}\uv\}$, $\cR_2^*=\frac{1-\xi}{\frac{\uv}{v_2}+\frac{v_2}{\xi\uv}-2}$ and $x_1=\frac{v_2-\xi\uv}{\frac{\xi\uv^2}{v_2}-2\xi \uv+v_2}.$
    \item If $\xi\uv> \omega$, then $v_1=\omega$, $v_2=\max\{\frac{\omega}{\xi},\sqrt{\omega\uv}\}$, $\cR_2^*=\frac{1}{\frac{v_2}{\omega}+\frac{\overline{v}}{v_2}-1}$ and $x_1 = \frac{v_2^{2}}{v_2^{2}-\omega v_{2}+\omega\overline{v}}$ 
\end{enumerate}
\QED\endproof

\proof{Proof of \Cref{thm:quantile-inf}.} To prove \Cref{thm:quantile-inf}, we first take the dual of Problem \eqref{eq:quantile-continuous} and then we prove \Cref{thm:quantile-inf} by verifying the feasibility of a pair of primal and dual solutions.
Denoting as $\Gamma(p,v)$ the dual variable corresponding to the constraints in Problem \eqref{eq:quantile-continuous}, the dual of Problem \eqref{eq:quantile-continuous} is defined as
\beqn 
\min_{\Gamma:[0,\uv]^2 \rightarrow \R_+, \gamma } & & \gamma \\
\mbox{s.t.} & & \int_{0}^{\uv} ( \mathbbm{1}[v\ge \omega] - \xi )\cdot \Gamma(p,v)dv\ge 0, \ \forall \ p\in[0,\uv]\\
& &\gamma -v \int_{p=0}^{\uv}  \int_{u=v}^{\uv}  \Gamma(p,u)dudp \ge 0, \ \forall \ v\in[0,\uv]\\
& & \int_{p=0}^{\uv} p\cdot \int_{v=p}^{\uv}  \Gamma(p,v)dvdp \ge 1.
\eeqn 
We first construct a primal feasible solution. Let $\hp=\min\{(2-\xi)\omega,\uv\}$.
A feasible solution to Problem \eqref{eq:quantile-continuous} is provided as follows. We postpone the verification of the feasibility later.
\begin{align}
\begin{aligned}
& r = \big(
\frac{\ln \omega-\ln (\xi\hp)}{(1-\xi)}+\frac{\hp-\omega}{(1-\xi)\omega} +\ln\uv-\ln\hp
\big)^{-1}\\
& \lm(p) = \begin{cases}
r\cdot \hp & \forall p\in[0,\xi\hp) \\
\frac{r\cdot(\hp-p)}{1-\xi} & \forall p\in[\xi\hp, \omega) \\
0 & \forall p\in[\omega,\uv] \\
\end{cases}  \quad \int_0^v \pi(u)u\, du = 
\begin{cases}
0 & \forall v\in[0,\xi\hp) \\
\frac{r\cdot (v-\xi\hp)}{1-\xi} & \forall v\in[\xi\hp, \omega) \\
r\cdot \hp & \forall v\in[\omega,\hp) \\
r\cdot v & \forall v\in[\hp,\uv] \\
\end{cases}
\end{aligned}
\label{eq:quantile-continuous-primal-sol}
\end{align}
At the same time, the feasible solution to the dual problem is characterized as 
\begin{align} 
\label{eq:quantile-continuous-dual-sol}
\begin{aligned}
 \gamma = \big(
\frac{\ln \omega-\ln (\xi\hp)}{(1-\xi)}+\frac{\hp-\omega}{(1-\xi)\omega} +\ln\uv-\ln\hp
\big)^{-1}\\
\Gamma\text{ is a two-dimensional distribution s.t.}
 \int_{u\in \{v\}}\Gamma(u,u)du= \frac{\gamma}{v^2} \quad \forall v\in[\xi\hp,\omega)\cup[\hp,\uv) &\quad\mbox{(line mass)}\\
\int_{u\in \{v\}} \Gamma(u,\omega)du=\frac{\xi}{1-\xi} \frac{\gamma}{v^2} \quad \forall v\in[\xi\hp,\omega) & \quad\mbox{(line mass)} \\
 \int_{u\in \{\uv\}} \Gamma(u,u)du=\gamma\big(\frac{1}{1-\xi}(\frac{1}{\omega}-\frac{1}{\hp})-\frac{1}{\hp}+\frac{1}{\uv} \big)=\begin{cases}
    \frac{\gamma}{\uv} & \mbox{ if } (2-\xi)\omega \le \uv\\
    \frac{\gamma}{1-\xi}(\frac{1}{\omega}-\frac{1}{\uv}) & \mbox{ if } (2-\xi)\omega > \uv
\end{cases} &\quad\mbox{(point mass)} 
\end{aligned}
\end{align}
Now that the primal solution defined in \eqref{eq:quantile-continuous-primal-sol} and dual solution defined in \eqref{eq:quantile-continuous-dual-sol} have the same objective value, we only need to prove the feasibility of both solutions in the following analysis.

We first prove the feasibility of the primal solution.
\begin{enumerate}
\item When $p\ge \omega$, $\lm(p) = 0$.
\begin{enumerate}
    \item If $v<p$, then the LFS of the constraint in primal problem \eqref{eq:quantile-continuous} is equal to 0 which is less than or equal to  the RHS which is non-negative.
    \item If $v \ge p$, then the LFS of the constraint in primal problem \eqref{eq:quantile-continuous} is $r\cdot p \le r\cdot v$. Notice that when $v\ge \omega$, we have $\int_0^v \pi(u)u\, du \ge r\cdot v$ according to the definition of the primal solution defined in \eqref{eq:quantile-continuous-primal-sol}. Thus, the constraints in \eqref{eq:quantile-continuous} are satisfied in this case.
\end{enumerate}
\item When $p\in[\xi\hp, \omega)$, $\lm(p) = \frac{r\cdot(\hp-p)}{1-\xi}$.
\begin{enumerate}
    \item If $v<p$, the LFS satisfies $(\mathbbm{1}[v\ge \omega] -\xi)\cdot \lm(p) =  -\xi\cdot  \frac{r\cdot(\hp-p)}{1-\xi} \le 0 \le \int_0^v \pi(u)u\, du$, so the constraints in Problem \eqref{eq:quantile-continuous} hold.
    \item If $v\in [p,\omega)$, the LFS satisfies $r\cdot p+(\mathbbm{1}[v\ge \omega] -\xi)\cdot \lm(p) =  r\cdot p -\xi\cdot  \frac{r\cdot(\hp-p)}{1-\xi}= \frac{r\cdot(p-\xi\hp)}{1-\xi} \le \frac{r\cdot(v-\xi\hp)}{1-\xi} =\int_0^v \pi(u)u\, du$, where the last inequality is because $p\le v$, and the last equality is due to the definition of $x$ in \eqref{eq:quantile-continuous-primal-sol} for $v\in [\xi\hp, \omega)$.
    \item If $v\in [\omega, \hp)$, the LFS satisfies $r\cdot p+(\mathbbm{1}[v\ge \omega] -\xi)\cdot \lm(p) =  r\cdot p +(1-\xi)\cdot  \frac{r\cdot(\hp-p)}{1-\xi}= r\hp =\int_0^v \pi(u)u\, du$, where the last equality is due to the definition of $x$ in \eqref{eq:quantile-continuous-primal-sol}.
    \item If $v\in [\hp, \uv)$, the LFS satisfies $r\cdot p+(\mathbbm{1}[v\ge \omega] -\xi)\cdot \lm(p) =  r\cdot p +(1-\xi)\cdot  \frac{r\cdot(\hp-p)}{1-\xi}= r\hp \le r\cdot v=\int_0^v \pi(u)u\, du$, where the last equality is due to the definition of $x$ in \eqref{eq:quantile-continuous-primal-sol}.
\end{enumerate}
    \item When $p\in[0,\xi\hp]$, $\lm(p) = r\cdot\hp$.
    \begin{enumerate}
        \item If $v<p$, we have that $v<p\le \xi\hp =\xi\cdot\min\{(2-\xi)\omega,\uv\}\le  \xi(2-\xi)\omega\le \omega$, where the last inequality is due to $ \xi(2-\xi)\le 1$. Then the LFS of the constraint in the primal problem \eqref{eq:quantile-continuous} satisfies $(\mathbbm{1}[v\ge \omega] -\xi)\cdot \lm(p) =  -\xi\cdot r\cdot \hp \le 0 \le \int_0^v \pi(u)u\, du$, where $\int_0^v \pi(u)u\, du$ is exactly the RHS of the constraint. Hence, the constraint is satisfied in this case.
        \item If $v\in [p, \xi\hp]$, we have that $r\cdot p+  \mathbbm{1}[v\ge \omega] \cdot \lm(p)-\xi\cdot \lm(p)= r\cdot (p - \xi\cdot \hp) \le 0  \le \int_0^v \pi(u)u\, du$. Hence, the constraint is satisfied.
        \item If $v\ge \xi\hp$, then the LFS is $r\cdot p +  \mathbbm{1}[v\ge \omega] \cdot \lm(p)-\xi\cdot \lm(p) = r\cdot p +r\cdot \hp \cdot (\mathbbm{1}[v\ge \omega] -\xi)\le r\cdot \xi\cdot \hp +\lm(\xi\hp)(\mathbbm{1}[v\ge \omega] -\xi)   \le \int_0^v \pi(u)u\, du$, where the first inequality is because $\lm(\xi\hp) = r\cdot \hp$ and $p\le \xi\hp$. The last inequality is due to the proof for $p\in[\xi\hp, \omega)$.
    \end{enumerate}
\end{enumerate}
Hence, we have shown the first two constraints in problem \eqref{eq:quantile-continuous} are satisfied for all $p\in [0,\uv]$ and $v\in [0,\uv]$. The only constraint we need to verify is the last constraint $\int_0^{\uv} \pi(v)\, dv = 1$. According to the definition in \eqref{eq:quantile-continuous-primal-sol}, $x$ contains a  a point mass $\frac{r\cdot (\hp-\omega)}{(1-\xi)\omega}$ at $\omega$, together with a density function defined as
\beqn
\pi(v) = \begin{cases}
0 & \forall v\in[0,\xi\hp) \\
\frac{r}{(1-\xi)v} & \forall v\in[\xi\hp, \omega) \\
0 & \forall v\in(\omega,\hp) \\
\frac{r}{v} & \forall v\in[\hp,\uv]. \\
\end{cases}
\eeqn 
Hence, we have that 
\begin{align*}
\int_0^{\uv} \pi(v)\, dv &= r\big(
\int_{\xi\hp}^{\omega} \frac{1}{(1-\xi)v}dv +
\int_{\hp}^{\uv} \frac{1}{v}dv +\frac{\hp-\omega}{(1-\xi)\omega}
\big) =r\big(
\frac{\ln \omega-\ln(\xi\hp)}{(1-\xi)} +
\ln\frac{\uv}{\hp}+ \frac{ \hp-\omega}{(1-\xi)\omega}
\big) \\
&= \big(
\frac{\ln \omega-\ln (\xi\hp)}{(1-\xi)}+\frac{\hp-\omega}{(1-\xi)\omega} +\ln\uv-\ln\hp
\big)^{-1}\cdot \big(
\frac{\ln \omega-\ln(\xi\hp)}{(1-\xi)} +
\ln\frac{\uv}{\hp}+ \frac{\hp-\omega}{(1-\xi)\omega}
\big) =1
\end{align*}
which completes the proof of the feasibility of the primal solution defined in \eqref{eq:quantile-continuous-primal-sol}.
Next, we are going to prove the feasibility of the dual solution. 

We first consider the first constraint in the dual problem. We consider $p$ in the following three scenarios.
\begin{enumerate}
    \item For any $p\ge \omega$, according to the definition in  \eqref{eq:quantile-continuous-dual-sol}, we have $\Gamma(p,v)=0$ for all $v\le \omega$, so $\int_{0}^{\uv} ( \mathbbm{1}[v\ge \omega] - \xi )\cdot \Gamma(p,v)dv = \int_{\omega}^{\uv} ( 1 - \xi )\cdot \Gamma(p,v)dv\ge 0$.
    \item  When $p< \xi\hp$, we have that $\Gamma(p,v)=0$ for all $v$, so $\int_{0}^{\uv} ( \mathbbm{1}[v\ge \omega] - \xi )\cdot \Gamma(p,v)dv= 0$.
    \item   When $p\in [\xi\hp,\omega)$, we have $\int_{0}^{\uv} ( \mathbbm{1}[v\ge \omega] - \xi )\cdot \Gamma(p,v)dv = -\xi\cdot \frac{\gamma}{p^2}  +(1-\xi) \frac{\xi}{1-\xi} \frac{\gamma}{p^2}= 0$. 
\end{enumerate}
Thus, the first constraint in the dual problem is always satisfied. Let us proceed to consider the second constraint. 
\begin{enumerate}
    \item For any $v\ge \hp$, $\int_{p=0}^{\uv}  \int_{u=v}^{\uv}  \Gamma(p,u)dudp \le \int_{v}^{\uv}  \frac{\gamma}{p^2} dp + \frac{\gamma}{\uv} = \gamma\cdot (\frac{1}{v}-\frac{1}{\uv})+\frac{\gamma}{\uv} =\frac{\gamma}{v}$, where the first inequality is because when  $ (2-\xi)\omega > \uv$ we always have $\frac{\gamma}{1-\xi}(\frac{1}{\omega}-\frac{1}{\uv})\le \frac{\gamma}{1-\xi}(\frac{2-\xi}{\uv}-\frac{1}{\uv}) = \frac{\gamma}{\uv} $ . Hence, the second constraint is satisfied since $\gamma-v\cdot \frac{\gamma}{v} = 0$.
    \item For any $v\in (\omega, \hp)$, $\int_{p=0}^{\uv}  \int_{u=v}^{\uv}  \Gamma(p,u)dudp \le \frac{\gamma}{\hp}$. Hence, the second constraint is satisfied since $\gamma-v\cdot \frac{\gamma}{\hp} >0$.
    \item When $v=\omega$, $\int_{p=0}^{\uv}  \int_{u=\omega}^{\uv}  \Gamma(p,u)dudp =\gamma( \frac{1}{1-\xi}(\frac{1}{\omega}-\frac{1}{\hp})-\frac{1}{\hp}+\frac{1}{\uv} )+\int_{\xi\hp}^{\omega}\frac{\xi}{1-\xi} \frac{\gamma}{v^2}dv + \int_{\hp}^{\uv} \frac{\gamma}{v^2}dv
    = \frac{\gamma}{1-\xi}(\frac{1}{\omega}-\frac{1}{\hp})-\frac{\gamma}{\hp}+\frac{\gamma}{\uv} +\frac{\xi\gamma}{1-\xi}(\frac{1}{\xi\hp}-\frac{1}{\omega})+\gamma (\frac{1}{\hp}-\frac{1}{\uv}) =\frac{\gamma}{\omega}$. 
    Hence the second constraint here holds since $\gamma-\omega \int_{p=0}^{\uv}  \int_{u=\omega}^{\uv}  \Gamma(p,u)dudp = \gamma-\omega \frac{\gamma}{\omega}=0$.
    \item When $v<\omega$, we have that $\int_{p=0}^{\uv}  \int_{u=v}^{\omega}  \Gamma(p,u)dudp + \int_{p=0}^{\uv}  \int_{u=\omega}^{\uv}  \Gamma(p,u)dudp \le \int_{p=v}^{\omega}\frac{\gamma}{p^2}dp +\frac{\gamma}{\omega} =\frac{\gamma}{v} $. Hence, the constraint holds by $\gamma-v\frac{\gamma}{v}=0$.
\end{enumerate}
Moreover, the last constraint holds since 
\begin{align*}
\int_{p=0}^{\uv} p\cdot \int_{v=p}^{\uv}  \Gamma(p,v)dvdp &= \gamma \big(\int_{p=\xi\hp}^\omega  p\cdot \frac{1}{1-\xi}\cdot \frac{1}{p^2} dp + \int_{p=\hp}^{\uv} p\cdot \frac{1}{p^2}dp + \uv (\frac{1}{1-\xi}(\frac{1}{\omega}-\frac{1}{\hp})-\frac{1}{\hp}+\frac{1}{\uv} ) 
\big) \\
&= \gamma \big(\frac{\ln \omega-\ln (\xi\hp)}{(1-\xi)} +\ln\uv-\ln\hp + \uv (\frac{1}{1-\xi}(\frac{1}{\omega}-\frac{1}{\hp})-\frac{1}{\hp}+\frac{1}{\uv} ) 
\big) \\
\end{align*}
Notice that $\uv (\frac{1}{1-\xi}(\frac{1}{\omega}-\frac{1}{\hp})-\frac{1}{\hp}+\frac{1}{\uv}=
\begin{cases}
    \frac{\uv-\omega}{(1-\xi)\omega} & \mbox{ if } \hp = \uv\\
    1 & \mbox{ if } \hp = (2-\xi)\omega
 \end{cases}
$, which can be written as $ \frac{\hp-\omega}{(1-\xi)\omega}$ in a more compact form. Thus, we have that 
\beqn 
\int_{p=0}^{\uv} p\cdot \int_{v=p}^{\uv}  \Gamma(p,v)dvdp = \big(
\frac{\ln \omega-\ln (\xi\hp)}{(1-\xi)}+\frac{\hp-\omega}{(1-\xi)\omega} +\ln\uv-\ln\hp
\big)^{-1} \cdot \big(
\frac{\ln \omega-\ln (\xi\hp)}{(1-\xi)}+\frac{\hp-\omega}{(1-\xi)\omega} +\ln\uv-\ln\hp
\big)=1
\eeqn 
which completes our proof of the feasibility of the dual problem.

Finally, since the primal and dual problems have the same objective value, we have completed the proof of the optimality and feasibility of the primal solutions given in \eqref{eq:quantile-continuous-primal-sol}, which completes the proof of \Cref{thm:quantile-inf}.
\QED
\endproof
\proof{Proof of \Cref{cor:quantile-dominance}.
}According to \Cref{thm:quantile-inf}, the allocation probability is shown as 
$$q(v)=\int_0^v \pi(u)\, du = 
\begin{cases}
0 & \forall v\in[0,\xi\hp) \\
\frac{r}{1-\xi}\ln\frac{v}{\xi\hp} & \forall v\in[\xi\hp, \omega) \\
1-r\ln(\uv/\hp) & \forall v\in[\omega,\hp) \\
1-r\ln(\uv/v) & \forall v\in[\hp,\uv]
\end{cases}.$$ 
We denote the $q^{\dag}$ the allocation function under conversion rate $\xi^{\dag}$ and denote $q^{\ddag}$ the allocation function under conversion rate $\xi^{\ddag}$. Since  $\hp^{\dag}=\min\{(2-\xi^{\dag})\omega,\uv\}\ge\min\{(2-\xi^{\ddag})\omega,\uv\}=\hp^{\ddag}$, we compare $q$ in the following cases.
\begin{enumerate}
    \item when $v\in[\hp^{\dag},\uv]$, since $r^{\dag}\le r^{\ddag}$, we have that $q^{\dag}(v)=1-r^{\dag}\ln(\uv/v)\ge 1-r^{\ddag}\ln(\uv/v)=q^{\ddag}(v)$.
    \item when $v\in[\hp^{\ddag},\hp^{\dag})$, it is straightforward that $q^{\ddag}(v)\le q^{\ddag}(\hp^{\dag})\le q^{\dag}(\hp^{\dag})=q^{\dag}(v)$.
    \item when $v\in[\omega,\hp^{\ddag})$,  we have that $q^{\dag}(v)=1-r^{\dag}\ln(\uv/\hp^{\dag})\ge  1-r^{\ddag}\ln(\uv/\hp^{\ddag})=q^{\ddag}(v)$ since $r^{\dag}\le r^{\ddag}$ and $\hp^{\dag}\ge \hp^{\ddag}$.
    \item when $v\le \omega$, the derivative of $q(v)$ with respect to $v$ is $\frac{r}{1-\xi}\frac{1}{v}$. Since $r^{\ddag}\ge r^{\dag}$ and $\xi^{\ddag}\ge \xi^{\dag}$, we have that the derivative of $q^{\ddag}$ is greater than that of $q^{\dag}$. Since $q^{\ddag}(\omega)\le q^{\dag}(\omega)$, it is straightforward that $q^{\ddag}(v)\le q^{\dag}(v)$ for any $v\le \omega$.
\end{enumerate}
Therefore, we have shown that for any $v\in[0,\uv]$, $q^{\ddag}(v)\le q^{\dag}(v)$, which completes our proof that the price density function under $\xi^{\ddag}$ stochastically dominates that under  $\xi^{\dag}$ .\QED
\endproof

\subsection{Proofs in \Cref{sec:alternative}}
\begin{proof}{Proof of \Cref{prop:general-mean}.}

Denoting $\mathsf{OBJ}_\alpha(\bpi)$ as $\Delta$, Problem \eqref{general} is equivalent to
\begin{align}
    \sup_{\bpi\in \bPi_n} & \Delta \nonumber\\
    \Delta & \le \inf_{\F\in \cF} \ \int_{0}^{\uv}\bigof{\int_0^v \pi(u)u\, du}\, d\F(v) -\alpha \cdot  p(1-\F(p-)) \quad \forall \ p\in [0,\uv]  \label{eq:general-rhs}
\end{align}
Considering the mean ambiguity set, the right-hand side of the constraint at index $p$ becomes
\begin{align*}
    \inf_{\F\in \cF} \   & \int_{0}^{\uv}\bigof{\int_0^v \pi(u)u\, du}\, d\F(v) -\alpha \cdot  p(1-\F(p-))\\
    \mbox{s.t. } & \int_{0}^{\uv} d\F(v) = 1\\
   &  \int_{0}^{\uv}(v-\mu) d\F(v) = 0
\end{align*}
By Sion's Minimax Theorem, the objective value of this problem is equal to that of its dual problem, i.e.,
\begin{align*}
\sup_{\lm_0,\lm_1} \  &  \lm_0 \\
\mbox{s.t.} \ & \lm_0 + \lm_1(v-\mu) \le \int_0^v \pi(u)u\, du -\alpha p \mathbbm{1}(v\ge p)   \quad \forall v\in [0,\uv]
\end{align*}
Embedding this into the right-hand side of \eqref{eq:general-rhs}, the problem becomes
\begin{align*}
    \sup_{\bpi\in \bPi_n, \lmb } \, & \Delta \nonumber\\
   \mbox{s.t.} \ &  \Delta  \le \int_0^v \pi(u)u\, du  - \alpha p \mathbbm{1}(v\ge p) - \lm(p)\cdot (v-\mu) \quad \forall p\in [0,\uv]\  v\in [0,\uv] 
\end{align*}
When price level set $\cV=\{v_1,\dots,v_n\}$ is fixed, the problem becomes 
\begin{subequations}
\begin{align}
\displaystyle
\addtocounter{equation}{-1}
\displaystyle
    \sup_{\bx\in \R_+^n, \lmb} \  & \Delta \label{eq:general-obj}\\
  \mbox{s.t.} \  &  \Delta  \ \le  \sum_{j=1}^{n}\bigof{v_jx_j\cdot \mathbbm{1}[v\ge v_j] } -\alpha p \mathbbm{1}(v\ge p) - \lm(p)\cdot (v-\mu) \quad \forall p\in [0,\uv]\  v\in [0,\uv] \label{eq:general-rhs2} \\
    & \sum_{j=1}^n x_j = 1
\end{align}
\end{subequations}
 The proof will be completed if we can show that $\lm_1(v_j-)$ is feasible for any $p\in [v_{j-1},v_j-)$ and the constraints are tight only at $v=0$ or $v=v_i-$ for $i=1,2,\dots,n+1$.
\begin{enumerate}
    \item  If $\lm_1(v_j-)\le 0$, then since $\sum_{l=1}^{n}\bigof{v_jx_j\cdot \mathbbm{1}[v\ge v_j] } - \lm_1(v_j-)\cdot (v-\mu) $ is increasing in $v$, the right hand side of \eqref{eq:general-rhs2}, i.e. $-\alpha p \mathbbm{1}(v\ge p) +\sum_{j=1}^{n}\bigof{v_jx_j\cdot \mathbbm{1}[v\ge v_j] } - \lm_1(v_j-)\cdot (v-\mu)$ is minimized at $p$ or $0$. Moreover, the feasibility at $v=0$ implies that $\Delta  \le  0 + \lm_1(v_j-)\cdot \mu$. Hence, for any $p\in [v_{j-1},v_j-)$, we only need to check the feasibility at $v=p$. If $\lm_1(v_j-)\le -\alpha$, then $ -\alpha p +\sum_{l=1}^{j-1}\bigof{v_lx_l} - \lm_1(v_j-)\cdot (p-\mu) =\lm_1(v_j-)\mu  +\sum_{l=1}^{j-1}\bigof{v_lx_l} - (\alpha+\lm_1(v_j-))\cdot p \ge \lm_1(v_j-)\mu \ge \Delta$; if $\lm_1(v_j-)>- \alpha $, then $ -\alpha p +\sum_{l=1}^{j-1}\bigof{v_lx_l} - \lm_1(v_j-)\cdot (p-\mu)=\lm_1(v_j-)\mu  +\sum_{l=1}^{j-1}\bigof{v_lx_l} - (\alpha+\lm_1(v_j-))\cdot p \ge  \lm_1(v_j-)\mu  +\sum_{l=1}^{j-1}\bigof{v_lx_l} - (\alpha+\lm_1(v_j-))\cdot v_j- \ge \Delta$, where the first inequality is due to $v_j-\ge p$ and the second inequality is due to the feasibility at $p = v_j-$. Besides, since $-\alpha p \mathbbm{1}(v\ge p) +\sum_{j=1}^{n}\bigof{v_jx_j\cdot \mathbbm{1}[v\ge v_j] } - \lm_1(v_j-)\cdot (v-\mu)$ is minimized at $p$ or $0$, the constraint can only be tight at $0$ or $v_j-$.
    \item If $\lm_1(v_j-) > 0$, then $-\alpha \cdot p \mathbbm{1}(v\ge p) - \lm_1(v_j-)\cdot (v-\mu)$ is decreasing in $v$, so $-\alpha p \mathbbm{1}(v\ge p) +\sum_{j=1}^{n}\bigof{v_jx_j\cdot \mathbbm{1}[v\ge v_j] } - \lm_1(v_j-)\cdot (v-\mu)$ can only be minimized at $v_i-$, for $i=1,\dots, n+1$. Hence, this $\lm_1(v_j-)$ is also feasible for other $p\in [v_{j-1},v_j-)$ due to the same proof idea in \Cref{thm:finitelp0}.  \QED 
\end{enumerate}
\end{proof}

\begin{proof}{Proof of \Cref{prop:maximin revenue meansupport lp}.}
When $\alpha=0$, after taking the dual of the inner problem, i.e., the right-hand side of problem \eqref{eq:general-rhs}, we have an equivalent formulation as follows.
\begin{align*}
\max_{\bpi, \lambda_0,\lambda_1,\dots, \lambda_K} & \ \lambda_0  \\
\mathrm{s.t.} \quad & 
\lambda_0+\sum_{k=1}^K \phi_k(v) \lambda_k \le \int_0^v \pi(u)u\, du \quad \forall \ v\in [0,\uv]
\end{align*}
Suppose the price level set is $\cV=\{v_1,\dots, v_n\}$ and our decision variables are $\bx=(x_1,\dots, x_n)$. Then an equivalent finite linear programming is as follows.
\begin{align*}
\max_{\bx, \lambda_0,\lambda_1,\dots, \lambda_K} & \ \lambda_0  \\
\mathrm{s.t.} \quad & 
\lambda_0+\sum_{k=1}^K \phi_k(v_j) \lambda_k \le \sum_{i=1}^{j-1} v_ix_i \quad \forall \ j=1,\dots,n+1
\end{align*}
Under the mean ambiguity set where $\phi(v) = v-\mu$, for any fixed $\cV$, the n-level pricing problem is equivalent to
\begin{align*}
\max_{\bx,\lambda_0,\lambda_1\ge 0} & \ \lambda_0  \\
\mathrm{s.t.} \quad & 
\lambda_0+\lambda_1\cdot (v_j-\mu) \le \sum_{i=1}^{j-1} v_ix_i\quad  \forall \ j=1,\dots,n+1
\end{align*} \QED
\end{proof}
\begin{proof}{Proof of \Cref{prop:maximin revenue meansuport closedform}.}
For $j=1$, the constraint is equivalent to $\lambda_0+\lambda_1(v_1-\mu)\le 0$. If $v_1\ge \mu$, then $\lambda_0\le \lambda_1(\mu-v_1)\le 0$, so we need to have $v_1<\mu$. The optimal payment function is a step function that is above a linear function $\lambda_0+\lambda_1\cdot (v_j-\mu) $. In order to maximize $\lambda_0$, the breakpoints of the payment function should be on the linear function $\lambda_0+\lambda_1\cdot (v-\mu) $. The optimal pricing probabilities are solved by the following linear equalities.
\begin{align*}
    0 &= \lambda_0+\lambda_1(v_1-\mu)\\
    v_1x_1 &= \lambda_1(v_2-v_1) \\
    \dots \\
    v_j x_j & = \lambda_1(v_{j+1}-v_j) \\
    \dots \\
    v_nx_n &= \lambda_1(\uv-v_n)
\end{align*}
Thus, we have that the optimal pricing probabilities satisfy $x_j = \lambda_1 \cdot \frac{v_{j+1}-v_j}{v_j}$ and $\lambda_1=\frac{\lambda_0}{\mu-v_1}$. Hence, by $\sum_{j=1}^n x_j =1$, we have that 
\begin{align*}
    1=\lambda_1 \cdot \sum_{j=1}^n \frac{v_{j+1}-v_j}{v_j} = \frac{\lambda_0}{\mu-v_1} \cdot \sum_{j=1}^n \frac{v_{j+1}-v_j}{v_j} 
\end{align*}
Therefore, the worst-case revenue is
$$
\lm_0 = (\mu-v_1) \cdot \bigof{\sum_{j=1}^n \frac{v_{j+1}-v_j}{v_j} }^{-1} =(\mu-v_1) \cdot \bigof{\sum_{j=1}^n \frac{v_{j+1}}{v_j} -n}^{-1} .
$$
For fixed $v_1$, taking the partial derivative of $\sum_{j=1}^n \frac{v_{j+1}}{v_j}$ with respective to $v_j$, we have that the optimal $v_j = \sqrt{v_{j-1}v_{j+1}}$ for all $j=2,3,\dots, n$. Thus, 
\begin{align*}
    \lambda_0 = (\mu-v_1)\cdot n^{-1}\cdot \bigof{ (\uv/v_1)^{1/n} -1 }^{-1}.
\end{align*}
The derivative of $\lm_0$ with respect to $v_1$ is 
\begin{equation*}
\frac{\partial \lm_0}{\partial v_1} =    \frac{v_1^{1/n}\,\left(\mu \,\uv^{1/n}-\uv^{1/n}\,v_1-n\,\uv^{1/n}\,v_1+n\,v_1^{1+1/n}\right)}{n^2 \cdot v_1\,{\left(\uv^{1/n}-v_1^{1/n}\right)}^2}
\end{equation*}
Since the denominator is nonnegative, this derivative is nonnegative if and only if $g_1(v_1)$ is nonnegative, where
\begin{align*}
  g_1(v_1) =  \mu - (n+1)\, v_1 + n v_1^{1+1/n}/\uv^{1/n} 
\end{align*}
Taking the derivative of $g_1$ with respect to $v_1$ again, $\frac{\partial g_1}{\partial v_1} =(n+1)\,((v/\mu)^{1/n}-1)<0$. Thus, $g_1(v_1)$ is decreasing in $v_1$.  Since $g_1(0) = \mu> 0$, $g(\mu) = n\mu ((\mu/\uv)^{1/n}  - 1 )< 0$ and $g_1$  is continuous, we have that there exists a unique solution $v_1^*\in (0,\mu)$ such that $g_1(v_1^*) =0$. Hence, $\lambda_0$ is first increasing in $v_1$ when $v_1\in [0,v_1^*]$ and then decreasing in $v_1$ when $v_1\in [v_1^*,\uv]$. Thus, the optimal pricing policy is 
\begin{align*}
    v_j = v_1\cdot \bigof{\uv/v_1}^{\frac{j-1}{n}}, \, x_j = \frac{1}{n},\quad  \forall j= 1,\dots, n.
\end{align*}
where $v_1$ is the unique solution to $ \frac{\mu}{v} + n \bigof{\frac{v}{\uv}}^{1/n}=n+1 $. 
\end{proof}

\begin{proof}{Proof of \Cref{prop:minimax regret meansupport1}.}
Based on \Cref{prop:general-mean}, the finite linear programming for the 1-level pricing under minimax absolute regret objective is derived as follows.
\begin{subequations}
\begin{align}
\displaystyle
\addtocounter{equation}{-1}
\displaystyle
    \inf_{v_1, \lmb\in \R^2} \  & \Delta \label{eq:regret-mean-1-obj}\\
 \mbox{s.t.}
 &   \Delta  \  \ge  v_1  + \lm_1 \cdot (\mu-v_1)  \label{eq:regret-mean-1-1}\\
  &  \Delta  \  \ge   \lm_1 \mu \label{eq:regret-mean-1-2}\\
   &   \Delta  \  \ge  - \lm_1 \cdot (\uv-\mu) \label{eq:regret-mean-1-3}\\
  &   \Delta  \  \ge   \lm_2 \cdot (\mu - v_1) \label{eq:regret-mean-1-4}\\
 &   \Delta  \  \ge  \uv -v_1 - \lm_2 \cdot (\uv-\mu) \label{eq:regret-mean-1-5}\\
  &  \Delta  \  \ge   \lm_2 \mu \label{eq:regret-mean-1-6}
\end{align}
\end{subequations}
 First, we show that optimal $\lm_1\le 0$. Assuming $\lm_1>0$, we will show that $\lm'_1 = 0$ can only potentially improve the objective. Since $- \lm_1' \cdot (\uv-\mu)=0$, this constraint does not affect the value of $\Delta$. Besides, due to $v_1  + \lm_1' \cdot (\mu-v_1)<v_1  + \lm_1 \cdot (\mu-v_1)$ and $\lm_1' \mu \le \lm_1 \mu $, adjusting $\lm_1>0$ to $\lm_1'=0$ does not affect the optimality of the solution. When $\lm_1\le 0$, constraint \eqref{eq:regret-mean-1-2} is redundant since it is implied by \eqref{eq:regret-mean-1-1} and \eqref{eq:regret-mean-1-3}. Since the right-hand side of \eqref{eq:regret-mean-1-1} is increasing in $\lm_1$ and that of \eqref{eq:regret-mean-1-3} is decreasing in $\lm_1$, in order to minimize $\Delta$, we set $\lm_1 = -\frac{v_1}{\uv-v_1}$. Hence, the first three constraints are implied by $\Delta \ge \frac{v_1\cdot(\uv-\mu)}{\uv-v_1}$  

  Second,  we show that the optimal $\lm_2\ge 0$. Suppose, for contradiction, $\lm_2<0$. Then $\lm_2'=0$ can only improve the objective: (i) $\lm'_2 \cdot (\mu - v_1)=0$ and $\lm_2 \mu =0$ so these two constraints will not affect the value of $\Delta$; (ii) $\uv -v_1 - \lm'_2 \cdot (\uv-\mu)<  \uv -v_1 - \lm_2 \cdot (\uv-\mu)$ so $\Delta$ can potentially decrease if $\lm_2'=0$. Hence, $\lm_2<0$ is suboptimal and can be dominated by $\lm_2'=0$. When $\lm_2\ge 0$, constraint \eqref{eq:regret-mean-1-4} is redundant due to \eqref{eq:regret-mean-1-6}. Thus, in order to minimize $\Delta$, $\lm_2 $ satisfies $\uv -v_1 - \lm_2 \cdot (\uv-\mu) = \lm_2 \mu$ which implies that $\lm_2=\frac{\uv -v_1}{\uv}$. Under this condition, $\Delta \ge \frac{(\uv -v_1)\mu}{\uv}$. 
  
Therefore, the optimal worst-case regret $\Delta =\max\{ \frac{v_1\cdot(\uv-\mu)}{\uv-v_1},\frac{(\uv -v_1)\mu}{\uv}\}$. The optimal $v_1$ is solved as $\frac{\uv\,\left(\mu +\uv-\sqrt{\left(\uv-\mu\right)\,\left(3\,\mu +\uv\right)}\right)}{2\,\mu }$
  \QED 
\end{proof}

\subsection{Detailed Proofs in \Cref{thm:meansupport-2}}
\begin{enumerate}
    \item When $v_2\ge \mu$, we have that $$r=\min\left\{\frac{v_1x_1+v_2x_2}{\uv}, \ \frac{v_1x_1}{v_2},\,\frac{(\mu-v_1)(v_1x_1+v_2x_2)}{v_1(\uv-v_1)}, \, \frac{x_1(\mu-v_1)}{v_2-v_1}\right\},$$ which implies
$$r=\min\left\{\frac{v_1 x_1}{v_2}, \frac{(\mu-v_1)x_1}{v_2-v_1}, \frac{(\mu-v_1)[(v_1-v_2)x_1+v_2]}{v_1(\overline{v}-v_1)}, \frac{(v_1-v_2)x_1+v_2}{\overline{v}} \right\}$$
Since $v_1\leq \mu \leq v_2$, the first two terms $\frac{v_1 x_1}{v_2}$ and $\frac{(\mu-v_1)x_1}{v_2-v_1}$ are increasing in $x_1$, and the last two terms $\frac{(\mu-v_1)[(v_1-v_2)x_1+v_2]}{v_1(\overline{v}-v_1)}$ and $\frac{(v_1-v_2)x_1+v_2}{\overline{v}}$ are decreasing in $v_1$. Note the comparison between the first two terms only depends on whether $\frac{v_1(v_2-v_1) }{v_2(\mu-v_1)}$ is greater than 1, and the comparison between the last two terms only depends on whether $\frac{(\mu-v_1)\uv}{v_1(\overline{v}-v_1)}$ is greater than 1. In addition, the optimal $x_1$ is achieved at the intersection of the smaller one within the two increasing terms with the smaller one within the two decreasing terms.
\begin{enumerate}
    \item Suppose $\frac{v_1}{v_2}\leq \frac{\mu-v_1}{v_2-v_1}$ and $\frac{\mu-v_1}{v_1(\overline{v}-v_1)} \ge \frac{1}{\overline{v}}$.
In this case, the optimal $x_1$ is achieved at the intersection of $\frac{v_1 x_1}{v_2}$ and $\frac{(v_1-v_2)x_1+v_2}{\overline{v}}$. Hence, we have that
$x_1=\frac{v_2^2}{v_2^2+\overline{v}v_1-v_1 v_2}=\frac{1}{1+\frac{v_1}{v_2}(\frac{\overline{v}}{v_2}-1)}\leq 1$.
Therefore, 
\begin{equation*}
    r=\frac{v_1}{v_2}x_1=\frac{v_1 v_2}{v_2^2+\overline{v}v_1-v_1 v_2}.
\end{equation*}

\item Suppose $\frac{v_1}{v_2}\leq \frac{\mu-v_1}{v_2-v_1}$ and $\frac{\mu-v_1}{v_1(\overline{v}-v_1)} < \frac{1}{\overline{v}}$. 
In this case, $x_1$ is obtained at the intersection of $\frac{v_1 x_1}{v_2}$ and $\frac{(\mu-v_1)[(v_1-v_2)x_1+v_2]}{v_1(\overline{v}-v_1)}$. Thus,
$
    x_1=\frac{v_2^2(\mu-v_1)}{v_1^2(\overline{v}-v_1)+v_2(\mu-v_1)(v_2-v_1)}
    =\frac{1}{\frac{v_1^2}{v_2^2}\cdot \frac{\overline{v}-v_1}{\mu-v_1}+1-\frac{v_1}{v_2}}
    =\frac{1}{\frac{v_1}{v_2}(\frac{v_1}{v_2}\cdot \frac{\overline{v}-v_1}{\mu-v_1}-1)+1}
$.
Since $\frac{\mu-v_1}{\overline{v}-v_1}<\frac{v_1}{\overline{v}}$, we have $\frac{v_1}{v_2}\cdot \frac{\overline{v}-v_1}{\mu-v_1}-1>\frac{v_1}{v_2}\cdot \frac{\uv}{v_1}-1=\frac{\overline{v}}{v_2}-1 \ge 0$, so $x_1<1$. Therefore, 
$$r=\frac{v_1}{v_2}\cdot x_1
    =\frac{v_1}{v_2} \cdot \frac{v_2^2(\mu-v_1)}{v_1^2(\overline{v}-v_1)+v_2(\mu-v_1)(v_2-v_1)}=\frac{v_1 v_2 (\mu-v_1)}{v_1^2(\overline{v}-v_1)+v_2(\mu-v_1)(v_2-v_1)}.$$
\item If $\frac{v_1}{v_2}>\frac{\mu-v_1}{v_2-v_1}$ and $\frac{\mu-v_1}{v_1(\overline{v}-v_1)} < \frac{1}{\overline{v}}$, $x_1$ is solved by the intersection of $\frac{(\mu-v_1)x_1}{v_2-v_1}$ and $\frac{(\mu-v_1)[(v_1-v_2)x_1+v_2]}{v_1(\overline{v}-v_1)}$. Thus, we have:
$
    x_1=\frac{v_2(v_2-v_1)}{v_1(\overline{v}-v_1)+(v_2-v_1)^2}
    =\frac{1}{\frac{v_1}{v_2}\cdot \frac{\overline{v}-v_1}{v_2-v_1}+\frac{v_2-v_1}{v_2}}=\frac{1}{\frac{v_1}{v_2}\cdot \frac{\overline{v}-v_2}{v_2-v_1}+1} \leq 1.
$
Therefore, 
\begin{eqnarray*}
\begin{aligned}
    r&=\frac{\mu-v_1}{v_2-v_1}\cdot x_1=\frac{\mu-v_1}{v_2-v_1}\cdot \frac{v_2(v_2-v_1)}{v_1(\overline{v}-v_1)+(v_2-v_1)^2}=\frac{v_2(\mu-v_1)}{v_1(\overline{v}-v_1)+(v_2-v_1)^2}
\end{aligned}
\end{eqnarray*}
\item Suppose $\frac{v_1}{v_2}>\frac{\mu-v_1}{v_2-v_1}$ and $\frac{\mu-v_1}{v_1(\overline{v}-v_1)} \ge \frac{1}{\overline{v}}$. It follows  that
$\frac{v_1}{\overline{v}}(\overline{v}-v_1)\leq\mu-v_1<\frac{v_1}{v_2}(v_2-v_1)$
which implies that $\overline{v}<v_2$, contradicting to our assumption. Hence, this case does not exist.
\end{enumerate}
We summarize the closed-form competitive ratio under different cases as follows.
\begin{equation*}
r=\max
\begin{cases}r_a := \max\limits_{v_1,v_2}\, 
\frac{v_1 v_2}{v_2^2+\overline{v}v_1-v_1 v_2} & \text{s.t. }\frac{v_1}{v_2}\leq \frac{\mu-v_1}{v_2-v_1}, \frac{\mu-v_1}{\overline{v}-v_1}\geq \frac{v_1}{\overline{v}}\\
r_b := \max\limits_{v_1,v_2}\, \frac{v_1 v_2 (\mu-v_1)}{v_1^2(\overline{v}-v_1)+v_2(\mu-v_1)(v_2-v_1)} & \text{s.t. }\frac{v_1}{v_2}\leq \frac{\mu-v_1}{v_2-v_1}, \frac{\mu-v_1}{\overline{v}-v_1}<\frac{v_1}{\overline{v}}\\
r_c := \max\limits_{v_1,v_2}\, \frac{v_2(\mu-v_1)}{v_1(\overline{v}-v_1)+(v_2-v_1)^2} & \text{s.t. }\frac{v_1}{v_2}>\frac{\mu-v_1}{v_2-v_1}, \frac{\mu-v_1}{\overline{v}-v_1}<\frac{v_1}{\overline{v}}
\end{cases}
\end{equation*}

Now based on the closed-form competitive ratio above, we hope to jointly optimize $v_1$ and $v_2$. In the following analysis, we show that the competitive ratios achieved in cases (b) and (c) are dominated by those achieved in case (a). 
\begin{itemize}
    \item For case (a), 
$r=\frac{v_1 v_2}{v_2^2+\overline{v}v_1-v_1 v_2}=\frac{1}{\frac{v_2}{v_1}+\frac{\overline{v}}{v_2}-1}$.
For fixed $v_1$, $r$ is maximized when $v_2=\max\{\mu,\sqrt{v_1 \overline{v}}\}$. Embedding $v_2=\max\{\mu,\sqrt{v_1 \overline{v}}\}$, we have that $r=\min\{\frac{1}{\frac{\mu}{v_1}+\frac{\uv}{\mu}-1}, \,\frac{1}{2\sqrt{\frac{\overline{v}}{v_1}}-1}\}$, which is increasing in $v_1$. 
From $\frac{\mu-v_1}{\overline{v}-v_1}\geq \frac{v_1}{\overline{v}}$, we have $v_1\in [0, \overline{v}-\sqrt{\overline{v}^2-\mu \overline{v}}]$. Now we discuss whether $v_1=\overline{v}-\sqrt{\overline{v}^2-\mu \overline{v}}$ satisfies the following constraints.
$
(1)\, v_1=\overline{v}-\sqrt{\overline{v}^2-\mu \overline{v}}\leq \mu, \,
(2)\, \frac{v_1}{v_2}\leq \frac{\mu-v_1}{v_2-v_1}.
$
First, by $\mu \leq \overline{v}$, it is straightforward to have that $v_1-\mu=\overline{v}-\sqrt{\overline{v}^2-\mu \overline{v}}-\mu=\sqrt{\uv-\mu}(\sqrt{\uv-\mu}-\sqrt{\uv})\le 0$. 
Second, $\frac{v_1}{v_2}\leq \frac{\mu-v_1}{v_2-v_1}$ is equivalent to $(\mu-2v_1)v_2\geq -v_1^2$. Since $v_1=\overline{v}-\sqrt{\overline{v}^2-\mu \overline{v}}\ge\frac{\mu}{2}$, and $\frac{v_1^2}{2v_1-\mu}\ge \mu$, we only need to prove that when $v_1=\overline{v}-\sqrt{\overline{v}^2-\mu \overline{v}}$, $\sqrt{v_1 \overline{v}}\leq \frac{v_1^2}{2v_1-\mu}$, which is equivalent to
$\overline{v}(2v_1-\mu)^2\leq v_1^3$.
Embedding $v_1=\overline{v}-\sqrt{\overline{v}^2-\mu \overline{v}}$, the inequality is simplified to
$4\overline{v}^2-5\mu \overline{v}+\mu^2\leq (4\overline{v}-3\mu)\sqrt{\overline{v}^2-\mu \overline{v}}$.
Since
$(4\overline{v}^2-5\mu \overline{v}+\mu^2)^2-(4\overline{v}-3\mu)^2(\overline{v}^2-\mu \overline{v})=-\mu^3(\overline{v}-\mu)\leq0$,
we have that $v_2=\max\{\mu,\sqrt{v_1 \overline{v}}\}\le\frac{v_1^2}{2v_1-\mu}$. Therefore, the optimal price levels in case (a) is $v_1=\overline{v}-\sqrt{\overline{v}^2-\mu \overline{v}}$, $v_2=\max\left\{\mu, \sqrt{v_1\overline{v}}\right\}$, and the corresponding competitive ratio $r=\min\{\frac{1}{\frac{\mu}{v_1}+\frac{\uv}{\mu}-1}, \,\frac{1}{2\sqrt{\frac{\overline{v}}{v_1}}-1}\}$.

\item For case (b), $r=\frac{v_1 v_2(\mu-v_1)}{v_1^2(\overline{v}-v_1)+v_2(\mu-v_1)(v_2-v_1)}=\frac{1}{\frac{v_1}{v_2}\cdot \frac{\overline{v}-v_1}{\mu-v_1}+\frac{v_2}{v_1}-1}$. For fixed $v_1$,  $r$ is maximized at $v_2=v_1\sqrt{\frac{\overline{v}-v_1}{\mu-v_1}}$. 
Hence, letting $v_2=v_1\sqrt{\frac{\overline{v}-v_1}{\mu-v_1}}$, we have that $r=\frac{1}{2\sqrt{\frac{\overline{v}-v_1}{\mu-v_1}}-1}=\frac{1}{2\sqrt{\frac{\overline{v}-\mu}{\mu-v_1}+1}-1}$, which is decreasing in $v_1$.
Notice that compared to case (a), the only different condition for case (b) is that $\frac{\mu-v_1}{\overline{v}-v_1}<\frac{v_1}{\overline{v}}$, which implies that  $v>\uv-\sqrt{\uv^2-\mu\uv}$. It follows that for case (b), $r$ is maximized when $v_1\to \uv-\sqrt{\uv^2-\mu\uv}$, which is equal to the optimal $v_1$ in case (a).  Therefore, the optimal $r$ in case (b) is bounded from above by that in case (a).
\item For case (c), given that $\frac{v_1}{v_2}>\frac{\mu-v_1}{v_2-v_1}$, we have
$
r_c=\frac{v_2(\mu-v_1)}{v_1(\overline{v}-v_1)+(v_2-v_1)^2}
=\frac{v_1v_2(\mu-v_1)}{v_1^2(\overline{v}-v_1)+v_1(v_2-v_1)^2}
<\frac{v_1v_2(\mu-v_1)}{v_1^2(\overline{v}-v_1)+v_2(\mu-v_1)(v_2-v_1)}
$.
Therefore, we know that $r_c<r_b$ when $v_1$ and $v_2$ take the values from the feasible region in case (c). From case (b), we know $r_b$ takes the optimum at  $v_1=\overline{v}-\sqrt{\overline{v}^2-\mu \overline{v}}$ and then the constraint for $v_2$ is $\mu<v_2\leq \frac{v_1^2}{2v_1-\mu}=\overline{v}$. Hence, the maximum value of $r_c$ is less than the maximum value of $r_b$. Thus, we have that $r_c\le r_b\le r_a$.

\end{itemize}
Summarizing the cases above, we have that when $v_2\ge \mu$, the optimal $v_1=\overline{v}-\sqrt{\overline{v}^2-\mu \overline{v}}$, $v_2=\max\left\{\mu, \sqrt{v_1\overline{v}}\right\}$, and the corresponding competitive ratio $r=\min\{\frac{1}{\frac{\mu}{v_1}+\frac{\uv}{\mu}-1}, \,\frac{1}{2\sqrt{\frac{\overline{v}}{v_1}}-1}\}$.

\item 
When $v_2<\mu$, we have:
$$r=\min\left\{\frac{\mu-v_2}{\overline{v}-v_2}x_2+\frac{v_1 x_1}{v_2}, \frac{(\mu-v_1)(v_1 x_1+v_2 x_2)}{v_1(\overline{v}-v_1)}, \frac{v_1 x_1+v_2 x_2}{\overline{v}} \right\}$$
Since $x_2=1-x_1$, we have:
$$r=\min\left\{(\frac{v_1}{v_2}-\frac{\mu-v_2}{\overline{v}-v_2})x_1+\frac{\mu-v_2}{\overline{v}-v_2}, \frac{(\mu-v_1)[(v_1-v_2)x_1+v_2]}{v_1(\overline{v}-v_1)}, \frac{(v_1-v_2)x_1+v_2}{\overline{v}}\right\}$$
Similar to the first scenario, the monotonicity of all the terms can be separately discussed in the following cases.
\begin{enumerate}
    \item Suppose $\frac{v_1}{v_2}\leq \frac{\mu-v_2}{\overline{v}-v_2}$. In this case, all of the three terms are decreasing in $x_1$, so the optimal $x_1=0$ and  
$r=\min\left\{\frac{\mu-v_2}{\overline{v}-v_2}, \frac{v_2}{v_1}\cdot \frac{\mu-v_1}{\overline{v}-v_1}, \frac{v_2}{\overline{v}} \right\}$.
Note that $f=\frac{\mu-t}{\overline{v}-t}=1-\frac{\overline{v}-\mu}{\overline{v}-t}$ is decreasing in $t$, so $\frac{\mu-v_2}{\overline{v}-v_2}\leq \frac{\mu-v_1}{\overline{v}-v_1} \leq \frac{v_2}{v_1} \cdot \frac{\mu-v_1}{\overline{v}-v_1}$. Therefore,
$r=\min\left\{\frac{\mu-v_2}{\overline{v}-v_2}, \frac{v_2}{\overline{v}}\right\}$.
Since $\frac{\mu-v_2}{\overline{v}-v_2}$ is decreasing in $v_2$ and $\frac{v_2}{\overline{v}}$ is increasing in $v_2$, the optimal $v_2$ is equal to $\overline{v}-\sqrt{\overline{v}^2-\mu \overline{v}}<\mu$, and then 
$r=\frac{\overline{v}-\sqrt{\overline{v}^2-\mu \overline{v}}}{\overline{v}} $.
\item Suppose $\frac{v_1}{v_2}>\frac{\mu-v_2}{\overline{v}-v_2}$ and $\frac{\mu-v_1}{\overline{v}-v_1} \geq \frac{v_1}{\overline{v}}$. In this case, the first term is increasing in $x_1$ and the rest of the terms are decreasing in $x_1$. Besides, since $\frac{(\mu-v_1)(v_1-v_2)}{v_1(\overline{v}-v_1)}\leq \frac{v_1-v_2}{\overline{v}}$, $x_1$ is determined by the intersection of $(\frac{v_1}{v_2}-\frac{\mu-v_2}{\overline{v}-v_2})x_1+\frac{\mu-v_2}{\overline{v}-v_2}$ and $\frac{(v_1-v_2)x_1+v_2}{\overline{v}}$. Letting the two terms be equal, we have:
$
x_1=\frac{v_2[v_2(\overline{v}-v_2)-\overline{v}(\mu-v_2)]}{\overline{v}[v_1(\overline{v}-v_2)-v_2(\mu-v_2)]+v_2(v_2-v_1)(\overline{v}-v_2)}
=\frac{v_2(2v_2\overline{v}-v_2^2-\mu \overline{v})}{(v_2-v_1)(2v_2\overline{v}-v_2^2-\mu \overline{v})+v_1\overline{v}(\overline{v}-\mu)}
$.
Now we prove that $x_1\le 1$. Since $\frac{v_1}{v_2}>\frac{\mu-v_2}{\overline{v}-v_2}$, the denominator of the expression for $x_1$ is positive. Thus, $x_1\le 1$ is equivalent to 
$v_2[v_2(\overline{v}-v_2)-\overline{v}(\mu-v_2)]\le \overline{v}[v_1(\overline{v}-v_2)-v_2(\mu-v_2)]+v_2(v_2-v_1)(\overline{v}-v_2)$.
Collecting the terms, we have $v_1(\overline{v}-v_2)^2\ge 0$. This inequality always holds, so $x_1\le 1$.
Since $x_1\le 1$, we have:
\begin{equation*}
r=\frac{(v_1-v_2)x_1+v_2}{\overline{v}}=\frac{v_1 v_2(\overline{v}-\mu)}{(v_2-v_1)(2v_2\overline{v}-v_2^2-\mu \overline{v})+v_1\overline{v}(\overline{v}-\mu)}
\end{equation*}
\item Suppose $\frac{v_1}{v_2}>\frac{\mu-v_2}{\overline{v}-v_2}$ and $\frac{\mu-v_1}{\overline{v}-v_1} < \frac{v_1}{\overline{v}}$. Since $\frac{(\mu-v_1)(v_1-v_2)}{v_1(\overline{v}-v_1)}>\frac{v_1-v_2}{\overline{v}}$, $x_1$ is determined by the intersection of $(\frac{v_1}{v_2}-\frac{\mu-v_2}{\overline{v}-v_2})x_1+\frac{\mu-v_2}{\overline{v}-v_2}$ and $\frac{(\mu-v_1)[(v_1-v_2)x_1+v_2]}{v_1(\overline{v}-v_1)}$. Making the two terms equal, we have:
$
x_1=\frac{v_2[v_2(\overline{v}-v_2)(\mu-v_1)-v_1(\overline{v}-v_1)(\mu-v_2)]}{v_1(\overline{v}-v_1)[v_1(\overline{v}-v_2)-v_2(\mu-v_2)]+v_2(v_2-v_1)(\overline{v}-v_2)(\mu-v_1)}
=\frac{v_2[v_2(\overline{v}-v_2)(\mu-v_1)-v_1(\overline{v}-v_1)(\mu-v_2)]}{v_1^2(\overline{v}-v_2)(\overline{v}-v_1)-v_1 v_2(\overline{v}-v_1)(\mu-v_2)+v_2^2(\overline{v}-v_2)(\mu-v_1)-v_1 v_2(\overline{v}-v_2)(\mu-v_1)}=\frac{1}{1+\frac{v_1(\overline{v}-v_2)[v_1(\overline{v}-v_1)-v_2(\mu-v_1)]}{v_2[v_2(\overline{v}-v_2)(\mu-v_1)-v_1(\overline{v}-v_1)(\mu-v_2)]}}
$
Since $\frac{\mu-v_1}{\overline{v}-v_1}<\frac{v_1}{\overline{v}}$, we have $v_1(\overline{v}-v_1)>\overline{v}(\mu-v_1)>v_2(\mu-v_1)$. Thus, $v_1(\overline{v}-v_2)[v_1(\overline{v}-v_1)-v_2(\mu-v_1)]>0$.
In addition, since
$
v_2(\overline{v}-v_2)(\mu-v_1)-v_1(\overline{v}-v_1)(\mu-v_2)
=(v_2-v_1)[\mu(\overline{v}-v_1)-v_2(\mu-v_1)]
>v_2(v_2-v_1)(\overline{v}-\mu)\geq 0
$, we have that $x_1<1$. Therefore, 

\begin{eqnarray*}
\begin{aligned}
r=\frac{\mu-v_1}{v_1(\overline{v}-v_1)} \cdot [(v_1-v_2)x_1+v_2]=\frac{v_1 v_2(\overline{v}-\mu)(\mu-v_1)}{v_1^2(\overline{v}-v_2)(\overline{v}-v_1)-v_1 v_2(\overline{v}-v_1)(\mu-v_2)+v_2^2(\overline{v}-v_2)(\mu-v_1)-v_1 v_2(\overline{v}-v_2)(\mu-v_1)}
\end{aligned}
\end{eqnarray*}
\normalsize
\end{enumerate}
Now we optimize $v_1,v_2$ in cases (a), (b) and (c). From the following analysis, we conclude that the competitive ratios in case (a) and (c) are dominated by the competitive ratio defined in case (b).
\begin{itemize}
    \item For case (a), $r=\frac{\overline{v}-\sqrt{\overline{v}^2-\mu \overline{v}}}{\overline{v}}$. Now we compare the competitive ratio in case (a) with that in case (b). We hope to show that the ratio in case (a), $\frac{\overline{v}-\sqrt{\overline{v}^2-\mu \overline{v}}}{\overline{v}}$ is less than or equal to that in case (b), $ \frac{v_1}{\frac{(v_2-v_1)[v_2(\overline{v}-v_2)-\overline{v}(\mu-v_2)]}{v_2(\overline{v}-\mu)}+\frac{v_1\overline{v}}{v_2}}$, which is implied by
$\frac{(v_2-v_1)[v_2(\overline{v}-v_2)-\overline{v}(\mu-v_2)]}{v_2(\overline{v}-\mu)}+\frac{v_1\overline{v}}{v_2}<\overline{v}$.
This is equivalent to $(\overline{v}-v_2)^2>0$, which always holds.
    \item For case (b), we have
$r=\frac{v_1 v_2(\overline{v}-\mu)}{(v_2-v_1)(2v_2\overline{v}-v_2^2-\mu \overline{v})+v_1\overline{v}(\overline{v}-\mu)}=\frac{v_2(\overline{v}-\mu)}{\frac{v_2(2v_2\overline{v}-v_2^2-\mu \overline{v})}{v_1}+(\overline{v}-v_2)^2}$.
In case (b), to make $x_1\geq0$, we have $2v_2\overline{v}-v_2^2-\mu\overline{v}\geq0$. 
Thus, for fixed $v_2\ge v_1$, $r$ is increasing in $v_1$. From $\frac{\mu-v_1}{\overline{v}-v_1}\geq \frac{v_1}{\overline{v}}$, we have $v_1\in [0, \overline{v}-\sqrt{\overline{v}^2-\mu \overline{v}}]$, so the optimal $v_1=\overline{v}-\sqrt{\overline{v}^2-\mu \overline{v}}$. 
 We provide an equivalent representation of $r$ that $r=\frac{v_1(\overline{v}-\mu)}{-v_2^2+(2\overline{v}+v_1)v_2-\overline{v}(\mu+2v_1)+\frac{v_1\overline{v}^2}{v_2}}$. 
 If we denote the denominator by $f=-v_2^2+(2\overline{v}+v_1)v_2-\overline{v}(\mu+2v_1)+\frac{v_1\overline{v}^2}{v_2}$, then the derivative of $f$ with respect to $v_2$ is 
$f^{'}=\frac{-2v_2^3+(2\overline{v}+v_1)v_2^2-v_1\overline{v}^2}{v_2^2}=-\frac{(v_2-\overline{v})(2v_2^2-v_1 v_2-v_1 \overline{v})}{v_2^2}$.
Thus, $f$ is decreasing in $v_2\in[0, \frac{v_1+ \sqrt{v_1^2+8v_1\overline{v}}}{4}]$ and increasing in $v_2\in[\frac{v_1+ \sqrt{v_1^2+8v_1\overline{v}}}{4}, \overline{v}]$. Hence $r$ is maximized at $v_2=\frac{v_1+ \sqrt{v_1^2+8v_1\overline{v}}}{4}$. 
Now we prove that $v_1=\overline{v}-\sqrt{\overline{v}^2-\mu \overline{v}}$, $v_2=\frac{v_1+ \sqrt{v_1^2+8v_1\overline{v}}}{4}$ satisfy $\frac{v_1}{v_2}>\frac{\mu-v_2}{\overline{v}-v_2}$, which is equivalent to prove $v_2^2-(v_1+\mu)v_2+v_1\overline{v}>0$. When $v_1=\overline{v}-\sqrt{\overline{v}^2-\mu \overline{v}}$, we have $
\Delta=(v_1+\mu)^2-4v_1\overline{v}
=-2\overline{v}^2+\mu \overline{v}+\mu^2+2(\overline{v}-\mu)\sqrt{\overline{v}^2-\mu \overline{v}}=-(\overline{v}-\mu)[2\overline{v}+\mu-2\sqrt{\overline{v}^2-\mu \overline{v}}]<0$.
Hence $v_2^2-(v_1+\mu)v_2+v_1\overline{v}>0$. Therefore, in this case , the optimal solution becomes $v_1=\overline{v}-\sqrt{\overline{v}^2-\mu \overline{v}}$, $v_2=\frac{v_1+ \sqrt{v_1^2+8v_1\overline{v}}}{4}$, $r=\frac{v_1 v_2(\overline{v}-\mu)}{(v_2-v_1)(2v_2\overline{v}-v_2^2-\mu \overline{v})+v_1\overline{v}(\overline{v}-\mu)}$.

    \item For case (c), we have that \begin{eqnarray*}
\begin{aligned}
r&=\frac{v_1 v_2(\overline{v}-\mu)(\mu-v_1)}{v_1^2(\overline{v}-v_2)(\overline{v}-v_1)-v_1 v_2(\overline{v}-v_1)(\mu-v_2)+v_2^2(\overline{v}-v_2)(\mu-v_1)-v_1 v_2(\overline{v}-v_2)(\mu-v_1)}\\
&=\frac{v_1 v_2(\overline{v}-\mu)(\mu-v_1)}{(v_2-v_1)[v_2(\overline{v}-v_2)(\mu-v_1)]+v_1(\overline{v}-v_1)[v_1(\overline{v}-v_2)-v_2(\mu-v_2)]}\\
&=\frac{v_1 v_2(\overline{v}-\mu)}{(v_2-v_1)[v_2(\overline{v}-v_2)]+\frac{v_1(\overline{v}-v_1)}{\mu-v_1}\cdot[v_1(\overline{v}-v_2)-v_2(\mu-v_2)]}
\end{aligned}
\end{eqnarray*}
Since $\frac{\mu-v_1}{\overline{v}-v_1}<\frac{v_1}{\overline{v}}$, we have:
\begin{eqnarray*}
r<\frac{v_1 v_2(\overline{v}-\mu)}{(v_2-v_1)[v_2(\overline{v}-v_2)]+\overline{v}[v_1(\overline{v}-v_2)-v_2(\mu-v_2)]}=\frac{v_1 v_2(\overline{v}-\mu)}{(v_2-v_1)(2v_2\overline{v}-v_2^2-\mu \overline{v})+v_1\overline{v}(\overline{v}-\mu)}
\end{eqnarray*}
Then for fixed $v_2$, the partial derivative of $r$ with respect to $v_1$ is
\begin{equation*}
\frac{\partial r}{\partial v_1}=\frac{v_2(\overline{v}-\mu)}{y^2}\cdot \left\{v_1^2(\mu-\overline{v})[v_1(\overline{v}-v_2)-v_2(\mu-v_2)]+(\mu-v_1)(\overline{v}-v_2)[v_2^2(\mu-v_1)-v_1^2(\overline{v}-v_1)] \right\}
\end{equation*}
where $y=v_1^2(\overline{v}-v_2)(\overline{v}-v_1)-v_1 v_2(\overline{v}-v_1)(\mu-v_2)+v_2^2(\overline{v}-v_2)(\mu-v_1)-v_1 v_2(\overline{v}-v_2)(\mu-v_1)$. Now we prove that $r$ is decreasing in $v_1$, i.e., $\frac{\partial r}{\partial v_1}\le 0$. If this is true, then the optimal $v_1=\overline{v}-\sqrt{\overline{v}^2-\mu \overline{v}}$, equal to $v_1$ in case (b), which implies that $r$ in case (c) is no greater than $r$ in case (b).
Let $h=v_1^2(\mu-\overline{v})[v_1(\overline{v}-v_2)-v_2(\mu-v_2)]+(\mu-v_1)(\overline{v}-v_2)[v_2^2(\mu-v_1)-v_1^2(\overline{v}-v_1)]$, which is the second term in $\frac{\partial r}{\partial v_1}$. In order to prove $\frac{\partial r}{\partial v_1}\le 0$, we hope to show $h\le 0$ for all $v_1,v_2,\mu$ in the feasible region. 
Since
\begin{eqnarray*}
\begin{aligned}
\frac{\partial h}{\partial \mu}=-[{v_{2}}^2\,\left(\mu -v_{1}\right)+{v_{1}}^2\,\left(v_{1}-\bar{v}\right)](v_{2}-\bar{v})-{v_{1}}^2[v_{2}(\mu -v_{2})+v_{1}(v_{2}-\bar{v})]
-{v_{1}}^2\,v_{2}\,\left(\mu -\bar{v}\right)-{v_{2}}^2\,\left(\mu -v_{1}\right)\,\left(v_{2}-\bar{v}\right)
\end{aligned}
\end{eqnarray*}
which is a linear function of $\mu$ with coefficient $-v_2^2(v_2-\bar{v})-v_1^2v_2-v_1^2v_2-v_2^2(v_2-\bar{v})=2v_2((\bar{v}-v_2)v_2-v_1^2)$ and a constant term $v_{1}\,\left(-2\,{v_{1}}^2\,v_{2}+2\,{v_{1}}^2\,\bar{v}+v_{1}\,{v_{2}}^2+2\,v_{1}\,v_{2}\,\bar{v}-v_{1}\,{\bar{v}}^2+2\,{v_{2}}^3-2\,{v_{2}}^2\,\bar{v}\right)$. Now we prove that $h\le 0$ in the following two scenarios.
\begin{itemize}
    \item If $2v_2((\bar{v}-v_2)v_2-v_1^2)\ge 0$, then $h$ is first decreasing and then increasing in $\mu$. Then the maximum point takes at either $\mu=v_1$ or $\mu = 2v_1-\frac{v_1^2}{\bar{v}}$. First, $\mu=v_1$, we have that 
$h={v_{1}}^2(v_{1}-\bar{v})({v_{2}}^2-2v_{1}v_{2}+v_{1}\bar{v}) = {v_{1}}^2(v_{1}-\bar{v})\,({v_{2}}^2-2\,v_{1}\,v_{2}+v_{1}^2 + v_1(\bar{v}-v_1))\le 0$.
Second, since $\mu\le 2v_1-\frac{v_1^2}{\bar{v}}$, when 
$\mu$ takes the maximum value, we have 
$h = -\frac{{v_{1}}^2\,{\left(v_{1}-\bar{v}\right)}^2\,\left({v_{1}}^2\,v_{2}-4\,v_{1}\,v_{2}\,\bar{v}+2\,v_{1}\,{\bar{v}}^2+{v_{2}}^3\right)}{{\bar{v}}^2}$.
Now we only need to show that $\left({v_{1}}^2\,v_{2}-4\,v_{1}\,v_{2}\,\bar{v}+2\,v_{1}\,{\bar{v}}^2+{v_{2}}^3\right)\ge 0$, which is proved by
\begin{eqnarray*}
 v_1^2 v_2-4v_1 v_2 \overline{v}+2v_1\overline{v}^2+v_2^3
=v_2(v_1^2+v_2^2-2v_1v_2)+2v_2^2 v_1-4v_1 v_2 \overline{v}+2v_1\overline{v}^2
=v_2(v_2-v_1)^2+2v_1(\overline{v}-v_2)^2\geq 0.
\end{eqnarray*}

\item If $2v_2((\bar{v}-v_2)v_2-v_1^2)< 0$, then it means that the derivative of $\mu$ is decreasing in $\mu$. 
When $\mu =\bar{v}$, we have that $\frac{\partial h}{\partial \mu} =(v_2 - \overline{v})(-2v_1^3 + v_1^2v_2 + \overline{v}v_1^2 + 2v_1v_2^2 - 2\overline{v}v_2^2)
$.
Now we show that $\frac{\partial h}{\partial \mu} =(v_2 - \overline{v})(-2v_1^3 + v_1^2v_2 + \overline{v}v_1^2 + 2v_1v_2^2 - 2\overline{v}v_2^2)$ is always nonnegative when $(\bar{v}-v_2)v_2-v_1^2< 0$:
{\footnotesize
\begin{eqnarray*}
 -2v_1^3+v_1^2 v_2+\overline{v}v_1^2+2v_1 v_2^2-2\overline{v}v_2^2
=v_1^2(\overline{v}+v_2-2v_1)-2v_2^2(\overline{v}-v_1)
\leq2v_1^2(\overline{v}-v_1)-2v_2^2(\overline{v}-v_1)
=2(\overline{v}-v_1)(v_1^2-v_2^2)\leq 0.
\end{eqnarray*}}
Then since $\frac{\partial h}{\partial \mu}$ is decreasing in $\mu$, it implies that $\frac{\partial h}{\partial \mu}\ge 0$ for all feasible $\mu$. Thus, we just need to verify the $h\le 0$ value when $\mu = 2v_1-\frac{v_1^2}{\bar{v}}$, which is the same case in the previous case. 
\end{itemize}

Now that we have proved that the competitive ratio $r$ is decreasing in $v_1$, the optimal $v_1=\overline{v}-\sqrt{\overline{v}^2-\mu \overline{v}}$, which is equal to the $v_1$ in case (b). This implies that $r$ in case (c) is no greater than $r$ in case (b).
\end{itemize}
\end{enumerate}
Now we summarize the discussions above. First, when 
$v_1=\overline{v}-\sqrt{\overline{v}^2-\mu\overline{v}}$, then $v_2=\sqrt{v_1\overline{v}}\geq \mu$ if and only if $\mu\leq\frac{(\sqrt{5}-1)\overline{v}}{2}$. Thus, when $\mu\leq\frac{(\sqrt{5}-1)\overline{v}}{2}$, if $v_2\ge \mu$, then the optimal $r$ is $\frac{1}{2\sqrt{\frac{\overline{v}}{v_1}}-1}$.
 Second, $v_2=\frac{v_1+\sqrt{v_1^2+8v_1\overline{v}}}{4}\le \mu$ if and only if $\mu\ge \frac{(\sqrt{17}-1)\overline{v}}{8}$. 
 Thus, when $\mu\ge \frac{(\sqrt{17}-1)\overline{v}}{8}$, if $v_2\le \mu$, then the optimal $r$ is $\frac{v_1 v_2(\overline{v}-\mu)}{(v_2-v_1)(2v_2\overline{v}-v_2^2-\mu \overline{v})+v_1\overline{v}(\overline{v}-\mu)}$. Then we discuss whether we should choose $v_2\ge \mu$ or $v_2\le \mu$. 
 When comparing the competitive ratios achieved under these two conditions, we find that 
$\frac{1}{2\sqrt{\frac{\overline{v}}{v_1}}-1} \geq \frac{v_1 v_2(\overline{v}-\mu)}{(v_2-v_1)(2v_2\overline{v}-v_2^2-\mu \overline{v})+v_1\overline{v}(\overline{v}-\mu)}$
if and only if
$\mu\leq 0.49 \overline{v}$. 
Notice that $\frac{(\sqrt{17}-1)\overline{v}}{8}< 0.49 \overline{v}  < \frac{(\sqrt{5}-1)\overline{v}}{2}$, which implies that when $\mu\leq 0.49 \overline{v}$, the optimal $v_2 = \sqrt{v_1\overline{v}}$ and $r= \frac{1}{2\sqrt{\frac{\overline{v}}{v_1}}-1}$, and when $\mu> 0.49 \overline{v}$, the optimal $v_2 = \frac{v_1+\sqrt{v_1^2+8v_1\overline{v}}}{4}$ and $r= \frac{v_1 v_2(\overline{v}-\mu)}{(v_2-v_1)(2v_2\overline{v}-v_2^2-\mu \overline{v})+v_1\overline{v}(\overline{v}-\mu)}$. Therefore, we have that the optimal competitive ratio in the 2-level pricing problem is
$r=\max\left\{\frac{1}{2\sqrt{\frac{\overline{v}}{v_1}}-1}, \frac{v_1 v_2(\overline{v}-\mu)}{(v_2-v_1)(2v_2\overline{v}-v_2^2-\mu \overline{v})+v_1\overline{v}(\overline{v}-\mu)}\right\}$, and the optimal second price $v_2 = \sqrt{v_1\overline{v}}$ when $\mu \le 0.49\uv$ and $v_2=\frac{v_1+\sqrt{v_1^2+8v_1\overline{v}}}{4}$ otherwise, which completes our proof. \hfill\QED

\end{document}